\PassOptionsToPackage{square,comma,numbers,sort&compress}{natbib} 
\documentclass[nohyperref]{article}

\usepackage{microtype}
\usepackage{booktabs} 
\usepackage{caption}
\usepackage{subcaption}
\usepackage{multirow}
\usepackage{url}
\usepackage{graphicx}
\usepackage{wrapfig}
\usepackage{lipsum}
\usepackage{hyperref}
\usepackage{tikz}

\usepackage{mathtools}
\usepackage{lineno}

\usepackage{color}
\usepackage{colortbl}

\usepackage[T1]{fontenc}

\newcommand{\benr}{\begin{eqnarray}}
\newcommand{\eenr}{\end{eqnarray}}
\newcommand{\benrr}{\begin{eqnarray*}}
\newcommand{\eenrr}{\end{eqnarray*}}
\newcommand{\ben}{\begin{equation}}
\newcommand{\een}{\end{equation}}
\newcommand{\benn}{\begin{equation*}}
\newcommand{\eenn}{\end{equation*}}

\usepackage{hyperref}

\usepackage[accepted]{icml2022}
\setcitestyle{square,numbers,comma}

\usepackage{amsmath}
\usepackage{amssymb}
\usepackage{mathtools}
\usepackage{amsthm}

\usepackage[capitalize,noabbrev]{cleveref}

\theoremstyle{plain}

\theoremstyle{definition}

\theoremstyle{remark}

\usepackage{amsmath,amsfonts,bm}
\usepackage{amssymb}

\def\eqref#1{equation~\ref{#1}}

\def\1{\bm{1}}

\def\rr{{\textnormal{r}}}
\def\rs{{\textnormal{s}}}

\def\ru{{\textnormal{u}}}

\def\rw{{\textnormal{w}}}

\def\ry{{\textnormal{y}}}
\def\rz{{\textnormal{z}}}

\def\rvu{{\mathbf{i}}}

\def\rvs{{\mathbf{s}}}

\def\rvu{{\mathbf{u}}}

\def\rvw{{\mathbf{w}}}

\def\rvz{{\mathbf{z}}}

\def\rmQ{{\mathbf{Q}}}

\DeclareMathAlphabet{\mathsfit}{\encodingdefault}{\sfdefault}{m}{sl}
\SetMathAlphabet{\mathsfit}{bold}{\encodingdefault}{\sfdefault}{bx}{n}

\def\gA{{\mathcal{A}}}

\def\gC{{\mathcal{C}}}
\def\gD{{\mathcal{D}}}

\def\gF{{\mathcal{F}}}
\def\gG{{\mathcal{G}}}

\def\gP{{\mathcal{P}}}
\def\gQ{{\mathcal{Q}}}

\def\gS{{\mathcal{S}}}

\def\gU{{\mathcal{U}}}

\def\gZ{{\mathcal{Z}}}

\def\sN{{\mathbb{N}}}

\newcommand{\E}{\mathbb{E}}

\newcommand{\R}{\mathbb{R}}

\newcommand{\Var}{\mathrm{Var}}

\DeclareMathOperator*{\argmax}{arg\,max}
\DeclareMathOperator*{\argmin}{arg\,min}

\newcommand{\bbox}{\hfill $\Box$}

\usepackage[textsize=tiny]{todonotes}

\icmltitlerunning{Policy Diagnosis via Measuring Role Diversity in Cooperative Multi-agent RL}

\begin{document}

\newcommand{\xd}[1]{\textcolor{red}{\em (xd: #1)}}
\newcommand{\xcl}{\textcolor{blue}}

\twocolumn[
\icmltitle{Policy Diagnosis via Measuring Role Diversity in Cooperative Multi-agent RL}

\icmlsetsymbol{equal}{*}

\begin{icmlauthorlist}
\icmlauthor{Siyi Hu}{yyy,uts}
\icmlauthor{Chuanlong Xie}{bnu,noah}
\icmlauthor{Xiaodan Liang}{sch}
\icmlauthor{Xiaojun Chang}{uts}
\end{icmlauthorlist}

\icmlaffiliation{yyy}{Monash University}
\icmlaffiliation{uts}{The ReLER Lab, University of Technology Sydney}
\icmlaffiliation{bnu}{Beijing Normal University}
\icmlaffiliation{noah}{Huawei Noah's Ark Lab}
\icmlaffiliation{sch}{Sun Yat-sen University}

\icmlcorrespondingauthor{Xiaojun Chang}{Xiaojun.Chang@uts.edu.au}

\icmlkeywords{Machine Learning, ICML}

\vskip 0.3in
]

\printAffiliationsAndNotice{}  

\begin{abstract}
Cooperative multi-agent reinforcement learning (MARL) is making rapid progress for solving tasks in a grid world and real-world scenarios, 
in which agents are given different attributes and goals, resulting in different behavior through the whole multi-agent task. 
In this study, we quantify the agent's behavior difference and build its relationship with the policy performance via {\bf Role Diversity}, a metric to measure the characteristics of MARL tasks. We define role diversity from three perspectives: action-based, trajectory-based, and contribution-based to fully measure a multi-agent task. Through theoretical analysis, we find that the error bound in MARL can be decomposed into three parts that have a strong relation to the role diversity. The decomposed factors can significantly impact policy optimization on three popular directions including parameter sharing, communication mechanism, and credit assignment. 
The main experimental platforms are based on {\bf Multiagent Particle Environment (MPE) }and {\bf The StarCraft Multi-Agent Challenge (SMAC).
Extensive experiments} 
clearly show that role diversity can serve as a robust measurement for the characteristics of a multi-agent cooperation task and help diagnose whether the policy fits the current multi-agent system for a better policy performance.
\end{abstract}

\vspace{-7pt}
\section{Introduction}
\vspace{-3pt}
\begin{figure*}[!h]
    \centering 
\begin{subfigure}{0.48\textwidth}
  \includegraphics[width=\linewidth]{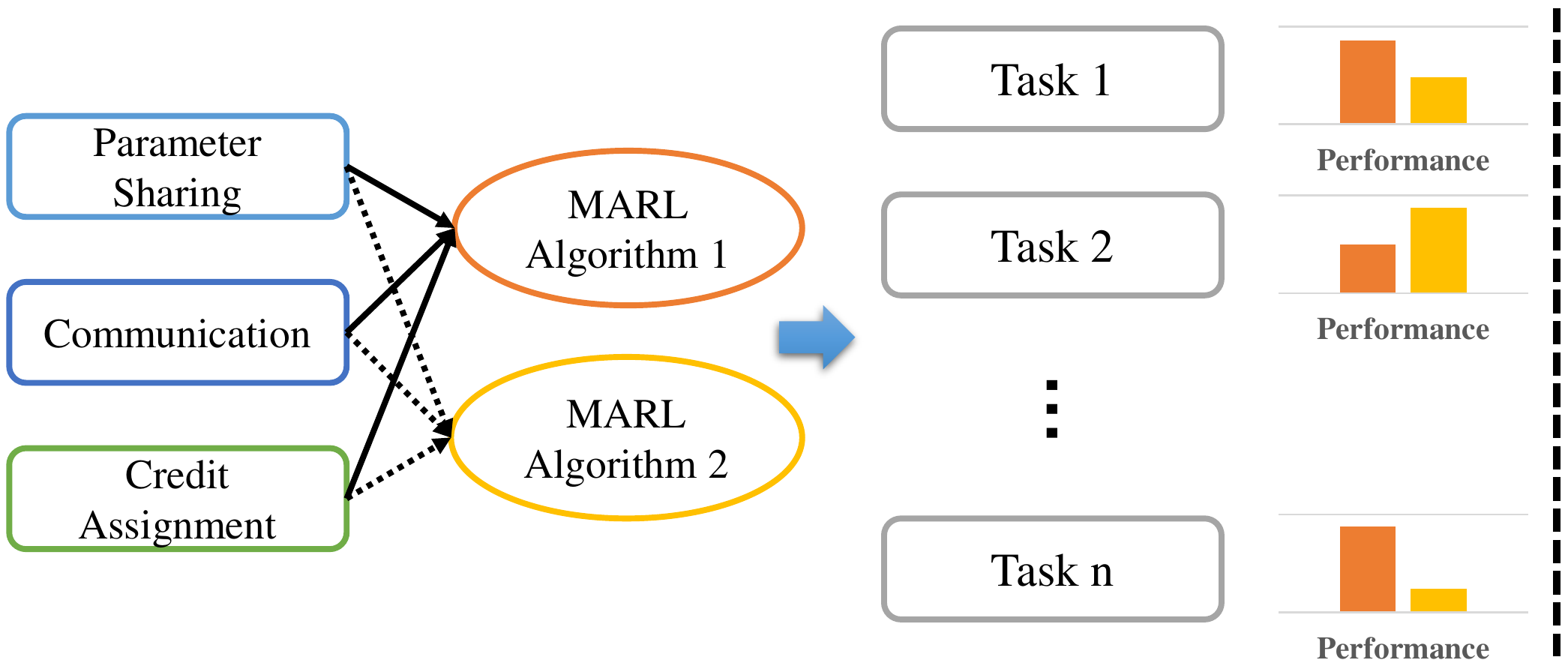}
  \caption{}
   \label{marl_framework:1}
\end{subfigure}\hfil 
\begin{subfigure}{0.48\textwidth}
  \includegraphics[width=\linewidth]{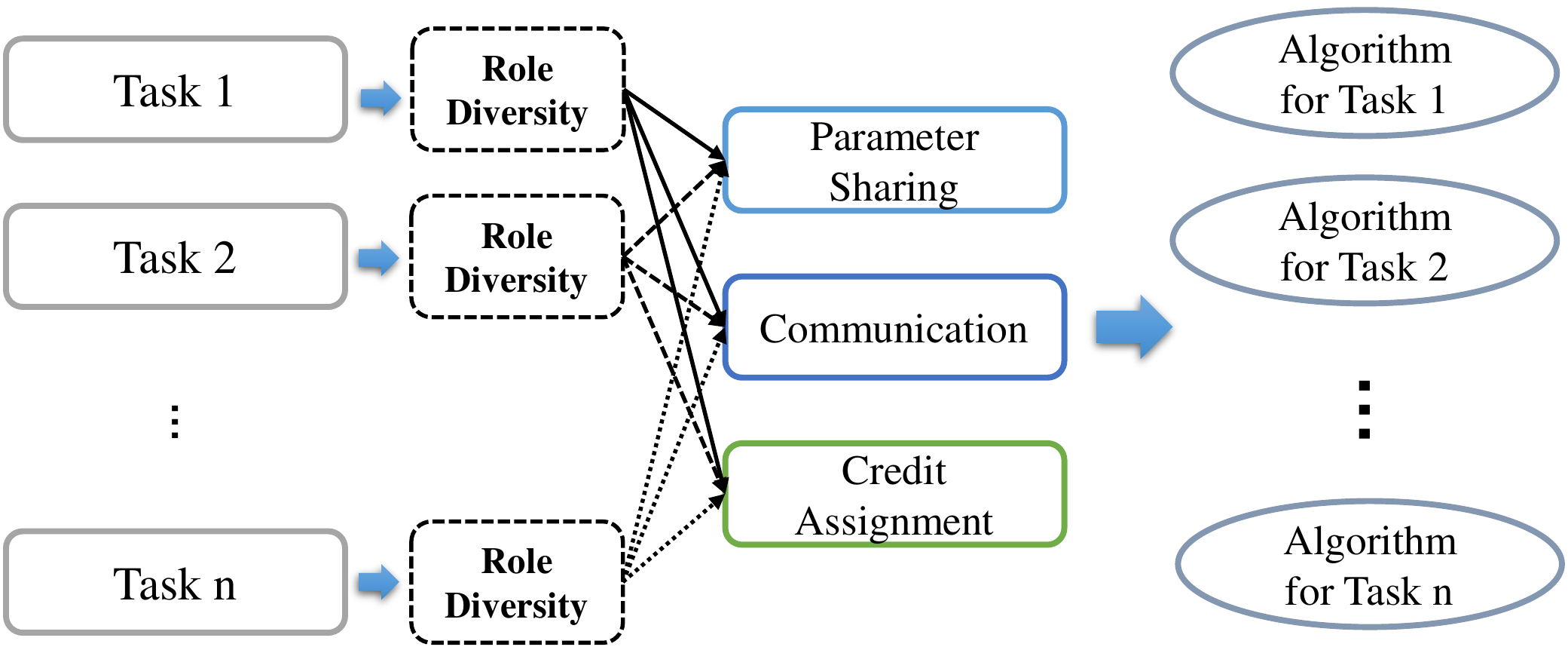}
  \caption{}
   \label{marl_framework:2}
\end{subfigure}\hfil 
\vspace{-7pt}
\caption{(a) Algorithms 1 \& 2 here are two examples to illustrate how the MARL policy is assembled using different training strategies. These strategies can cover three independent directions in MARL including parameter sharing, communication, and credit assignment. Note that the policy performance of algorithms 1 \& 2 varies from tasks $1$ to $n$. (b) Using Role Diversity to measure each task can help diagnose the misuse of a certain training strategy in current policy, and help determine a more suitable combination from different training strategies, which leads to better policy performance.}
\label{fig:marl_framework}
\vspace{-10pt}
\end{figure*}

Recently, multi-agent reinforcement learning (MARL) has attracted researchers attention due to its impressive achievements 
with super human-level intelligence in video games \cite{vinyals2019grandmaster,baker2019emergent,berner2019dota,ye2020mastering}, card games \cite{silver2017mastering,brown2019superhuman,li2020suphx,zha2021douzero}, and real-world applications \cite{zhou2020smarts,zhu2021main,zhong2021towards}. 
These achievements have benefited substantially from the success of single-agent reinforcement learning (RL) \cite{mnih2015human,hessel2018rainbow,haarnoja2018soft,schulman2015trust,schulman2017proximal} and rapid progress of MARL \cite{bacsar1998dynamic,littman2001friend,hu2003nash,yang2018mean,mguni2021learning, HuZCL21}. 

In MARL, one of the most attracting sub-tasks is the multi-agent cooperation tasks, where agents are required to achieve their common goals by collaborating with teammates \cite{lowe2017multi,zheng2017magent,samvelyan19smac}. However, the achievements on the cooperative MARL are more based on empirical results \cite{kurach2019google,bard2020hanabi,suarez2021neural,tuyls2021game} than theoretical analysis. And one key problem of cooperative MARL is how to fairly compare different algorithms as shown in Fig.~\ref{marl_framework:1}. 
Current researches focus on developing algorithms on the tasks they are good at but lack the study of why the performance declines on other tasks \cite{hostallero2019learning,wang2020rode,rashid2018qmix,wang2020rode,yu2021surprising}. 
Sometimes, even adopting the state-of-the-art algorithms does not guarantee an optimized performance \cite{sunehag2017value,rashid2018qmix,foerster2017counterfactual,wang2020rode,yu2021surprising}. 
This may be due to the varying characteristic (e.g agent's attributes and goals) of MARL tasks and scenarios, one single algorithm is not able to cover them all, which means we have to change the policy or the training strategy in order to guarantee that the policy we used fits for the current MARL task.

From this perspective, a metric to help measure the characteristic of different MARL tasks is desirable. 
This metric can be used to choose a better policy or training strategy for the current task, which is illustrated in Fig.~\ref{marl_framework:2}. 
We name this measurement as {\bf Role Diversity}, which aims at quantifying the difference of agents' behavior in a MARL task. 
We then analyze how the role diversity impacts learning performance both theoretically and empirically. 
For theoretical analysis, we decompose the estimation error
of the joint action-value function
to discuss how role diversity impacts the policy optimization process. 
The experiment further verifies the theoretical analysis that the role diversity is strongly related to the model performance and can serve as a good measurement of a MAS. 
As shown in Fig.~\ref{fig:marl_framework}, with the role diversity measurement of each task, we can diagnose the improper training strategies used in current policy or find a better training strategy combination.

More specifically, we define role distance and role diversity from three aspects: action-based, trajectory-based, and contribution-based. Each type of role diversity is measured by a unified variable called role distance. 
Theoretical analysis shows that different types of role diversity have a different impact on decomposed estimation error terms including
algorithmic error, approximation error, and statistical error. 
Comprehensive experiments are conducted, covering three popular directions in MARL training strategies: parameter sharing, communication, and credit assignment to verify the utility of role diversity.
A set of guidelines is provided to help diagnose the shortcomings of the current policy and decide whether a better training strategy exists according to the role diversity measurement. In addition, role diversity can explain why the MARL algorithms performance varies on different MARL tasks enabling us to fairly compare the algorithm's performance.

The main experiments are conducted on MPE \cite{lowe2017multi}
and SMAC \cite{samvelyan19smac} benchmarks, covering a variety of scenarios. 
The impact of role diversity is evaluated on representative MARL algorithms including IQL \cite{tan1993multi}, IA2C \cite{mnih2016asynchronous}, VDN \cite{sunehag2017value}, QMIX \cite{rashid2018qmix}, MADDPG \cite{lowe2017multi}, and MAPPO \cite{yu2021surprising}, covering the most common used methods in MARL including independent learning methods, centralised policy gradient methods, and joint Q learning methods. 
The experiment results prove that the model performance of different algorithms and training strategies is largely dependent on the role diversity. 
As a brief conclusion, we give three following insights for guiding future works and helping policy diagnosis to avoid using improper training strategies: scenarios with large action-based role diversity prefer no parameter sharing strategy; communication is not needed in scenarios with trajectory-based role diversity less than $30$ percent; learnable credit assignment modules should be avoided when training on scenarios with contribution-based role diversity higher than $0.5$. Following this guidance, we can significantly improve the final performance by introducing a proper training strategy into the current policy. In some cases, the performance can even be doubled when we give the best combination of training strategies compared to the original one.


The key contributions of this study are as follows: 
First, a new comprehensive definition of the role and role diversity is proposed to measure MARL tasks.
Second, a theoretical analysis of how role diversity impacts MARL policy optimization with estimation error decomposition is built.
Third, in the experiment, role diversity is proven to be strongly related to performance variance when choosing different training strategies, including parameter sharing, communication, and credit assignment on the MARL benchmarks.
Finally, based on both theoretical analysis and experimental results, a set of guidelines for current policy diagnosis and better training strategy selection are provided. These guidelines can help guarantee a better performance for cooperative MARL.

\vspace{-5pt}
\section{Related Work}
\vspace{-1pt}

Research on the development of cooperative MARL algorithms is mainly focusing on three directions: parameter sharing, communication, and credit assignment.

\textbf{Parameter Sharing} is a common technique in MARL. The most adopted approach is to fully share the model parameters among the agents \cite{sunehag2017value,rashid2018qmix,hostallero2019learning,wang2020rode}.
In this way, the policy optimization can benefit from the shared experience buffer with samples from all different agents, providing a higher data efficiency. 
However, it has also been noted in recent works that parameter sharing is not always a good choice \cite{christianos2021scaling,papoudakis2021benchmarking,terry2020revisiting}. 
In some scenarios, a selective parameter sharing strategy, or even no parameter sharing, can significantly benefit agent performance and surpass the full parameter sharing. 
However, the question of why different parameter sharing strategies have different impacts on different scenarios remains open. 
In this study, we find that the role diversity can serve as a strong signal for selecting the parameter sharing strategy.

\textbf{Communication} mechanism is an intrinsic part of the multi-agent system (MAS) framework \cite{sukhbaatar2016learning,jiang2018learning,lowe2017multi,kim2019learning,lazaridou2020emergent}. 
It provides the current agent with essential information of other agents to form a better joint policy, which substantially impacts the final performance. 
In some cases, communication restrictions exist, which hinder us from freely choosing communication methods \cite{lowe2017multi,samvelyan19smac}; in most cases, however, the communication is available and it is optional on when to communicate and how to ingest the shared information \cite{kim2019learning,singh2018learning}. 
We present a comprehensive study on the relationship between role diversity and information sharing via communication mechanisms and demonstrate that role diversity determines the necessity of communication.

\textbf{Credit Assignment} attracts most attention in cooperative MARL research. 
Most algorithms adopt Q-learning or policy gradient as the basic policy optimization method, combined with an extra value decomposition module \cite{sunehag2017value,rashid2018qmix,hostallero2019learning,wang2020rode} or shared critic function \cite{lowe2017multi,foerster2017counterfactual,yu2021surprising} to optimize the individual policy. 
Some other works find that leveraging the reward signal is unnecessary; however, optimizing the individual policy independently (independent learning, IL) can still get a strong joint policy \cite{tan1993multi,papoudakis2021benchmarking}. 
It then becomes slightly difficult to decide which credit assignment method (including IL) is better as there is no single method in cooperative MARL that is robust and always outperforms others (compared to PPO \cite{schulman2017proximal} or SAC \cite{haarnoja2018soft} in single-agent RL) on different tasks. 
In this study, we contend that role diversity is the key factor that impacts the performance of different credit assignment strategies.

Generally speaking, a good policy for MARL should consider all \textbf{Parameter Sharing}, \textbf{Communication}, and \textbf{Credit Assignment} strategies. The misuse of any one of them can bring significant performance degradation, known as the barrel effect. In the next section, we present the role definition from three aspects and propose the measurement of different types of role diversity to measure a MARL task. And by measuring each task, we can help diagnose whether the current policy has any shortcomings or should be replaced by a more suitable one.

\begin{figure*}[!h]
    \centering 
\begin{subfigure}{0.9\textwidth}
   \includegraphics[width=\linewidth]{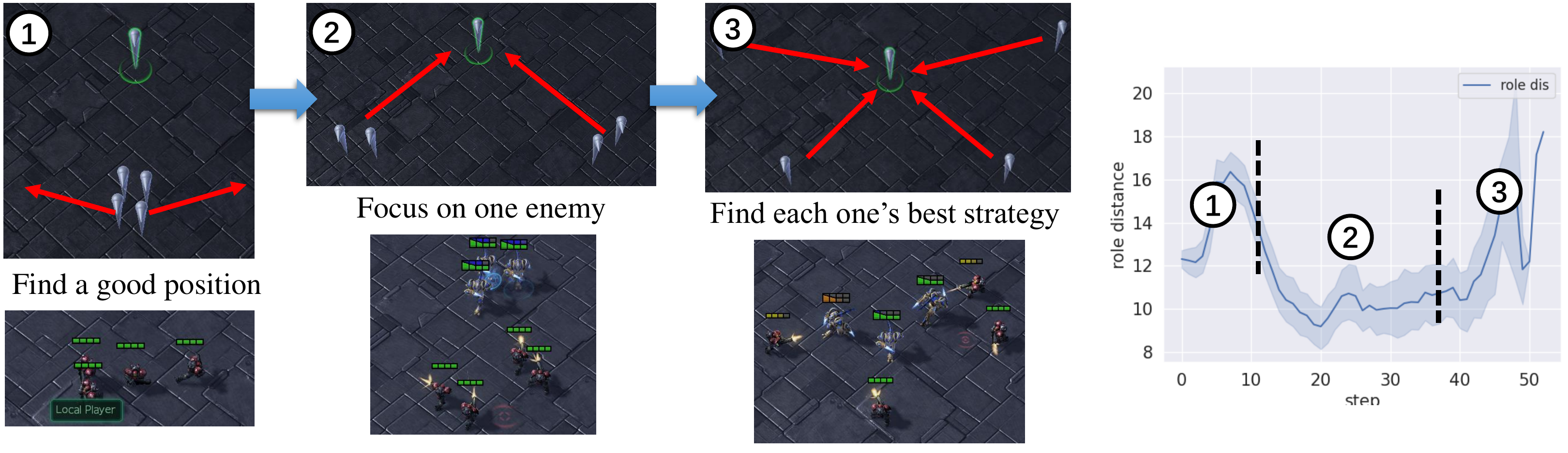}
   \caption{Action-based role diversity in one episode}
   \label{fig_spec:poli_role_case}
\end{subfigure}\hfil 
\vspace{7pt}
\medskip
\begin{subfigure}{0.5\textwidth}
  \includegraphics[width=\linewidth]{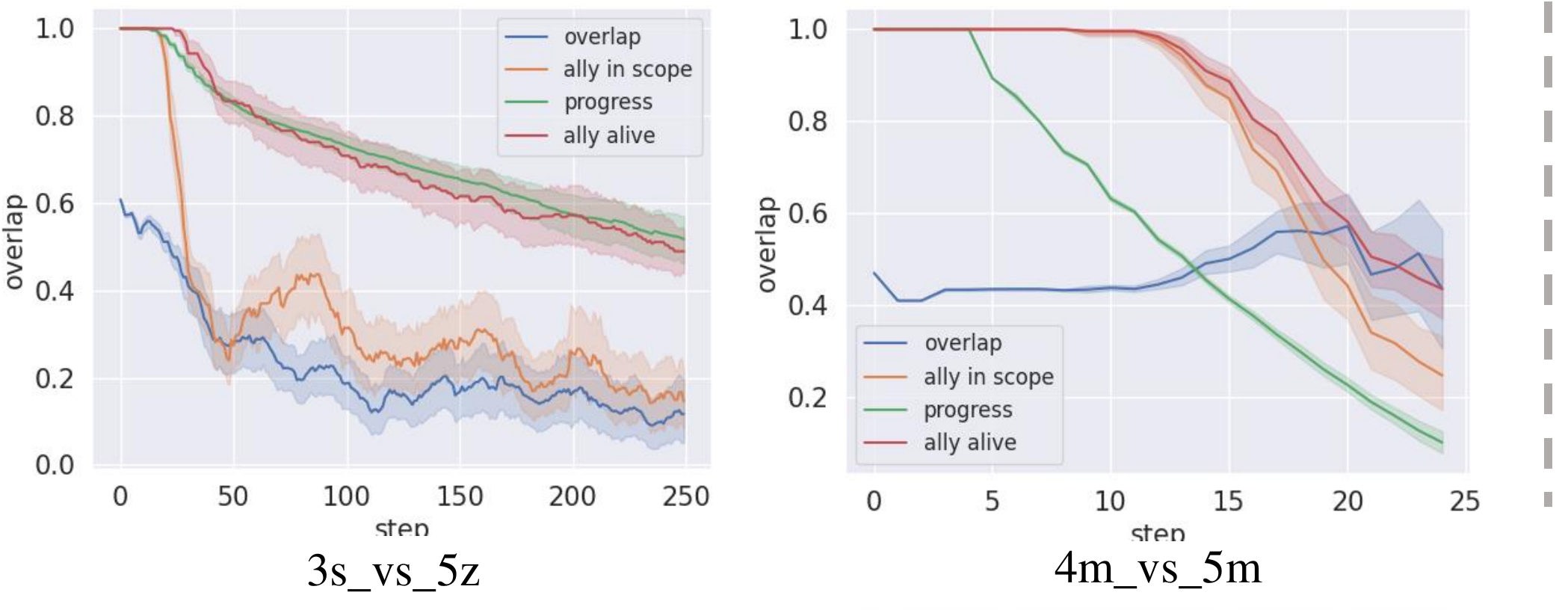}
  \caption{Trajectory-based role diversity in one episode}
   \label{fig_spec:traj_role_case}
\end{subfigure}\hfil 
\begin{subfigure}{0.5\textwidth}
  \includegraphics[width=\linewidth]{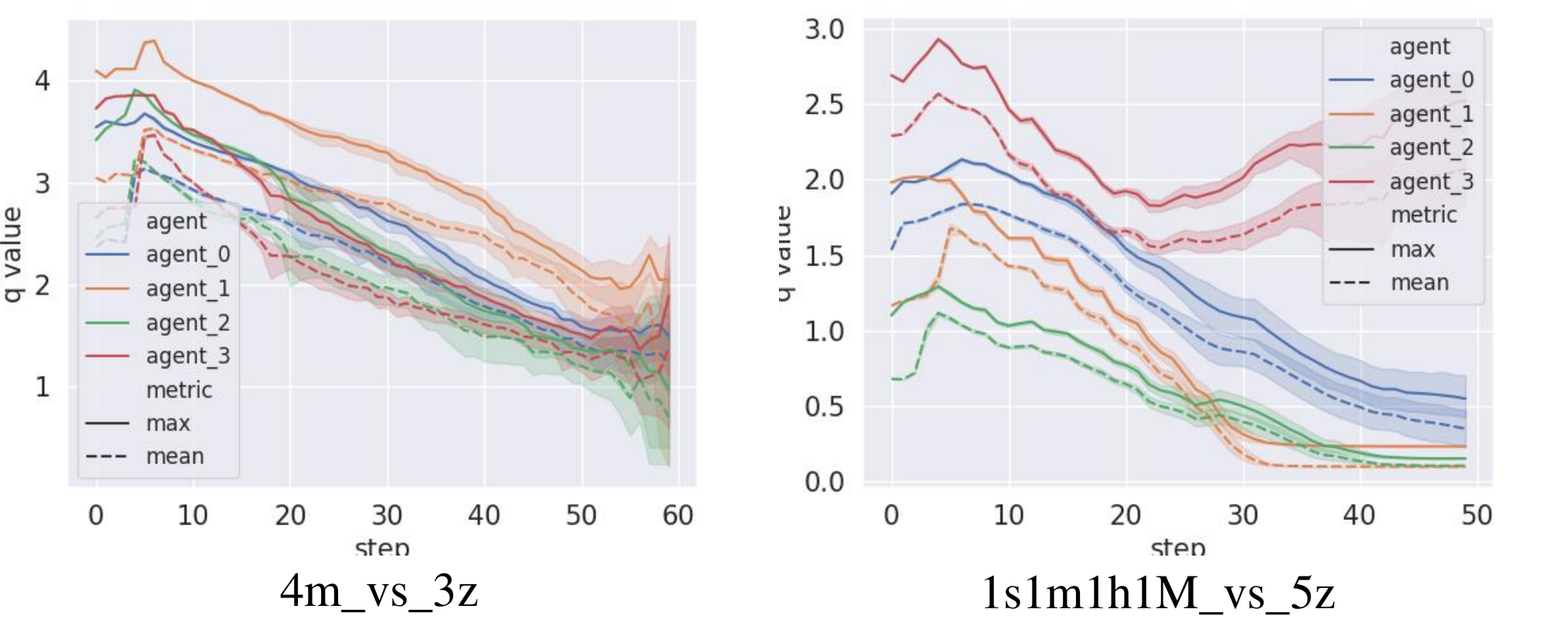}
  \caption{Contribution-based role diversity in one episode}
   \label{fig_spec:type_role_case}
\end{subfigure}\hfil 
\vspace{-13pt}
\caption{(a) An illustration of how action-based role difference varies in one game(4m\_vs\_3z). A detailed explanation of how it varies can be found in Sec.~\ref{sec:Policy Based Role}. (b) Instance curve of the trajectory-based role difference in different battle scenarios according to observation overlap. Trajectory-based role difference is larger in 3s\_vs\_5z but smaller in 4m\_vs\_5m. (c) Contribution-based role diversity in different battle scenarios represented by Q value. The contribution-based role diversity is larger in 1s1m1h1M\_vs\_5z but smaller in 4m\_vs\_3z.}
\label{fig_spec:role_case}
\vspace{-15pt}
\end{figure*}

\vspace{-5pt}
\section{Role Diversity}\label{sec:Role Diversity}
\vspace{-1pt}

Melees are strong and bulky, wizards are fragile and facile.
Using role to measure the characteristic of different agents in the MARL context has been proven to be effective in many recent works \cite{le2017coordinated,wang2020roma,wang2020rode,christianos2021scaling}. 
However, how to define an agent's role accurately and completely is still under exploration. 
In \citet{wang2020rode}, the role is defined as the higher-level option in the hierarchical RL framework \cite{kulkarni2016hierarchical}. In \citet{christianos2021scaling}, the role is defined as the environmental impact similarity of a random policy. These definitions are focusing on certain aspects of the agent's behavior differences but cannot fully measure them. In this study, we attempt to define the role in a more comprehensive way from three aspects: action-based, trajectory-based, and contribution-based. 
With our refined role, role distance between different agents can be measured. 
And by using role distance to further get the role diversity for each MARL task, a strong relationship between the role diversity and the MARL optimization process can be established. 
In this way, the role diversity can serve as a good measurement for policy diagnosis in cooperative MARL tasks.

\vspace{-7pt}
\subsection{Action-based Role}\label{sec:Policy Based Role}
\vspace{-3pt}

In MARL, different agents output different actions based on their current status. As common sense would indicate, actions taken at the same timestep can indicate different roles \cite{wang2020rode}. 
However, there are many exceptions. Consider two soccer players passing a ball to each other repeatedly \cite{kurach2019google}, although the action is different at each time step, the roles of these two soccer players can be very similar before who finally give the shot. Therefore, it is not sufficient to distinguish the role difference based on a single timestep. Instead, we contend that this difference should be defined based on a period. As this role is purely based on action distribution $\pi$, we refer to it as an action-based role.

Specifically, we define the action-based role as the statistics of the actions' frequency over a period, which is $n$ steps backward and forward from the current timestep. Here, $n$ is the time interval that is half the length of the total time. More details can be found in Fig.~\ref{fig:poli_based_role}, where we provide a real scenario from SMAC.  
Action-based role difference can be represented as follows:
\begin{equation}\label{eq:policy based role}
\setlength{\abovedisplayskip}{5pt}
r_{T}^{a}=\frac{1}{2n+1}\sum_{t=T-n}^{T+n}\pi_{t}^{a}, 
\setlength{\belowdisplayskip}{5pt}
\end{equation}
where $T$ represents the current timestep, $n$ is the time interval, $a$ is the agent index, $\pi$ is the policy distribution. We adopt symmetrical KL divergence to measure the distance of different action-based roles. The role distance of two agents $a_0$ and $a_1$ can be computed as follows:
\begin{equation}\label{eq:policy diversity}
\setlength{\abovedisplayskip}{5pt}
d_{(act)T}^{a_0,a_1}=KL(r_{T}^{a_0}|r_{T}^{a_1})+KL(r_{T}^{a_1}|r_{T}^{a_0}),
\setlength{\belowdisplayskip}{5pt}
\end{equation}
where $d_{(act)T}^{a_0,a_1}$ represents the action-based role distance at timestep $T$ between two agents, and {\it KL} represents Kullback–Leibler divergence.

\vspace{-7pt}
\subsection{Trajectory-Based Role}\label{sec:Trajectory Based Role}
\vspace{-3pt}

An action-based role only considers the role diversity from the action distribution perspective. 
A slight action difference may not diversify the action-based role; however, this difference can be enlarged by time, which eventually results in a de facto behavior difference. 
The most significant phenomenon caused by this is the agent's trajectory. 
Again, consider a soccer game where players repeatedly pass a ball to each other; these players may have a similar action-based role, but their trajectories are different. 
This difference is enhanced by the partial observation setting that exists in many popular cooperative MARL environments \cite{suarez2021neural,samvelyan19smac,zheng2017magent} as the partial observed input from different agents shares a less common pattern when the vision scope is smaller. 
Therefore, trajectory-based role diversity is an important supplement to action-based role diversity.

Generally speaking, we can define the trajectory-based role as the record of the agent's movement. 
However, the determination of the extent to which two trajectories differ is not a straightforward matter. 
To measure this difference, we use an indirect metric called observation overlap percentage. Using observation overlap to measure the trajectory-based role difference is: {\it I}. easy to compute, {\it II}. able to utilize a constant scale from 0 to 1 and {\it III}. strongly related to observation scope, which means that two trajectories can have varying role distances. 
The trajectory difference between two agents $a_0$ and $a_1$ is annotated as $d_{(traj)T}^{a_0,a_1}$
which is the observation overlap percentage of the total area between agent $a_0$ and agent $a_1$ at timestep $T$. 
An illustration of calculating the observation overlap percentage can be found in Appx.~\ref{app:obs overlap smac}. 
More scenarios to show the generalization of trajectory-based role can be found in Appx.~\ref{app:obs overlap}.

\vspace{-7pt}
\subsection{Contribution-Based Role}\label{sec:Contribution Based Role}
\vspace{-3pt}

To help ensure the authenticity of the modern MAS, agents are initialized with different attributes in recently proposed MARL environments \cite{samvelyan19smac,suarez2021neural,lowe2017multi,kurach2019google}. 
For instance, in \cite{kurach2019google}, the roles of forward and goalkeeper are quite different and characterized by different observation scopes, action spaces, and reward functions. 
The type differences are easy to notice but hard to define, as is the role distance between them. 

Here, we use an indirect variable called contribution to measure the type difference. In cooperative MARL, the final target is to obtain an optimal joint policy consisting of $A$ individual policies. 
To do this, the credit assignment strategy is proposed in \cite{sunehag2017value,rashid2018qmix,foerster2017counterfactual} to leverage the reward signals to each agent to achieve individual policy optimization. 
A good credit assignment strategy should be able to leverage the reward signal in a manner equal to each agent's contribution to the global reward. 
In this way, the Q value (Q function in off-policy RL) or state value (critic function in on-policy RL) of each agent can be estimated based on the leveraged reward signal. 
From this perspective, the value can be regarded as the agent's contribution to the team. Generally speaking, we use the Q value or state value to measure the contribution of a single agent. Contribution-based role diversity between two agents can be computed as follows:
\begin{equation}\label{eq:contribution based role}
\setlength{\abovedisplayskip}{5pt}
d_{(cont)T}^{a_0,a_1} = \frac{d_{T}^{v_{a_0},v_{a_1}}}{max(v_{a_i},v_{a_j})} \; i,j \in A.
\setlength{\belowdisplayskip}{5pt}
\end{equation}
Here, $v$ is the Q value or state value of the policy output and $d_{T}^{v_{a_0},v_{a_1}}$ is the absolute value difference between the agents' output. In addition, we use the max value difference in all $A$ agents to keep the range of contribution-based role diversity from 0 to 1.  

\vspace{-7pt}
\subsection{Distance to Diversity}
\vspace{-3pt}

Role diversity can be counted as the mean value of the role distance between all different agent pairs. Different role distances are used to measure different role diversity aspects. For instance, in a $A$ agents MAS, the role diversity can be calculated as:
\begin{equation}\label{eq:traj based role diversity}
\setlength{\abovedisplayskip}{5pt}
D_{T} = \frac{1}{\sum_0^{A-1}}\sum_{a_0=0}^{A}\sum_{a_1=a_0}^{A}d_{T}^{a_0,a_1}.
\setlength{\belowdisplayskip}{5pt}
\end{equation}
Here $d_{T}^{a_0,a_1}$ can be any of the $d_{(act)T}^{a_0,a_1}$,$d_{(traj)T}^{a_0,a_1}$, and $d_{(cont)T}^{a_0,a_1}$ to get corresponding action-based, trajectory-based, and contribution-based role diversity. 

More detailed role definition, analysis on connection between different role types, and application can be found in Appx.~\ref{app:obs overlap},~\ref{app:Different Types of Role} and~\ref{app: connection}. More specific scenario-based illustration can be found in Fig.~\ref{fig_abs:role_mpe} and Fig.~\ref{fig:role_smac}.

\vspace{-5pt}
\section{Policy Diagnosis}
\vspace{-1pt}

With role diversity to measure a MARL task, we can now diagnose whether we have chosen the proper training strategies to get the optimal policy.
In this section, we provide case studies on how different types of role diversity vary in one episode. We give a brief introduction on how to find the shortcomings of current training strategies based on different types of role diversity. Popular MARL training strategies including parameter sharing, communication, and credit assignment are discussed. The theoretical evidence of the diagnosis process can be found in Sec.~\ref{sec:3} and the experimental proof can be found in Sec.~\ref{Experiment}.

The action-based role diversity case is provided in fig.~\ref{fig_spec:poli_role_case}. We take a real battle scenario (4m\_vs\_3z) from SMAC, and find three stages including \textit{\textbf{Find a good position}}, \textit{\textbf{Focus on enemy}} and \textit{\textbf{Find each one's best strategy}} in this battle scenario. 
In stage 1, agents try to find their own best location; the role diversity is large. In stage 2, agents focus on the same enemy target; the policies become similar and the role diversity is decreased. In stage 3, the formation of the agents is broken up by the enemies. Action-based role diversity again increases as each agent is required to find its own best strategy to deal with its current situation. In this case, sharing or not sharing the model parameter can significantly impact policy optimization. In stage 1 and stage 3 where the role diversity is big, agents prefer to have their individual policy. While in stage 2 where the role diversity is small, sharing the policy is a better choice.

The trajectory-based role diversity instance is provided as the observation overlap percentage curves in Fig.~\ref{fig_spec:traj_role_case} in two scenarios including 3s\_vs\_5z and 4m\_vs\_5m. With the game progressing, the observation overlap curve becomes significantly different in these scenarios. In 3s\_vs\_5z, the better training strategy is not sharing the observation or communicating with other agents as the trajectory-based role diversity is big. While in 4m\_vs\_5m, sharing the observation helps the policy learning as the trajectory-based role diversity is small.

The contribution-based role diversity instance can be found in in Fig.~\ref{fig_spec:type_role_case}. We provide the Q value (mean \& max) curves to demonstrate that the contribution-based role diversity can vary a lot in different scenarios. In 4m\_vs\_3z, the contribution-based role diversity is significantly smaller than that in 1s1m1h1M\_vs\_5z. Therefore, it is a wise choice to have a learnable credit assignment module in 4m\_vs\_3z's policy. In the experiment, we find that tackling 1s1m1h1M\_vs\_5z with a learnable credit assignment module can result in a serious performance drop.

\vspace{-5pt}
\section{Theoretical Analysis}\label{sec:3}
\vspace{-1pt}

This section presents a simple example to illustrate that the role diversity can serve as policy diagnostic criteria since it affects the Q-function estimation error.
Suppose each agent makes individual observations and the learning procedure of all agents is independent. 
We provide finite-sample analysis for the estimation error of the joint Q-function
and identify the terms corresponding to the role diversity. 
We denote $Q_{tot}^*$ and $Q^*_i$ as the optimal joint and individual Q-function respectively and write  $\|\cdot\|_{p, \mu}$ as the $L_p$ norm with respect to a probability measure $\mu.$
Motivated by \cite{sunehag2017value} and \cite{rashid2018qmix}, we consider a simple case:
\[
\setlength{\abovedisplayskip}{5pt}
Q_{tot}^* \approx \rvw^\top \rmQ^*  \,\,
\text{with} \,\,
\rmQ^* = (Q^*_1,\ldots, Q^*_n),
\setlength{\belowdisplayskip}{5pt}
\]
where $n$ is the number of agents and $\rvw$ is a $n$-length weight vector. 
Here, the credit assignment function is a weighted sum of $Q^*_i$ with non-negative weights.
We then study the excess risk as follows: 
\[
\setlength{\abovedisplayskip}{5pt}
Err = \|Q^*_{tot} - \hat \rvw^\top \rmQ_t\|_{1,\mu} - \|Q_{tot}^* - (\rvw^*)^\top \rmQ^*\|_{1,\mu},
\setlength{\belowdisplayskip}{5pt}
\]
where $\rvw^*$ and $\hat \rvw$ are the optimal and estimated weights and $\rmQ_t$ is the output of FQI algorithm at the iteration $t.$ 
We further denote $\gQ$ as the space of individual Q-functions and write $\omega(\gQ) = \sup_{Q \in \gQ} \inf_{Q'\in \gQ} \|Q'-T Q\|_{2,\nu}^2.$
In Sec.~\ref{sec:separate}, we prove that
\small
\begin{equation}\label{eq31}
\begin{split}
Err  \leq & \sqrt{n} \|\rvw^* - \hat \rvw\| \left\|\sqrt{\Var_n \big(\rmQ^*\big)}\right\|_{1,\mu}  +  \frac{4 \gamma^{T+1}}{(1-\gamma)^2} M \\
     & + \frac{4 \phi_{\mu,\nu} \gamma}{(1-\gamma)^2} \sqrt{\omega(\gQ)}  + O(\sqrt{\frac{\ln N_0}{N}}), 
\end{split}
\end{equation}
\normalsize
where $N$ is the sample size and $T$ is the number of iterations.
In addition, $\phi_{\mu,\nu}$ is the concentration coefficient  and $N_0$ represents the $1/N$-covering number of $\gQ.$ 
Please refer to Sec.~\ref{sec:D1} for the detailed definitions.

The first term on the RHS of (\ref{eq31}) reflects the benefit of credit assignment that is strong related to the {\bf Contribution-Based Role} (Sec.~\ref{sec:Contribution Based Role}).
When $\Var_n(\rmQ^*)$ is non-negligible, minimizing $\|\rvw^* - \hat \rvw\|$ can significantly decrease the excess risk.
The term that involves $\omega(\gQ)$ stands for the approximation error caused by functional approximation in $\gQ.$ 
It depends on the concentration of the sample and the scale of the hypothetical space.
The remaining two terms are statistical error and  algorithmic error. If the sample size is sufficiently large and the learning time is long enough, they can be arbitrarily small.
In Sec.~\ref{sec:rate}, we assume $Q$ is a sparse ReLU network and $TQ$ is a composition of H$\ddot{\text{o}}$lder smooth functions, and analyze the convergence rate as $N, T \to \infty.$

Next, we demonstrate that the variance term $\Var_n(\rmQ^*)$ is related to both the {\bf Action-Based Role} (Sec.~\ref{sec:Policy Based Role}) and the {\bf Trajectory-Based Role} (Sec.~\ref{sec:Trajectory Based Role}). We consider the case in which all agents share one Q-function and denote the optimal share Q-function as $\bar Q^*.$ 
Sec.~\ref{sec:share} proves that
\small
\begin{equation}\label{eq32}
\begin{split}
Err  \leq & \sqrt{n}  \|\rvw^* - \hat \rvw\| \left\|\sqrt{ \Var_n \big(\bar \rmQ^*(\rvz,\rvu) \big)}\right\|_{1,\mu} \\
& + \left\| \sum_{i=1}^n \rw_i^* (Q^*_i - \bar Q^*) \right\|_{1,\mu} + \frac{4 \phi_{\mu,\nu} \gamma}{(1-\gamma)^2} \sqrt{\omega(\gQ)}\\
& + O(\sqrt{\frac{\ln N'_0}{nN}}) +  \frac{4 \gamma^{T+1}}{(1-\gamma)^2} M,
\end{split}
\end{equation}
\normalsize
where $N'_0$ is the $1/(nN)$-covering number of $\gQ$ and
\[
\setlength{\abovedisplayskip}{5pt}
\bar \rmQ^*(\rvz, \rvu) = \big( \bar Q^*(\rz_1, \ru_1), \bar Q^*(\rz_2, \ru_2), \ldots, \bar Q^*(\rz_n, \ru_n) \big).
\setlength{\belowdisplayskip}{5pt}
\]
The second term on the RHS of (\ref{eq32}) stands for the bias caused by parameter sharing. If all $Q^*_i$ are the same, the bias will disappear. Therefore, the {\bf Action-Based Role} is related to this bias. Second, the variance $\Var_n(\bar \rmQ^*)$ here is caused by the trajectory diversity. To reduce this term, we should ensure that all agents have similar observations. 
In addition, the {\bf Trajectory-Based Role} measures the concentration of all agents' support set. It is therefore natural to group the highly overlapped agents into one sub-joint agent via communication mechanism.  
This can be compared to the separate case in (\ref{eq31}), where approximation error and the learning error are the same. 
In Sec.~\ref{sec:rate}, we show that the parameter sharing improves the convergence rate of the statistical error via sample pooling, while the communication decreases the convergence rate by
activating more input variables.

\begin{table*}[!h]
\small
\caption{
 Performance of three parameter sharing strategies on different scenarios. {\it Warm-up} refers to the reward value point where the strategies start to differentiate. {\it +} represents the additional reward gained based on warm-up performance. The left side and right side of the {\it /} represent the reward gained at the half training steps and the full training steps respectively. The best performance in each scenario is marked in bold red. Role Diversity column is marked with gradient grey. Deeper the grey, the larger the role diversity. Detailed analysis can be found in Sec.~\ref{exp:Parameter Sharing}. More scenarios and parameter sharing strategies for comparison can be found in Appx.~\ref{app:Parameter Sharing}.
 }
 \vspace{-8pt}
\begin{center}
\resizebox{1.0\textwidth}{!}{
\setlength{\tabcolsep}{1.5em}
{\renewcommand{\arraystretch}{1.0}
\begin{tabular}{c|cccccc}
\hline
\hline
Benchmark & Scenario & Role Diversity & Warm-up & No shared & Partly shared & Shared \\ 
\hline
& SimpleSpread  &  {\cellcolor[gray]{.95}} 14.1  & -598.3 & {\color[HTML]{000000} +137.0 / +142.9} & {\color[HTML]{000000} +149.0 / +176.4}         & {\color[HTML]{FE0000} \textbf{+154.1 / +198.0}} \\ 
& Tag           &  {\cellcolor[gray]{.90}} 17.8  & 3.8    & {\color[HTML]{000000} +43.4 / +57.3}   & {\color[HTML]{FE0000} \textbf{+47.0 / +60.9}}  & {\color[HTML]{000000} +48.8 / +59.2}            \\ 
& Adversary          & {\cellcolor[gray]{.85}} 18.3           & 10.7    & {\color[HTML]{000000} +5.2 / +5.7}             & {\color[HTML]{FE0000} \textbf{+6.2 / +6.6}}    & {\color[HTML]{000000} +5.4 / +5.9}              \\ 
& DoubleSpread-2     & {\cellcolor[gray]{.91}} 17.6           & 7.3     & {\color[HTML]{FE0000} \textbf{+47.8 / +53.2}}  & {\color[HTML]{000000} +28.6 / +34.6}           & {\color[HTML]{000000} +3.6 / +15.9}             \\ 
\multirow{-5}{*}{MPE}   & DoubleSpread-4     & {\cellcolor[gray]{.80}} 19.5           & 22.0    & {\color[HTML]{FE0000} \textbf{+29.5 / +192.4}} & {\color[HTML]{000000} +12.0 / +91.3}           & {\color[HTML]{000000} +11.4 / +5.3}             \\ \hline
& 2m                 & {\cellcolor[gray]{.95}} 3.1 / 12.2     & 6.0     & +9.2 / +11.1                                   & {\color[HTML]{000000} +15.5 / +15.6}           & {\color[HTML]{FE0000} \textbf{+18.1 / +17.6}}   \\ 
& 4m\_vs\_4z         & {\cellcolor[gray]{.90}} 3.3 / 19.3     & 4.4     & {\color[HTML]{000000} + 8.8 / +12.7}           & {\color[HTML]{FE0000} \textbf{+ 10.5 / +14.7}} & {\color[HTML]{000000} +5.4 / +8.4}              \\ 
& 4m\_vs\_3z         & {\cellcolor[gray]{.85}} 3.8 / 12.1     & 7.2     & +12.4 / +12.1                                  & {\color[HTML]{FE0000} \textbf{+12.5 / +12.5}}  & {\color[HTML]{000000} +11.9 / +12.3}            \\ 
& 1c1s1z\_vs\_1c1s3z & {\cellcolor[gray]{.80}}8.7 / 22.0     & 11.8    & {\color[HTML]{FE0000} \textbf{+4.1 / +6.1}}    & +3.7 / +5.9                                    & {\color[HTML]{000000} +2.7 / +5.4}              \\ 
\multirow{-5}{*}{SMAC} & 1s1m1h1M\_vs\_5z   & {\cellcolor[gray]{.75}} 6.2 / 22.5     & 6.2     & {\color[HTML]{FE0000} \textbf{+6.4 / +9.1}}    & +4.2 / +8.5                                    & {\color[HTML]{000000} +3.7 / +6.1}              \\ 
\hline
\hline
\end{tabular}
 }
 }
 \end{center}

 \vspace{-10pt}
 \label{table:parameter sharing vdn}
 \end{table*}

\vspace{-5pt}
\section{Experiment}\label{Experiment}
\vspace{-1pt}

In this section, we mainly demonstrate how model performance varies with role diversity and how to adjust the training strategy in the context of cooperative MARL. The experimental results show the following: 1. that the performance of different parameter sharing strategies is strongly related to the {\bf Action-Based Role} (Sec.~\ref{exp:Parameter Sharing}). 2. that the benefit brought by different communication mechanisms can be easily affected by the {\bf  Trajectory-Based Role} (Sec.~\ref{exp:Communicate: When and How}). 3. that the performance of the credit assignment method, or the centralized training strategy, is largely dependent on the {\bf Contribution-Based Role} (Sec.~\ref{exp:Credit Assignment}). 4. that the choice of training strategies should be determined by the scale of role diversity for different scenarios. The main experimental platforms are MPE \cite{lowe2017multi} and SMAC \cite{samvelyan19smac}. Extensions are made to fulfill the requirements of parameter sharing and the communication mechanism, these include separated training of policy in Sec.~\ref{exp:Parameter Sharing} and information exchange among agents in Sec.~\ref{exp:Communicate: When and How}. All results come from eight random seeds. All the role diversity values and curves come from our baseline policy: VDN\cite{lowe2017multi}, no parameter sharing, and no communication for robust performance and training efficiency. More details and experimental results can be found in the appendix.

\vspace{-7pt}
\subsection{Parameter Sharing}\label{exp:Parameter Sharing}
\vspace{-3pt}

\begin{figure*}[!h]
    \centering 
\begin{subfigure}{0.24\textwidth}
  \includegraphics[width=\linewidth]{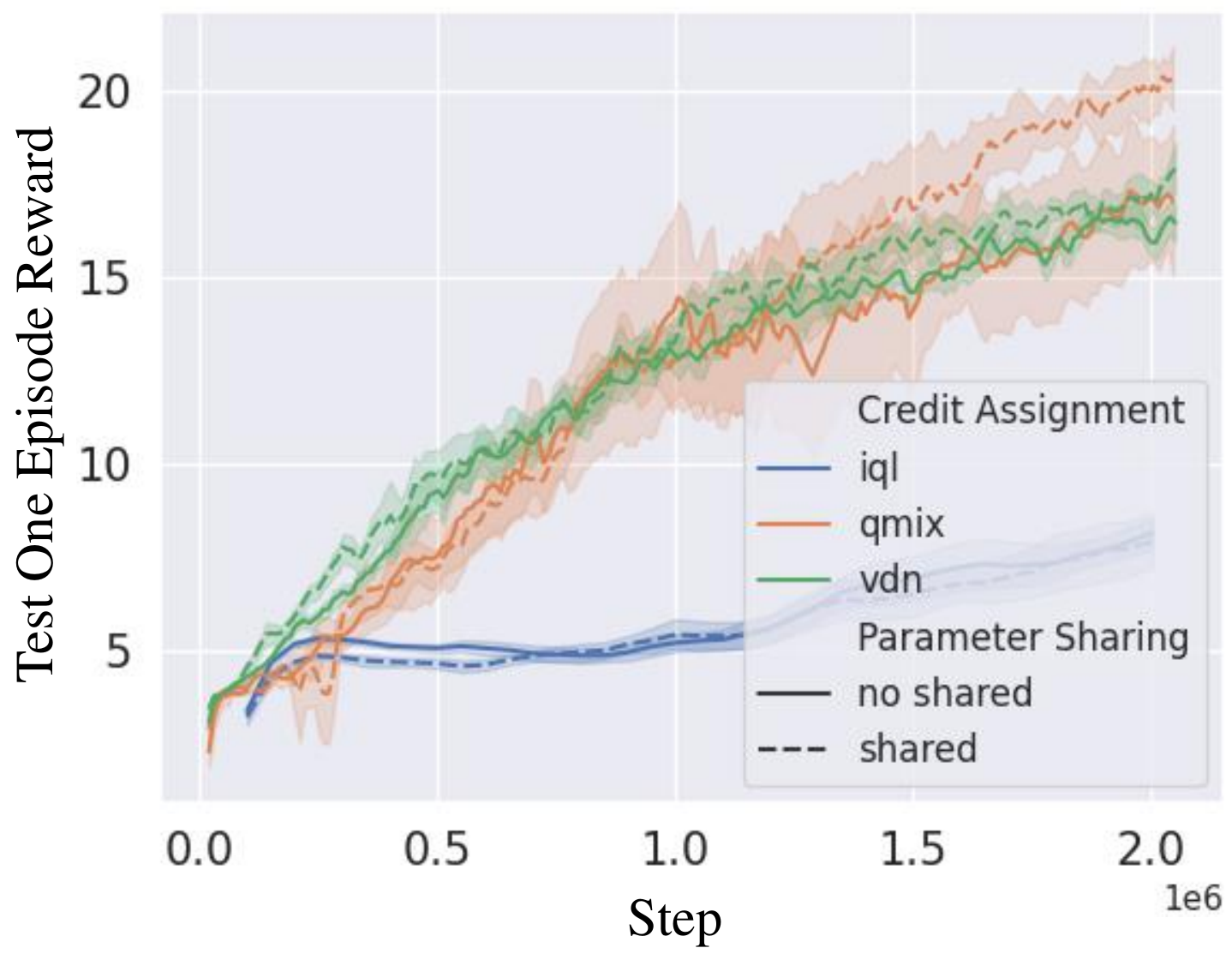}
  \caption{3s\_vs\_5z}
  \label{chart2:3s_vs_5z_Q}
\end{subfigure}\hfil 
\begin{subfigure}{0.25\textwidth}
  \includegraphics[width=\linewidth]{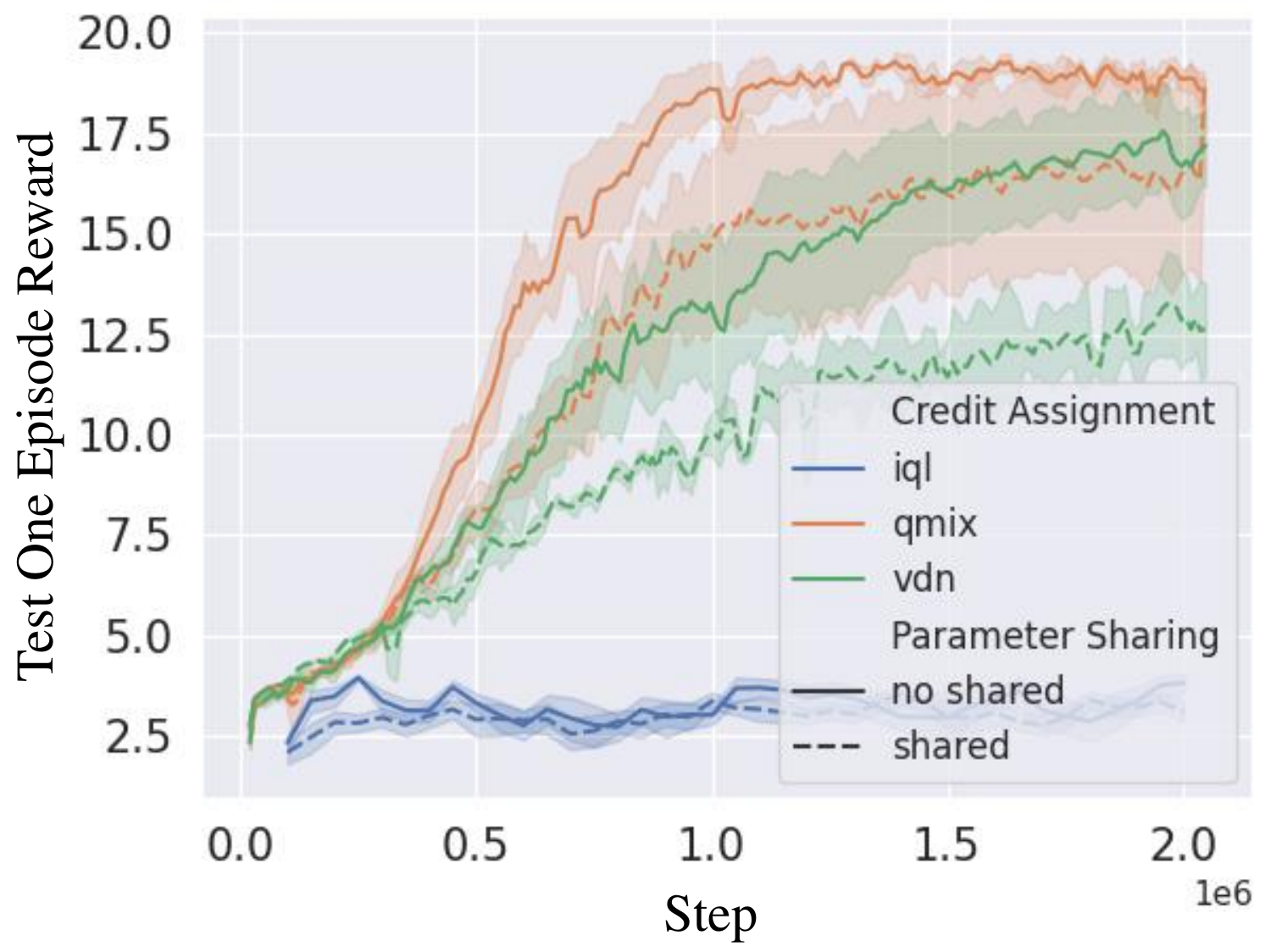}
  \caption{4m\_vs\_4z}
  \label{chart2:4m_vs_4z_Q}
\end{subfigure}\hfil 
\begin{subfigure}{0.24\textwidth}
  \includegraphics[width=\linewidth]{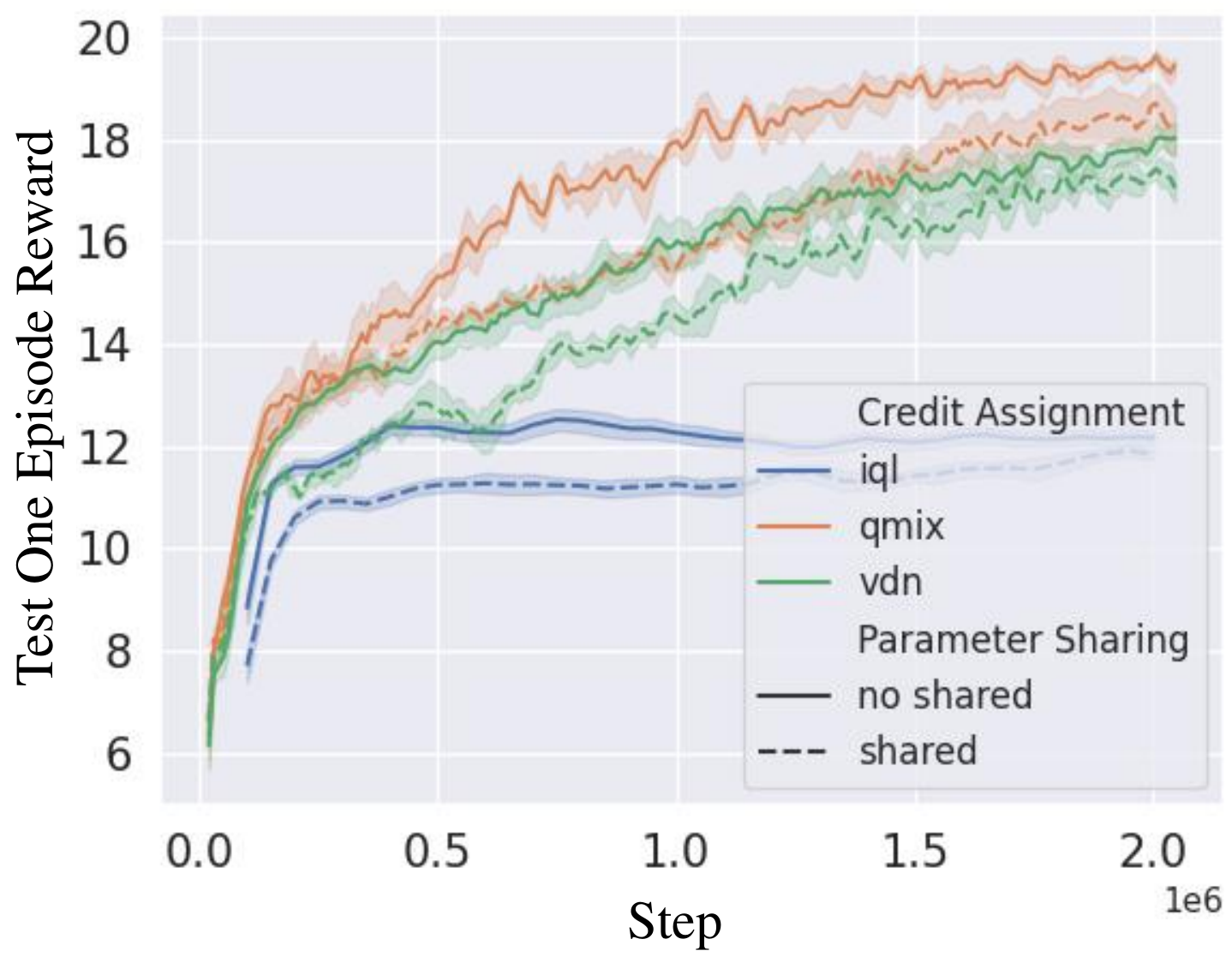}
  \caption{1c1s1z\_vs\_1c1s3z}
  \label{chart2:1c1s1z_vs_1c1s3z_Q}
\end{subfigure}\hfil 
\begin{subfigure}{0.24\textwidth}
  \includegraphics[width=\linewidth]{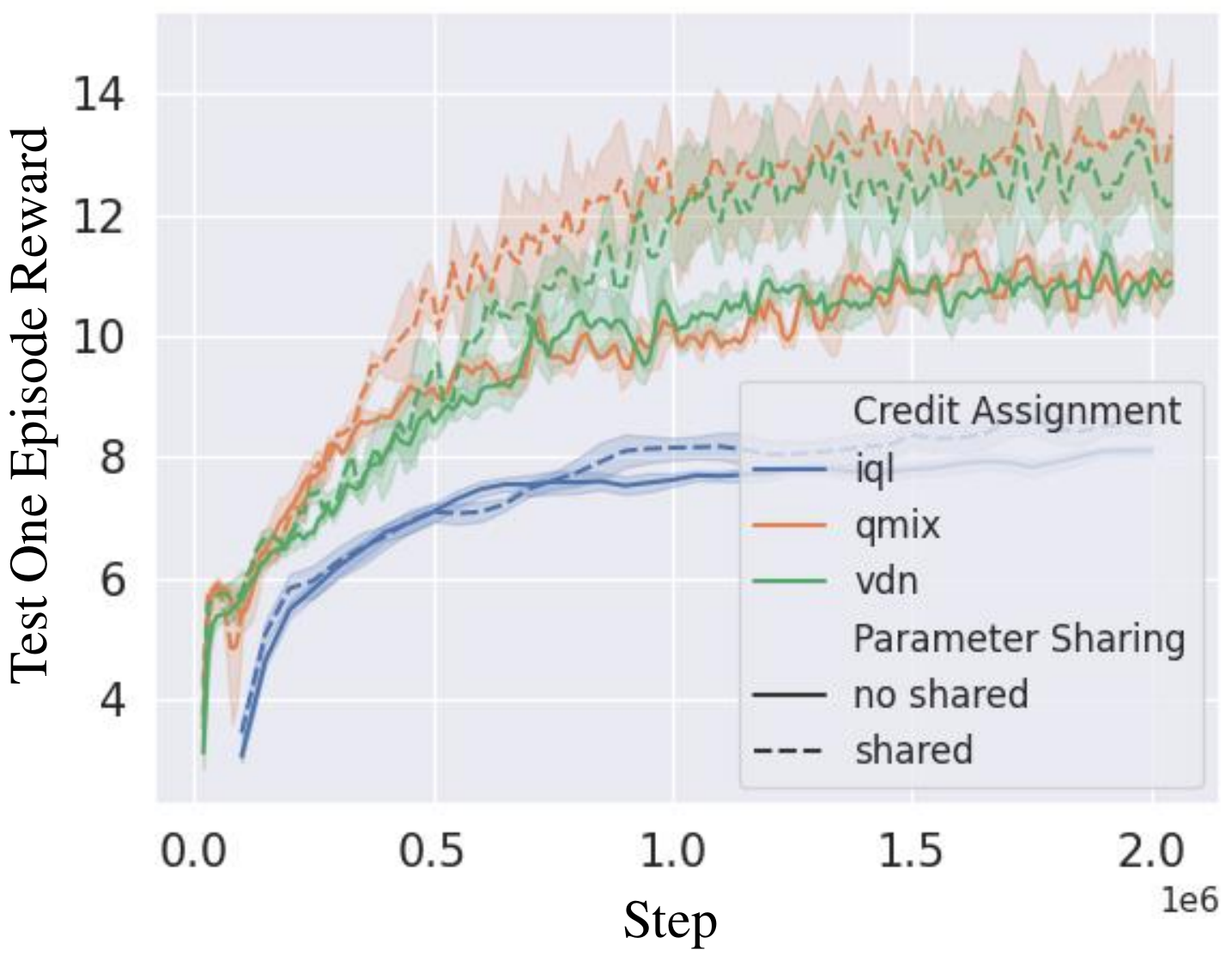}
  \caption{4m\_vs\_5m}
  \label{chart2:4m_vs_5m_Q}
\end{subfigure}\hfil 
\medskip
\begin{subfigure}{0.24\textwidth}
  \includegraphics[width=\linewidth]{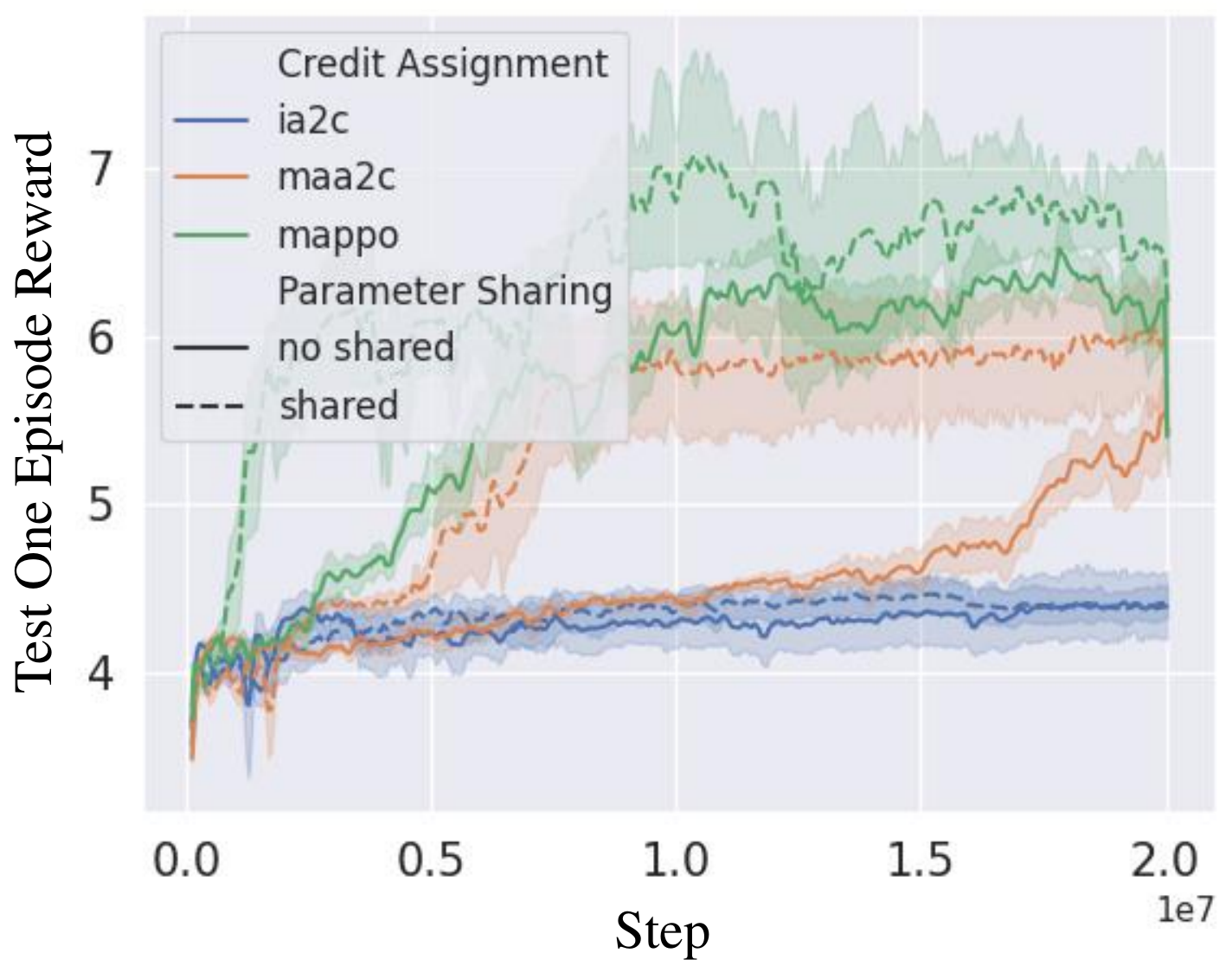}
  \caption{3s\_vs\_5z}
  \label{chart2:3s_vs_5z_P}
\end{subfigure}\hfil 
\begin{subfigure}{0.24\textwidth}
  \includegraphics[width=\linewidth]{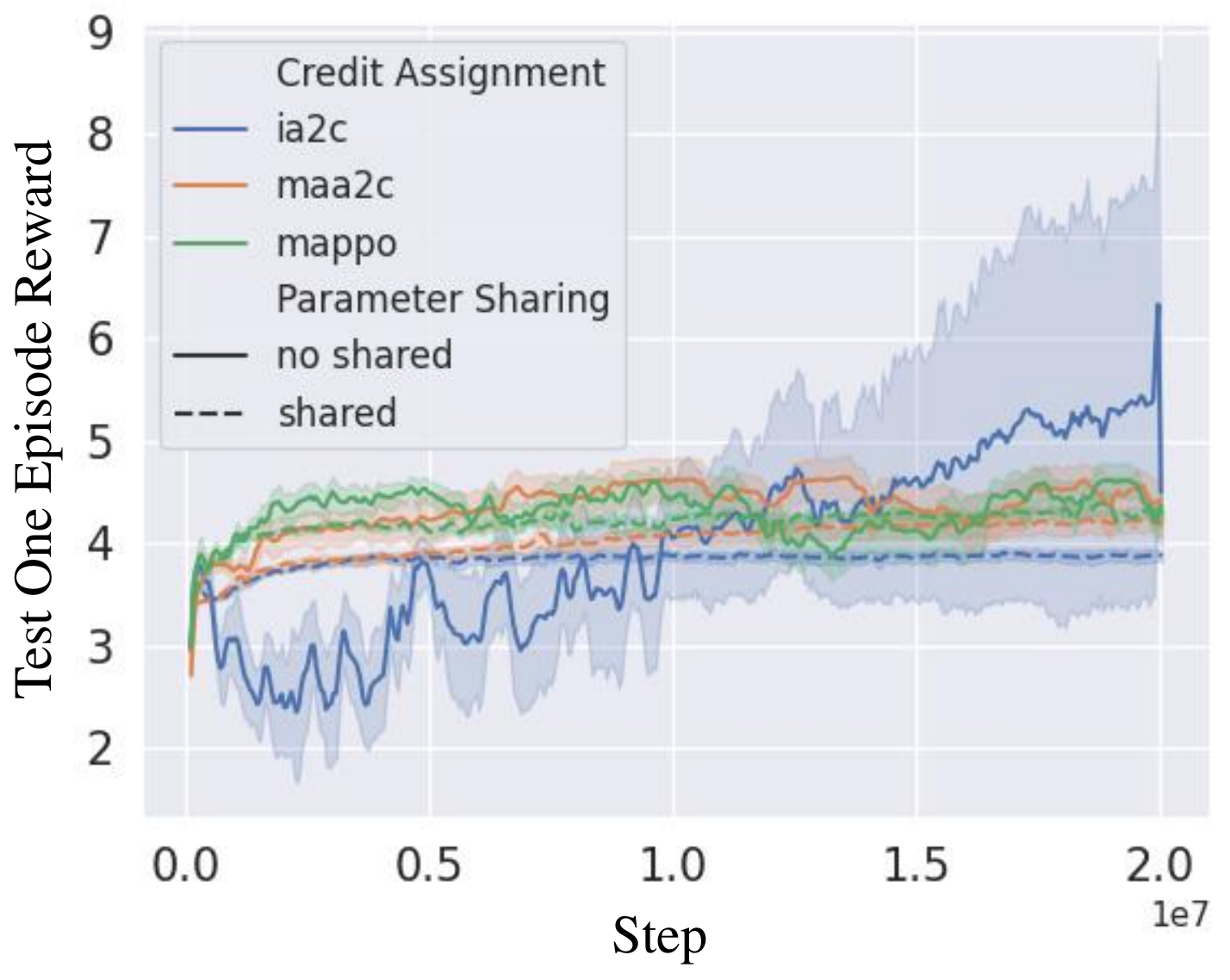}
  \caption{4m\_vs\_4z}
  \label{chart2:4m_vs_4z_P}
\end{subfigure}\hfil 
\begin{subfigure}{0.24\textwidth}
  \includegraphics[width=\linewidth]{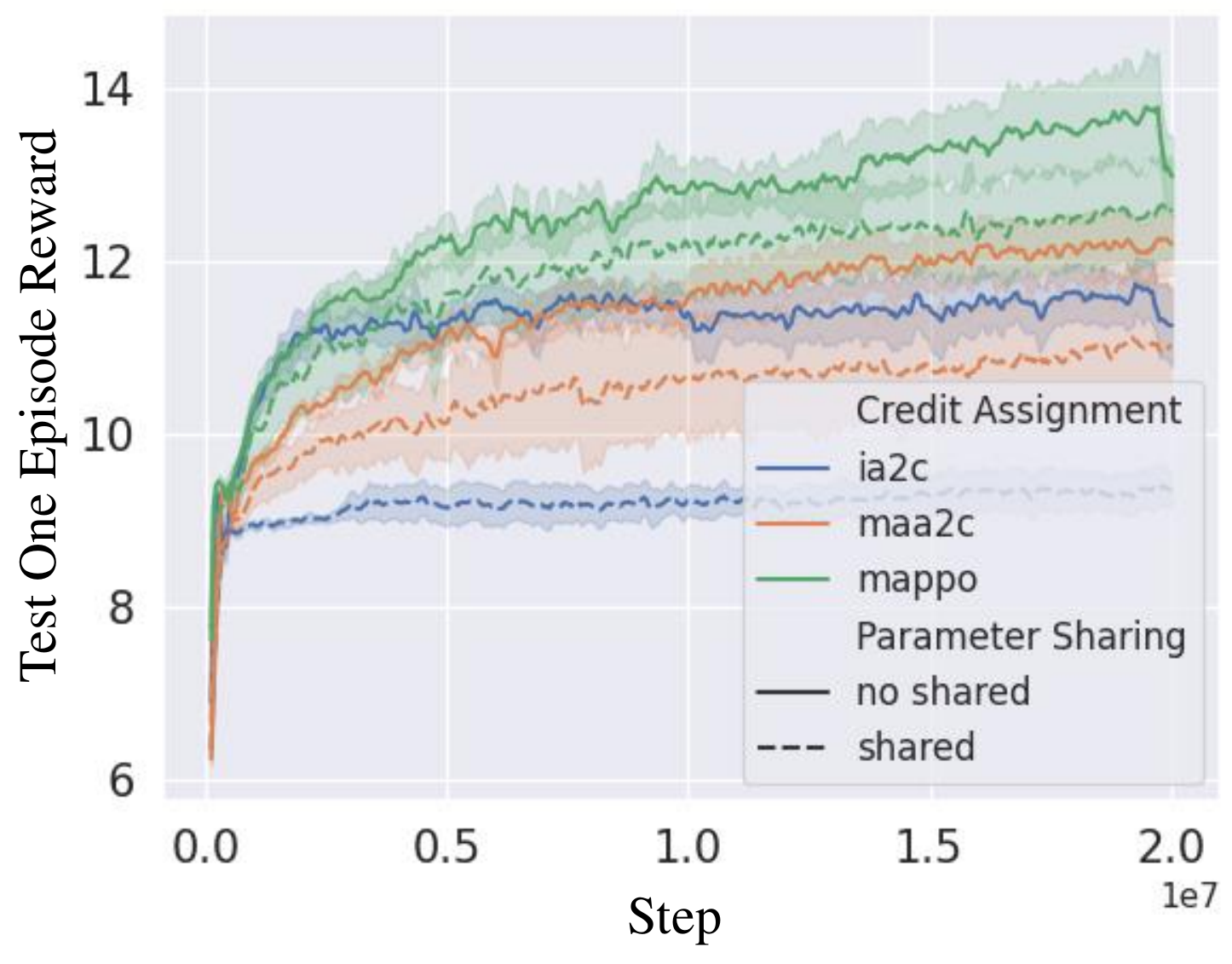}
  \caption{1c1s1z\_vs\_1c1s3z}
  \label{chart2:1c1s1z_vs_1c1s3z_P}
\end{subfigure}\hfil 
\begin{subfigure}{0.24\textwidth}
  \includegraphics[width=\linewidth]{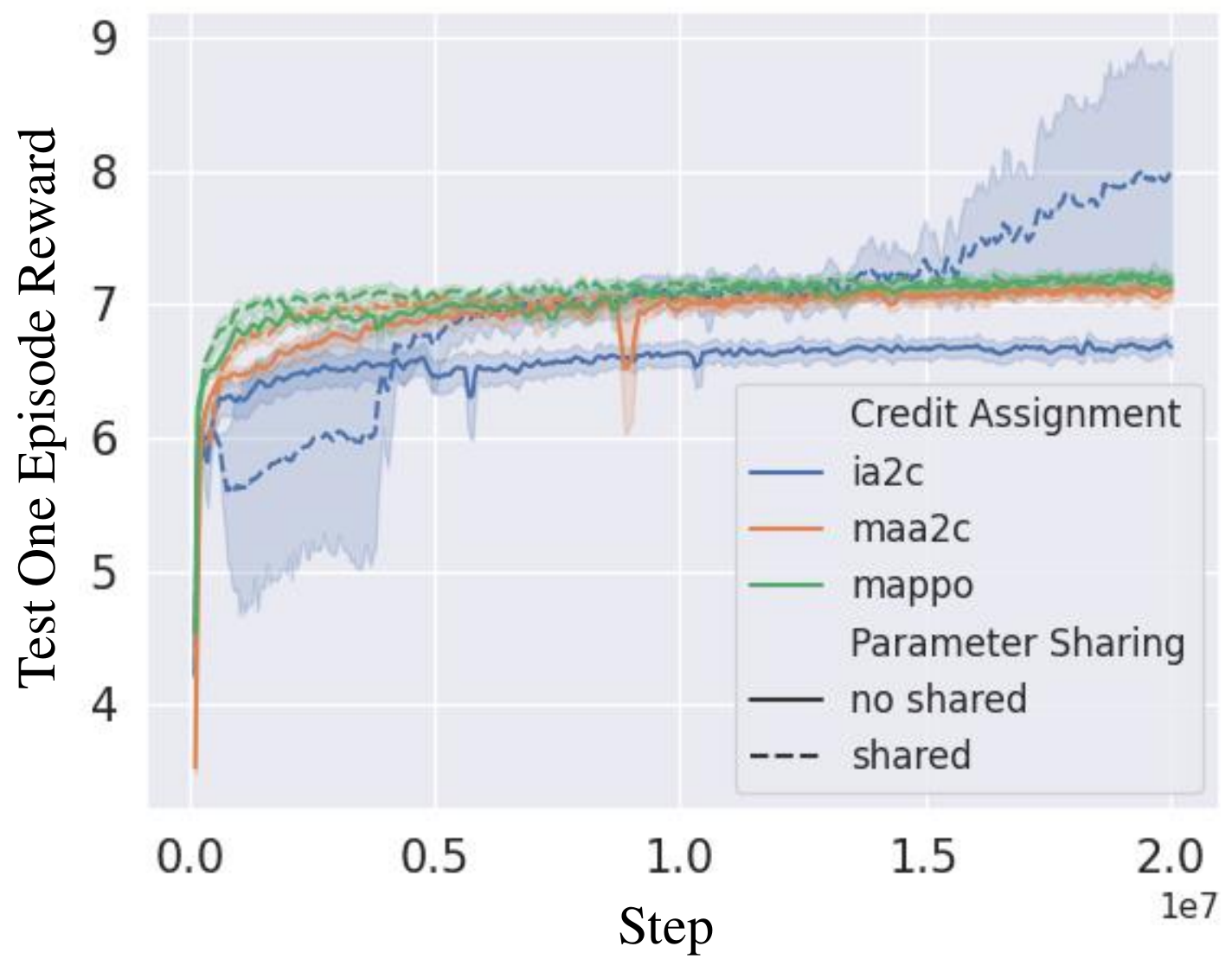}
  \caption{4m\_vs\_5m}
  \label{chart2:4m_vs_5m_P}
\end{subfigure}\hfil 
 \vspace{-13pt}
\caption{Performance curves include Q value-based(first row) and policy gradient-based(second row) credit assignment with {\it Shared} and {\it No shared} parameter sharing strategies.}
\label{fig:P and Q}
\vspace{-15pt}
\end{figure*}

Action-based role influences the convergence speed and final performance of different parameter sharing strategies in cooperative MARL. The scenarios we choose from the MPE and SMAC benchmarks are simple but diverse, covering action-based role diversity that ranges from small to large. In table.~\ref{table:parameter sharing vdn} and Fig.~\ref{fig:P and Q}, we provide action-based role diversity and the model performance curve on these chosen scenarios. For the SMAC benchmark, we adopt two metrics to count $r_{T}^{u}$ in Eq.~\ref{eq:policy based role}. The first metric is {\bf real action diversity}, which treats each action as an independent. The second way is {\bf semantic action diversity}, which distributes the actions to different groups according to their semantic type (e.g. move \& attack). There is no semantic action-based role in scenarios chosen from the MPE benchmark as all the actions are of the same semantic type (move). The action-based role diversity is then calculated according to Eq.~\ref{eq:policy diversity}. The base MARL credit assignment strategy we choose in table.~\ref{table:parameter sharing vdn} is {\bf VDN\cite{sunehag2017value}}, combined with a fully/partly/no parameter sharing strategy. For details of how the partial parameter sharing strategy works, please refer to the Appx.~\ref{app:Parameter Sharing}. We also evaluate other popular credit assignment strategies including {\bf IQL\cite{tan1993multi}, IA2C, MADDPG\cite{lowe2017multi} MAPPO\cite{yu2021surprising}, MAA2C} and {\bf QMIX\cite{rashid2018qmix}} combined with a fully/no parameter sharing strategy in Fig.~\ref{fig:P and Q}. {\bf IA2C} and {\bf MAA2C} are the extension of {\bf A2C\cite{mnih2016asynchronous}} on multi-agent scenarios.

Table.~\ref{table:parameter sharing vdn} outlines the performance of three parameter training strategies including {\it No shared}, {\it Partly shared} and {\it Shared} with the base credit assignment method VDN\cite{sunehag2017value}. Detailed framework and settings can be found in Appx.~\ref{app:Parameter Sharing}. As the action-based role diversity increases, the performance of the {\it Shared} strategy is degraded in terms of both convergence speed (half training steps) and the final reward (full training steps). One interesting phenomenon also emerges: the same agent type (e.g. 3s\_vs\_5z, 4m\_vs\_4z) does not always indicate small action-based role diversity, and vice versa (e.g. 1s1m1h1M\_vs\_3z), which means it is hard to define the role before identifying an adequate policy. Fig.~\ref{fig:P and Q} shows the model performances of two parameter sharing strategies including {\it No shared} and {\it Shared} with different credit assignment methods. For policy gradient-based methods, we extend the training steps from the standard 2M to 20M (10 times) as the convergence speed of policy gradient-based methods (e.g. MAPPO, MAA2C) is slower than Q value-based methods. From Fig.~\ref{fig:P and Q}, we find that different credit assignment methods have a slight impact on parameter sharing strategies but the trend in which no parameter sharing strategy achieves performance improvement continues to be present as the action-based role diversity increases. 

Here we conclude that scenarios with large action-based role diversity prefer no parameter sharing strategy, and vice versa. A suitable parameter sharing strategy helps obtain faster convergence speed and higher final performance.

\vspace{-7pt}
\subsection{Communication}\label{exp:Communicate: When and How}
\vspace{-3pt}

 \begin{table*}[t]
\small
 \caption{
 Different vision scopes (6-9-18) impact the model performance. The smallest scope is {\it 6}, which is the attack range of agents. The best performance is marked in red. Vision scope columns are marked with gradient red. The deeper the red is, the better the policy performance. Detailed analysis can be found in Sec.~\ref{exp:Communicate: When and How}.
 }
 \vspace{-8pt}
 \begin{center}
 \resizebox{1.0\textwidth}{!}{
 \setlength{\tabcolsep}{1.3em}
 {\renewcommand{\arraystretch}{1.0}
\begin{tabular}{|c|ccc|c|ccc|}
\hline
scenario & obs overlap & scope & performance & scenario & obs overlap  & scope  & performance  \\ 
\hline
& {\cellcolor[gray]{.8}} & \cellcolor{red!5} 6  & 15.6 / 19.5 / 19.5  &  & {\cellcolor[gray]{.8}} &  \cellcolor{red!5} 6  & 6.3 / 9.2 / 10.4  \\ 
1s1m1h1M\_vs\_3z  & {\cellcolor[gray]{.8}} 0.41 & \cellcolor{red!15}  9  & 16.4 / 19.5 /19.7  & 4m\_vs\_5m &{\cellcolor[gray]{.8}}  0.47 & \cellcolor{red!15}  9 & 6.5 / 10.1 / 10.9 \\ 
&  {\cellcolor[gray]{.8}}  & \cellcolor{red!25} 18 & {\color[HTML]{FE0000} 16.1 / 19.6 / 19.9} & & {\cellcolor[gray]{.8}} & \cellcolor{red!25}  18  & {\color[HTML]{FE0000} 6.8 / 10.9 / 11.1}   \\ 
\hline
 & {\cellcolor[gray]{.9}} &  \cellcolor{red!15} 6   & 8.4 / 15.3 / 18.8   &   &  {\cellcolor[gray]{.9}}  &  \cellcolor{red!5} 6 & {\color[HTML]{000000} 11.5 / 15.1 / 17.6} \\ 
1s1m1h1M\_vs\_4z & {\cellcolor[gray]{.9}} 0.25 &  \cellcolor{red!25} 9   & {\color[HTML]{FE0000} 8.4 / 15.7 / 19.7}  & 1c1s1z\_vs\_1c1s3z & {\cellcolor[gray]{.9}} 0.40 &  \cellcolor{red!25} 9  & {\color[HTML]{FE0000} 12.3 / 16.0 / 17.8}  \\ 
 & {\cellcolor[gray]{.9}} &  \cellcolor{red!5} 18     & 7.8 / 11.8 / 15.9   &  & {\cellcolor[gray]{.9}} &  \cellcolor{red!15} 18     & 12.4 / 15.3 / 17.6 \\ 
\hline
& {\cellcolor[gray]{.95}} &  \cellcolor{red!25} 6  & {\color[HTML]{FE0000} 6.6 / 14.2 / 17.7}  &  & {\cellcolor[gray]{.95}}  &  \cellcolor{red!25} 6 & {\color[HTML]{FE0000} 6.0 / 15.1 / 17.5}   \\ 
1s1m1h1M\_vs\_5z & {\cellcolor[gray]{.95}} 0.18 &  \cellcolor{red!15} 9  & 6.3 / 12.6 / 15.3  &  3s\_vs\_5z  &  {\cellcolor[gray]{.95}} 0.21  &  \cellcolor{red!15} 9  & 5.4 / 12.9 / 16.4  \\ 
 & {\cellcolor[gray]{.95}} &  \cellcolor{red!5} 18 & 5.9 / 8.9 / 10.4  & & {\cellcolor[gray]{.95}} &  \cellcolor{red!5} 18 & 5.2 / 9.0 / 12.1 \\ 
 \hline
\end{tabular}
 }
 }
 \end{center}
 \vspace{-20pt}

 \label{table:scope}
 \end{table*}

 \begin{table*}[!h]
\small
 \caption{
 Action-based role diversity influences the performance of different parameter sharing strategies on the MPE\cite{lowe2017multi} and SMAC\cite{samvelyan19smac} benchmarks. The best performance in each scenario is marked \checkmark with red cell color. Asterisks denote the algorithms that are not significantly different. Q value curves can be referred to Fig.~\ref{fig:q_diversity_curve_policy}.
 }
  \vspace{-8pt}
 \begin{center}
 \resizebox{1.0\textwidth}{!}{
 \setlength{\tabcolsep}{1.3em}
 {\renewcommand{\arraystretch}{1.0}
\begin{tabular}{|c|c|ccc|ccc|}
\hline
{\color[HTML]{000000} }                           & {\color[HTML]{000000} }                              & \multicolumn{3}{c|}{{\color[HTML]{000000} no shared}}                                                                                     & \multicolumn{3}{c|}{{\color[HTML]{000000} shared}}                                                                                        \\ \cline{3-8} 
\multirow{-2}{*}{{\color[HTML]{000000} scenario}} & \multirow{-2}{*}{{\color[HTML]{000000} Q diversity}} & {\color[HTML]{000000} vdn}                    & {\color[HTML]{000000} qmix}                   & {\color[HTML]{000000} iql}                & {\color[HTML]{000000} vdn}                    & {\color[HTML]{000000} qmix}                   & {\color[HTML]{000000} iql}                \\ \hline
{\color[HTML]{000000} 1c1s1z\_vs\_1c1s3z}         & \cellcolor[gray]{0.95}{\color[HTML]{000000} }                   & {\color[HTML]{000000} 12.3 / 15.9 / 17.9 }     & \cellcolor{red!20}{\color[HTML]{000000} 12.9 / 17.8 / 19.4 \checkmark} & {\color[HTML]{000000} 10.8 / 12.3 / 12.2} & {\color[HTML]{000000} 11.2 / 14.5 / 17.2 }     & {\color[HTML]{000000} 12.5 / 15.8 / 18.4 \checkmark \cellcolor{red!20}} & {\color[HTML]{000000} 9.8 / 11.2 / 11.9}  \\  
{\color[HTML]{000000} 3s\_vs\_5z}                 & \cellcolor[gray]{0.95}{\color[HTML]{000000} }                              & {\color[HTML]{000000} 5.4 / 12.9 / 16.4 }      & {\color[HTML]{000000} 4.6 / 13.5 / 17.0 \checkmark \cellcolor{red!20}}  & {\color[HTML]{000000} 4.6 / 5.1 / 7.9}    & {\color[HTML]{000000} 6.0 / 13.6 / 17.2 }      & {\color[HTML]{000000} 4.2 / 12.9 / 20.0 \checkmark \cellcolor{red!20}}  & {\color[HTML]{000000} 4.3 / 5.3 / 7.8}    \\  
{\color[HTML]{000000} 4m\_vs\_4z}                 &  \multirow{-3}{*}{\cellcolor[gray]{.95}{\color[HTML]{000000} \textless 0.1} }         & {\color[HTML]{000000} 4.3 / 13.2 / 17.1 }      & {\color[HTML]{000000} 4.3 / 18.3 / 18.8 \checkmark \cellcolor{red!20}}  & {\color[HTML]{000000} 3.3 / 3.2 / 3.7}    & {\color[HTML]{000000} 4.6 / 9.8 / 12.8 }       & {\color[HTML]{000000} 4.3 / 14.8 / 16.5 \checkmark \cellcolor{red!20}}  & {\color[HTML]{000000} 2.6 / 3.2 / 3.2}    \\ \hline
{\color[HTML]{000000} 4m\_vs\_5m}                 & \cellcolor[gray]{0.85}{\color[HTML]{000000} }                              & {\color[HTML]{000000} 6.5 / 10.1 / 10.9 * \cellcolor{red!10}}  & {\color[HTML]{000000} 7.0 / 9.9 / 10.9 * \cellcolor{red!10}}   & {\color[HTML]{000000} 4.8 / 7.6 / 8.1}    & {\color[HTML]{000000} 6.8 / 11.9 / 12.6 * \cellcolor{red!10}}  & {\color[HTML]{000000} 6.9 / 12.4 / 13.3 * \cellcolor{red!10}}  & {\color[HTML]{000000} 5.1 / 8.1 / 8.5}    \\ 
{\color[HTML]{000000} 4m\_vs\_3z}                 & \multirow{-2}{*}{\cellcolor[gray]{0.85}{\color[HTML]{000000} 0.1-0.5}}           & {\color[HTML]{000000} 7.5 / 19.6 / 19.3 * \cellcolor{red!10}}  & {\color[HTML]{000000} 6.5 / 19.7 / 19.3 * \cellcolor{red!10}}  & {\color[HTML]{000000} 4.5 / 5.7 / 11.1}   & {\color[HTML]{000000} 6.3 / 19.1 / 19.5 * \cellcolor{red!10}}  & {\color[HTML]{000000} 6.1 / 19.7 / 19.7 * \cellcolor{red!10}}  & {\color[HTML]{000000} 4.2 / 4.5 / 5.7}    \\ \hline
{\color[HTML]{000000} 1s1m1h1M\_vs\_3z}           & \cellcolor[gray]{0.75}{\color[HTML]{000000} }                              & {\color[HTML]{000000} 16.4 / 19.6 / 19.6 \checkmark \cellcolor{red!20}} & {\color[HTML]{000000} 6.5 / 7.5 / 7.8}        & {\color[HTML]{000000} 11.1 / 16.9 / 19.2 } & {\color[HTML]{000000} 16.1 / 19.6 / 19.8 \checkmark \cellcolor{red!20}} & {\color[HTML]{000000} 9.9 / 9.8 / 8.9}        & {\color[HTML]{000000} 12.2 / 17.9 / 19.6 }  \\ 
{\color[HTML]{000000} 1s1m1h1M\_vs\_4z}           & \cellcolor[gray]{0.75}{\color[HTML]{000000} }                              & {\color[HTML]{000000} 8.4 / 16.0 / 19.8 \checkmark \cellcolor{red!20}}  & {\color[HTML]{000000} 4.9 / 5.1 / 6.1}        & {\color[HTML]{000000} 7.4 / 9.0 / 10.7}    & {\color[HTML]{000000} 8.1 / 13.5 / 18.2 \checkmark \cellcolor{red!20}}  & {\color[HTML]{000000} 5.5 / 5.0 / 5.0}        & {\color[HTML]{000000} 7.1 / 8.5 / 8.5}    \\ 
{\color[HTML]{000000} 1s1m1h1M\_vs\_5z}           & \multirow{-3}{*}{\cellcolor[gray]{0.75}{\color[HTML]{000000} \textgreater 0.5}}           & {\color[HTML]{000000} 6.3 / 12.6 / 15.3 \checkmark \cellcolor{red!20}}  & {\color[HTML]{000000} 4.2 / 4.2 / 3.6}        & {\color[HTML]{000000} 5.5 / 6.1 / 6.5}     & {\color[HTML]{000000} 6.2 / 9.9 / 12.3 \checkmark \cellcolor{red!20}}   & {\color[HTML]{000000} 4.0 / 2.5 / 4.2}        & {\color[HTML]{000000} 5.4 / 6.3 / 6.3}     \\ \hline
\end{tabular}
 }
 }
 \end{center}

 \vspace{-20pt}
 \label{table:contribution}
 \end{table*}

More information is better. This principle is common sense in many areas including computer vision or natural language processing. With a bigger dataset and more detailed and accurate annotations, the model can be better optimized. However, in reinforcement learning, the data is sampled using the current policy; in most cases, this policy learning starts with a randomly initialized neural network. In this way, the provision of more information may introduce a burden for the policy optimization and can moreover further degrade the sampled data quality, which gives rise to a vicious circle. Therefore, it is critical to determine when and how to accept the extra information provided via communication mechanism in the context of cooperative MARL. As discussed in Sec.~\ref{sec:3}, the pattern of different agents' support sets for policy optimization can determine whether or not the extra information is needed. Notably, the similarity of these support sets is largely dependent on the trajectories of different agents, corresponding to the trajectory-based role diversity defined in Sec.~\ref{sec:Trajectory Based Role}. Small trajectory-based role diversity corresponds to a similar support set pattern, which means that forming a concentrated input is preferred for policy optimization. Experiment results in table.~\ref{table:communication} and Fig.~\ref{fig:Histogram} further prove that scenarios with larger observation overlap are more suitable for communication. Detailed setting of communication mechanism can be found in Appx.~\ref{app:Communication Framework}

To prove that small trajectory-based role diversity prefers obtaining extra information via communication and vice versa, we conduct extra experiments to study the relationship between the pattern of the input observation (support set) and the model performance by shrinking the vision scope ($r$ in eq.~\ref{eq:observation overlap}). The results can be found in table~\ref{table:scope}. The results show that the model performance is strongly related to the visual scope and the trajectory-based role diversity determines whether the small or large vision is preferred. Small trajectory-based role diversity prefers large vision scope, indicating the similar pattern of support set is better. Large trajectory-based role diversity prefers small vision scope, which means enlarging the pattern difference benefits the policy optimization. This further proves that extra information provided by communication forms a similar pattern of support set which is preferred in scenarios with small trajectory-based role diversity.

\vspace{-7pt}
\subsection{Credit Assignment}\label{exp:Credit Assignment}
\vspace{-3pt}

The performance of different credit assignment methods is strongly related to the contribution-based role. We mainly focus on three representative Q value-based MARL algorithm: VDN\cite{sunehag2017value}, QMIX\cite{rashid2018qmix} and IQL\cite{tan1993multi}, and compare their performance on different scenarios that have different contribution-based role diversity measured by Q value according to Eq.~\ref{eq:contribution based role}. The result can be found in table.~\ref{table:contribution}. In small Q diversity ($\simeq 0.5$) scenarios, QMIX significantly outperforms VDN with both shared and no shared strategies. With the increase of Q diversity, the performance of QMIX starts to degrade. In scenarios where agents have significantly different Q value distribution (Fig.~\ref{fig:q_diversity_curve_policy} 1s1m1h1M\_vs\_3/4/5z), VDN significantly outperforms QMIX. As for IQL, the performance is not as good as VDN and QMIX in most scenarios. However, IQL is not sensitive to Q diversity and can perform well in easy scenarios like 1s1m1h1M\_vs\_3z. Combined with theoretical analysis in Sec.~\ref{sec:3}, we can conclude that QMIX is not suitable for a large contribution-based role diversity scenario because of the additional value decomposition module, which is a {\it sum} function in VDN and a {\it learnable neural network} in QMIX. The neural network fails to minimize the approximation error for $Q_{tot}$, and is an extra burden when the reward function (or the contribution to the global reward) is diverse. IQL has no such problem as it treats $Q_{tot}$ as the individual Q value. From this part, we conclude that using credit assignment methods with a learnable value decomposition module should be avoided in scenarios with large contribution-based role diversity. 




\vspace{-7pt}
\subsection{Policy Diagnosis Guideline}\label{sec:guide}
\vspace{-3pt}

Based on our theoretical and experimental analysis, we contend that role diversity is a strong candidate metric to aid in diagnosing the policy shortcomings and selecting the proper training strategy for cooperative MARL. If the action-based role diversity is large, we should choose a no parameter sharing strategy, and vice versa. If the trajectory-based role diversity is large, we should avoid communication or other information sharing, and vice versa. If the contribution-based role diversity is large, the fixed credit assignment method or independent learning method is preferred, and vice versa. In this way, we can avoid the possible performance bottlenecks in the context of cooperative MARL. 


\section{Conclusion}\label{sec:Conclusion}

In this paper, we define the role and role diversity to measure a cooperative MARL task and help diagnose the current policy. We claim that a strong relationship between the role diversity and model performance exists and we prove it through both theoretical analyses on MARL error bound decomposition and experiments conducted on MARL benchmarks. The experiment results clearly show that the role diversity significantly impacts the model performance of different training strategies and this effect is ubiquitous in various environments and algorithms. Based on this, we provide a policy diagnosis guideline for a better policy in cooperative MARL.




\nocite{langley00}

\bibliography{example_paper}

\begin{thebibliography}{65}
\providecommand{\natexlab}[1]{#1}
\providecommand{\url}[1]{\texttt{#1}}
\expandafter\ifx\csname urlstyle\endcsname\relax
  \providecommand{\doi}[1]{doi: #1}\else
  \providecommand{\doi}{doi: \begingroup \urlstyle{rm}\Url}\fi

\bibitem[Anderson et~al.(2018)Anderson, Wu, Teney, Bruce, Johnson,
  S{\"{u}}nderhauf, Reid, Gould, and van~den Hengel]{anderson2018vision}
Anderson, P., Wu, Q., Teney, D., Bruce, J., Johnson, M., S{\"{u}}nderhauf, N.,
  Reid, I.~D., Gould, S., and van~den Hengel, A.
\newblock Vision-and-language navigation: Interpreting visually-grounded
  navigation instructions in real environments.
\newblock In \emph{{CVPR}}, 2018.

\bibitem[Anthony \& Bartlett(2002)Anthony and Bartlett]{anthony2009neural}
Anthony, M. and Bartlett, P.~L.
\newblock \emph{Neural Network Learning - Theoretical Foundations}.
\newblock Cambridge University Press, 2002.

\bibitem[Baker et~al.(2020)Baker, Kanitscheider, Markov, Wu, Powell, McGrew,
  and Mordatch]{baker2019emergent}
Baker, B., Kanitscheider, I., Markov, T.~M., Wu, Y., Powell, G., McGrew, B.,
  and Mordatch, I.
\newblock Emergent tool use from multi-agent autocurricula.
\newblock In \emph{{ICLR}}, 2020.

\bibitem[Bard et~al.(2020)Bard, Foerster, Chandar, Burch, Lanctot, Song,
  Parisotto, Dumoulin, Moitra, Hughes, et~al.]{bard2020hanabi}
Bard, N., Foerster, J.~N., Chandar, S., Burch, N., Lanctot, M., Song, H.~F.,
  Parisotto, E., Dumoulin, V., Moitra, S., Hughes, E., et~al.
\newblock The hanabi challenge: A new frontier for ai research.
\newblock \emph{Artificial Intelligence}, 280:\penalty0 103216, 2020.

\bibitem[Ba{\c{s}}ar \& Olsder(1998)Ba{\c{s}}ar and Olsder]{bacsar1998dynamic}
Ba{\c{s}}ar, T. and Olsder, G.~J.
\newblock \emph{Dynamic noncooperative game theory}.
\newblock SIAM, 1998.

\bibitem[Berner et~al.(2019)Berner, Brockman, Chan, Cheung, D{\k{e}}biak,
  Dennison, Farhi, Fischer, Hashme, Hesse, et~al.]{berner2019dota}
Berner, C., Brockman, G., Chan, B., Cheung, V., D{\k{e}}biak, P., Dennison, C.,
  Farhi, D., Fischer, Q., Hashme, S., Hesse, C., et~al.
\newblock Dota 2 with large scale deep reinforcement learning.
\newblock \emph{arXiv preprint arXiv:1912.06680}, 2019.

\bibitem[Brown \& Sandholm(2019)Brown and Sandholm]{brown2019superhuman}
Brown, N. and Sandholm, T.
\newblock Superhuman ai for multiplayer poker.
\newblock \emph{Science}, 365\penalty0 (6456):\penalty0 885--890, 2019.

\bibitem[Christianos et~al.(2021)Christianos, Papoudakis, Rahman, and
  Albrecht]{christianos2021scaling}
Christianos, F., Papoudakis, G., Rahman, M.~A., and Albrecht, S.~V.
\newblock Scaling multi-agent reinforcement learning with selective parameter
  sharing.
\newblock In \emph{{ICML}}, 2021.

\bibitem[Ernst et~al.(2005)Ernst, Geurts, and Wehenkel]{ernst05a}
Ernst, D., Geurts, P., and Wehenkel, L.
\newblock Tree-based batch mode reinforcement learning.
\newblock \emph{J. Mach. Learn. Res.}, 6:\penalty0 503--556, 2005.

\bibitem[Fan et~al.(2020)Fan, Wang, Xie, and Yang]{fan2020theoretical}
Fan, J., Wang, Z., Xie, Y., and Yang, Z.
\newblock A theoretical analysis of deep q-learning.
\newblock In \emph{{L4DC}}, 2020.

\bibitem[Farahmand et~al.(2010)Farahmand, Munos, and
  Szepesv{\'{a}}ri]{farahmand2010error}
Farahmand, A.~M., Munos, R., and Szepesv{\'{a}}ri, C.
\newblock Error propagation for approximate policy and value iteration.
\newblock In \emph{{NIPS}}, 2010.

\bibitem[Farahmand et~al.(2016)Farahmand, Ghavamzadeh, Szepesv{\'a}ri, and
  Mannor]{farahmand2016regularized}
Farahmand, A.-m., Ghavamzadeh, M., Szepesv{\'a}ri, C., and Mannor, S.
\newblock Regularized policy iteration with nonparametric function spaces.
\newblock \emph{The Journal of Machine Learning Research}, 17\penalty0
  (1):\penalty0 4809--4874, 2016.

\bibitem[Foerster et~al.(2018)Foerster, Farquhar, Afouras, Nardelli, and
  Whiteson]{foerster2017counterfactual}
Foerster, J.~N., Farquhar, G., Afouras, T., Nardelli, N., and Whiteson, S.
\newblock Counterfactual multi-agent policy gradients.
\newblock In \emph{{AAAI}}, 2018.

\bibitem[Haarnoja et~al.(2018)Haarnoja, Zhou, Abbeel, and
  Levine]{haarnoja2018soft}
Haarnoja, T., Zhou, A., Abbeel, P., and Levine, S.
\newblock Soft actor-critic: Off-policy maximum entropy deep reinforcement
  learning with a stochastic actor.
\newblock In \emph{{ICML}}, 2018.

\bibitem[Hessel et~al.(2018)Hessel, Modayil, Van~Hasselt, Schaul, Ostrovski,
  Dabney, Horgan, Piot, Azar, and Silver]{hessel2018rainbow}
Hessel, M., Modayil, J., Van~Hasselt, H., Schaul, T., Ostrovski, G., Dabney,
  W., Horgan, D., Piot, B., Azar, M., and Silver, D.
\newblock Rainbow: Combining improvements in deep reinforcement learning.
\newblock In \emph{{AAAI}}, 2018.

\bibitem[Hostallero et~al.(2019)Hostallero, Son, Kim, and
  Qtran]{hostallero2019learning}
Hostallero, W. J. K. D.~E., Son, K., Kim, D., and Qtran, Y.~Y.
\newblock Learning to factorize with transformation for cooperative multi-agent
  reinforcement learning.
\newblock In \emph{{ICML}}, 2019.

\bibitem[Hu \& Wellman(2003)Hu and Wellman]{hu2003nash}
Hu, J. and Wellman, M.~P.
\newblock Nash q-learning for general-sum stochastic games.
\newblock \emph{Journal of machine learning research}, 4\penalty0
  (Nov):\penalty0 1039--1069, 2003.

\bibitem[Hu et~al.(2021)Hu, Zhu, Chang, and Liang]{HuZCL21}
Hu, S., Zhu, F., Chang, X., and Liang, X.
\newblock Updet: Universal multi-agent {RL} via policy decoupling with
  transformers.
\newblock In \emph{ICLR}, 2021.

\bibitem[Jiang \& Lu(2018)Jiang and Lu]{jiang2018learning}
Jiang, J. and Lu, Z.
\newblock Learning attentional communication for multi-agent cooperation.
\newblock In \emph{{NeurIPS}}, 2018.

\bibitem[Kim et~al.(2019)Kim, Moon, Hostallero, Kang, Lee, Son, and
  Yi]{kim2019learning}
Kim, D., Moon, S., Hostallero, D., Kang, W.~J., Lee, T., Son, K., and Yi, Y.
\newblock Learning to schedule communication in multi-agent reinforcement
  learning.
\newblock In \emph{{ICLR}}, 2019.

\bibitem[Kulkarni et~al.(2016)Kulkarni, Narasimhan, Saeedi, and
  Tenenbaum]{kulkarni2016hierarchical}
Kulkarni, T.~D., Narasimhan, K., Saeedi, A., and Tenenbaum, J.
\newblock Hierarchical deep reinforcement learning: Integrating temporal
  abstraction and intrinsic motivation.
\newblock \emph{{NIPS}}, 2016.

\bibitem[Kurach et~al.(2020)Kurach, Raichuk, Stanczyk, Zajac, Bachem, Espeholt,
  Riquelme, Vincent, Michalski, Bousquet, and Gelly]{kurach2019google}
Kurach, K., Raichuk, A., Stanczyk, P., Zajac, M., Bachem, O., Espeholt, L.,
  Riquelme, C., Vincent, D., Michalski, M., Bousquet, O., and Gelly, S.
\newblock Google research football: {A} novel reinforcement learning
  environment.
\newblock In \emph{{AAAI}}, 2020.

\bibitem[Lazaric et~al.(2016)Lazaric, Ghavamzadeh, and
  Munos]{lazaric2016analysis}
Lazaric, A., Ghavamzadeh, M., and Munos, R.
\newblock Analysis of classification-based policy iteration algorithms.
\newblock \emph{Journal of Machine Learning Research}, 17:\penalty0 1--30,
  2016.

\bibitem[Lazaridou \& Baroni(2020)Lazaridou and Baroni]{lazaridou2020emergent}
Lazaridou, A. and Baroni, M.
\newblock Emergent multi-agent communication in the deep learning era.
\newblock \emph{arXiv preprint arXiv:2006.02419}, 2020.

\bibitem[Le et~al.(2017)Le, Yue, Carr, and Lucey]{le2017coordinated}
Le, H.~M., Yue, Y., Carr, P., and Lucey, P.
\newblock Coordinated multi-agent imitation learning.
\newblock In \emph{{ICML}}, 2017.

\bibitem[Li et~al.(2020)Li, Koyamada, Ye, Liu, Wang, Yang, Zhao, Qin, Liu, and
  Hon]{li2020suphx}
Li, J., Koyamada, S., Ye, Q., Liu, G., Wang, C., Yang, R., Zhao, L., Qin, T.,
  Liu, T.-Y., and Hon, H.-W.
\newblock Suphx: Mastering mahjong with deep reinforcement learning.
\newblock \emph{arXiv preprint arXiv:2003.13590}, 2020.

\bibitem[Lin(1992)]{lin1992self}
Lin, L.~J.
\newblock Self-improving reactive agents based on reinforcement learning,
  planning and teaching.
\newblock \emph{Mach. Learn.}, 8:\penalty0 293--321, 1992.

\bibitem[Littman(2001)]{littman2001friend}
Littman, M.~L.
\newblock Friend-or-foe q-learning in general-sum games.
\newblock In \emph{ICML}, 2001.

\bibitem[Lowe et~al.(2017)Lowe, Wu, Tamar, Harb, Abbeel, and
  Mordatch]{lowe2017multi}
Lowe, R., Wu, Y., Tamar, A., Harb, J., Abbeel, P., and Mordatch, I.
\newblock Multi-agent actor-critic for mixed cooperative-competitive
  environments.
\newblock In \emph{{NIPS}}, 2017.

\bibitem[Mguni et~al.(2021)Mguni, Wu, Du, Yang, Wang, Li, Wen, Jennings, and
  Wang]{mguni2021learning}
Mguni, D.~H., Wu, Y., Du, Y., Yang, Y., Wang, Z., Li, M., Wen, Y., Jennings,
  J., and Wang, J.
\newblock Learning in nonzero-sum stochastic games with potentials.
\newblock In \emph{{ICML}}, 2021.

\bibitem[Mnih et~al.(2015)Mnih, Kavukcuoglu, Silver, Rusu, Veness, Bellemare,
  Graves, Riedmiller, Fidjeland, Ostrovski, et~al.]{mnih2015human}
Mnih, V., Kavukcuoglu, K., Silver, D., Rusu, A.~A., Veness, J., Bellemare,
  M.~G., Graves, A., Riedmiller, M., Fidjeland, A.~K., Ostrovski, G., et~al.
\newblock Human-level control through deep reinforcement learning.
\newblock \emph{Nature}, 518\penalty0 (7540):\penalty0 529--533, 2015.

\bibitem[Mnih et~al.(2016)Mnih, Badia, Mirza, Graves, Lillicrap, Harley,
  Silver, and Kavukcuoglu]{mnih2016asynchronous}
Mnih, V., Badia, A.~P., Mirza, M., Graves, A., Lillicrap, T., Harley, T.,
  Silver, D., and Kavukcuoglu, K.
\newblock Asynchronous methods for deep reinforcement learning.
\newblock In \emph{{ICML}}, 2016.

\bibitem[Munos \& Szepesv{\'a}ri(2008)Munos and
  Szepesv{\'a}ri]{munos2008finite}
Munos, R. and Szepesv{\'a}ri, C.
\newblock Finite-time bounds for fitted value iteration.
\newblock \emph{Journal of Machine Learning Research}, 9\penalty0 (5), 2008.

\bibitem[Oliehoek et~al.(2016)Oliehoek, Amato, et~al.]{oliehoek2016concise}
Oliehoek, F.~A., Amato, C., et~al.
\newblock \emph{A concise introduction to decentralized POMDPs}, volume~1.
\newblock Springer, 2016.

\bibitem[Papoudakis et~al.(2021)Papoudakis, Christianos, Sch{\"a}fer, and
  Albrecht]{papoudakis2021benchmarking}
Papoudakis, G., Christianos, F., Sch{\"a}fer, L., and Albrecht, S.~V.
\newblock Benchmarking multi-agent deep reinforcement learning algorithms in
  cooperative tasks.
\newblock In \emph{{NeurIPS}}, 2021.

\bibitem[Rashid et~al.(2018)Rashid, Samvelyan, de~Witt, Farquhar, Foerster, and
  Whiteson]{rashid2018qmix}
Rashid, T., Samvelyan, M., de~Witt, C.~S., Farquhar, G., Foerster, J.~N., and
  Whiteson, S.
\newblock {QMIX:} monotonic value function factorisation for deep multi-agent
  reinforcement learning.
\newblock In \emph{{ICML}}, 2018.

\bibitem[Redmon et~al.(2016)Redmon, Divvala, Girshick, and
  Farhadi]{redmon2016you}
Redmon, J., Divvala, S., Girshick, R., and Farhadi, A.
\newblock You only look once: Unified, real-time object detection.
\newblock In \emph{{CVPR}}, 2016.

\bibitem[Ren et~al.(2015)Ren, He, Girshick, and Sun]{ren2015faster}
Ren, S., He, K., Girshick, R., and Sun, J.
\newblock Faster r-cnn: Towards real-time object detection with region proposal
  networks.
\newblock \emph{{NIPS}}, 28:\penalty0 91--99, 2015.

\bibitem[Riedmiller(2005)]{riedmiller2005neural}
Riedmiller, M.~A.
\newblock Neural fitted {Q} iteration - first experiences with a data efficient
  neural reinforcement learning method.
\newblock In \emph{{ECML}}, 2005.

\bibitem[Samvelyan et~al.(2019)Samvelyan, Rashid, de~Witt, Farquhar, Nardelli,
  Rudner, Hung, Torr, Foerster, and Whiteson]{samvelyan19smac}
Samvelyan, M., Rashid, T., de~Witt, C.~S., Farquhar, G., Nardelli, N., Rudner,
  T. G.~J., Hung, C.-M., Torr, P. H.~S., Foerster, J., and Whiteson, S.
\newblock {The} {StarCraft} {Multi}-{Agent} {Challenge}.
\newblock \emph{CoRR}, abs/1902.04043, 2019.

\bibitem[Scherrer et~al.(2015)Scherrer, Ghavamzadeh, Gabillon, Lesner, and
  Geist]{scherrer2015approximate}
Scherrer, B., Ghavamzadeh, M., Gabillon, V., Lesner, B., and Geist, M.
\newblock Approximate modified policy iteration and its application to the game
  of tetris.
\newblock \emph{Journal of Machine Learning Research}, 16:\penalty0 1629--1676,
  2015.

\bibitem[Schmidt-Hieber(2020)]{schmidt2020nonparametric}
Schmidt-Hieber, J.
\newblock Nonparametric regression using deep neural networks with relu
  activation function.
\newblock \emph{The Annals of Statistics}, 48\penalty0 (4):\penalty0
  1875--1897, 2020.

\bibitem[Schulman et~al.(2015)Schulman, Levine, Abbeel, Jordan, and
  Moritz]{schulman2015trust}
Schulman, J., Levine, S., Abbeel, P., Jordan, M., and Moritz, P.
\newblock Trust region policy optimization.
\newblock In \emph{{ICML}}, 2015.

\bibitem[Schulman et~al.(2017)Schulman, Wolski, Dhariwal, Radford, and
  Klimov]{schulman2017proximal}
Schulman, J., Wolski, F., Dhariwal, P., Radford, A., and Klimov, O.
\newblock Proximal policy optimization algorithms.
\newblock \emph{arXiv preprint arXiv:1707.06347}, 2017.

\bibitem[Silver et~al.(2017)Silver, Hubert, Schrittwieser, Antonoglou, Lai,
  Guez, Lanctot, Sifre, Kumaran, Graepel, et~al.]{silver2017mastering}
Silver, D., Hubert, T., Schrittwieser, J., Antonoglou, I., Lai, M., Guez, A.,
  Lanctot, M., Sifre, L., Kumaran, D., Graepel, T., et~al.
\newblock Mastering chess and shogi by self-play with a general reinforcement
  learning algorithm.
\newblock \emph{arXiv preprint arXiv:1712.01815}, 2017.

\bibitem[Singh et~al.(2019)Singh, Jain, and Sukhbaatar]{singh2018learning}
Singh, A., Jain, T., and Sukhbaatar, S.
\newblock Learning when to communicate at scale in multiagent cooperative and
  competitive tasks.
\newblock In \emph{{ICLR}}, 2019.

\bibitem[Suarez et~al.(2021)Suarez, Du, Zhu, Mordatch, and
  Isola]{suarez2021neural}
Suarez, J., Du, Y., Zhu, C., Mordatch, I., and Isola, P.
\newblock The neural {MMO} platform for massively multiagent research.
\newblock \emph{CoRR}, abs/2110.07594, 2021.

\bibitem[Sukhbaatar et~al.(2016)Sukhbaatar, Szlam, and
  Fergus]{sukhbaatar2016learning}
Sukhbaatar, S., Szlam, A., and Fergus, R.
\newblock Learning multiagent communication with backpropagation.
\newblock In \emph{{NIPS}}, 2016.

\bibitem[Sunehag et~al.(2018)Sunehag, Lever, Gruslys, Czarnecki, Zambaldi,
  Jaderberg, Lanctot, Sonnerat, Leibo, Tuyls, and Graepel]{sunehag2017value}
Sunehag, P., Lever, G., Gruslys, A., Czarnecki, W.~M., Zambaldi, V.~F.,
  Jaderberg, M., Lanctot, M., Sonnerat, N., Leibo, J.~Z., Tuyls, K., and
  Graepel, T.
\newblock Value-decomposition networks for cooperative multi-agent learning
  based on team reward.
\newblock In \emph{{AAMAS}}, 2018.

\bibitem[Tan(1993)]{tan1993multi}
Tan, M.
\newblock Multi-agent reinforcement learning: Independent vs. cooperative
  agents.
\newblock In \emph{{ICML}}, 1993.

\bibitem[Terry et~al.(2020)Terry, Grammel, Hari, Santos, and
  Black]{terry2020revisiting}
Terry, J.~K., Grammel, N., Hari, A., Santos, L., and Black, B.
\newblock Revisiting parameter sharing in multi-agent deep reinforcement
  learning.
\newblock \emph{arXiv preprint arXiv:2005.13625}, 2020.

\bibitem[Tuyls et~al.(2021)Tuyls, Omidshafiei, Muller, Wang, Connor, Hennes,
  Graham, Spearman, Waskett, Steel, et~al.]{tuyls2021game}
Tuyls, K., Omidshafiei, S., Muller, P., Wang, Z., Connor, J., Hennes, D.,
  Graham, I., Spearman, W., Waskett, T., Steel, D., et~al.
\newblock Game plan: What ai can do for football, and what football can do for
  ai.
\newblock \emph{Journal of Artificial Intelligence Research}, 71:\penalty0
  41--88, 2021.

\bibitem[Vinyals et~al.(2019)Vinyals, Babuschkin, Czarnecki, Mathieu, Dudzik,
  Chung, Choi, Powell, Ewalds, Georgiev, et~al.]{vinyals2019grandmaster}
Vinyals, O., Babuschkin, I., Czarnecki, W.~M., Mathieu, M., Dudzik, A., Chung,
  J., Choi, D.~H., Powell, R., Ewalds, T., Georgiev, P., et~al.
\newblock Grandmaster level in starcraft ii using multi-agent reinforcement
  learning.
\newblock \emph{Nature}, 575\penalty0 (7782):\penalty0 350--354, 2019.

\bibitem[Wang et~al.(2020{\natexlab{a}})Wang, Ren, Han, Ye, and
  Zhang]{wang2020towards}
Wang, J., Ren, Z., Han, B., Ye, J., and Zhang, C.
\newblock Towards understanding linear value decomposition in cooperative
  multi-agent q-learning.
\newblock \emph{arXiv preprint arXiv:2006.00587}, 2020{\natexlab{a}}.

\bibitem[Wang et~al.(2020{\natexlab{b}})Wang, Dong, Lesser, and
  Zhang]{wang2020roma}
Wang, T., Dong, H., Lesser, V.~R., and Zhang, C.
\newblock {ROMA:} multi-agent reinforcement learning with emergent roles.
\newblock In \emph{{ICML}}, 2020{\natexlab{b}}.

\bibitem[Wang et~al.(2021)Wang, Gupta, Mahajan, Peng, Whiteson, and
  Zhang]{wang2020rode}
Wang, T., Gupta, T., Mahajan, A., Peng, B., Whiteson, S., and Zhang, C.
\newblock {RODE:} learning roles to decompose multi-agent tasks.
\newblock In \emph{{ICLR}}, 2021.

\bibitem[Yang et~al.(2018)Yang, Luo, Li, Zhou, Zhang, and Wang]{yang2018mean}
Yang, Y., Luo, R., Li, M., Zhou, M., Zhang, W., and Wang, J.
\newblock Mean field multi-agent reinforcement learning.
\newblock In \emph{{ICML}}, 2018.

\bibitem[Ye et~al.(2020)Ye, Liu, Sun, Shi, Zhao, Wu, Yu, Yang, Wu, Guo,
  et~al.]{ye2020mastering}
Ye, D., Liu, Z., Sun, M., Shi, B., Zhao, P., Wu, H., Yu, H., Yang, S., Wu, X.,
  Guo, Q., et~al.
\newblock Mastering complex control in moba games with deep reinforcement
  learning.
\newblock In \emph{{AAAI}}, 2020.

\bibitem[Yu et~al.(2021)Yu, Velu, Vinitsky, Wang, Bayen, and
  Wu]{yu2021surprising}
Yu, C., Velu, A., Vinitsky, E., Wang, Y., Bayen, A., and Wu, Y.
\newblock The surprising effectiveness of mappo in cooperative, multi-agent
  games.
\newblock \emph{arXiv preprint arXiv:2103.01955}, 2021.

\bibitem[Zha et~al.(2021)Zha, Xie, Ma, Zhang, Lian, Hu, and
  Liu]{zha2021douzero}
Zha, D., Xie, J., Ma, W., Zhang, S., Lian, X., Hu, X., and Liu, J.
\newblock Douzero: Mastering doudizhu with self-play deep reinforcement
  learning.
\newblock In \emph{{ICML}}, 2021.

\bibitem[Zhang et~al.(2021)Zhang, Yang, Liu, Zhang, and Basar]{zhang2021finite}
Zhang, K., Yang, Z., Liu, H., Zhang, T., and Basar, T.
\newblock Finite-sample analysis for decentralized batch multi-agent
  reinforcement learning with networked agents.
\newblock \emph{IEEE Transactions on Automatic Control}, 2021.
\newblock \doi{10.1109/TAC.2021.3049345}.

\bibitem[Zheng et~al.(2018)Zheng, Yang, Cai, Zhou, Zhang, Wang, and
  Yu]{zheng2017magent}
Zheng, L., Yang, J., Cai, H., Zhou, M., Zhang, W., Wang, J., and Yu, Y.
\newblock Magent: {A} many-agent reinforcement learning platform for artificial
  collective intelligence.
\newblock In \emph{{AAAI}}, 2018.

\bibitem[Zhong et~al.(2021)Zhong, Sun, Luo, Yan, and Wang]{zhong2021towards}
Zhong, F., Sun, P., Luo, W., Yan, T., and Wang, Y.
\newblock Towards distraction-robust active visual tracking.
\newblock In \emph{{ICML}}, 2021.

\bibitem[Zhou et~al.(2020)Zhou, Luo, Villella, Yang, Rusu, Miao, Zhang, Alban,
  Fadakar, Chen, et~al.]{zhou2020smarts}
Zhou, M., Luo, J., Villella, J., Yang, Y., Rusu, D., Miao, J., Zhang, W.,
  Alban, M., Fadakar, I., Chen, Z., et~al.
\newblock Smarts: Scalable multi-agent reinforcement learning training school
  for autonomous driving.
\newblock \emph{arXiv preprint arXiv:2010.09776}, 2020.

\bibitem[Zhu et~al.(2021)Zhu, Hu, Zhang, Hong, Zhu, Chang, and
  Liang]{zhu2021main}
Zhu, F., Hu, S., Zhang, Y., Hong, H., Zhu, Y., Chang, X., and Liang, X.
\newblock Main: A multi-agent indoor navigation benchmark for cooperative
  learning.
\newblock 2021.

\end{thebibliography}
\bibliographystyle{icml2022}

\newpage
\appendix
\onecolumn

\section{Problem Formulation}

{\bf Multi-agent Reinforcement Learning}
A cooperative multi-agent task is a decentralized partially observable Markov decision process \cite{oliehoek2016concise} with a tuple $G=\left\langle \gS, \gA, \gU,P,r,\gZ,O,n,\gamma\right\rangle$. Let $\gS$ denote the global state of the environment, while $\gA$ represents the set of $n$ agents and $\gU$ is the action space. At each time step $t$, agent $a \in \gA \equiv \{1,...,n \}$  selects an action $u \in \gU$, forming a joint action $\rvu \in \gU^n$, which in turn causes a transition in the environment represented by the state transition function $ P(\rvs' \vert \rvs,\rvu): \gS\times \gU^n \times \gS \rightarrow\left [0,1\right] $. All agents share the same reward function $r(\rvs,\rvu): \gS\times \gU^n \rightarrow \R$ , while $\gamma \in \left [  0,1 \right )$ is a discount factor. For any state-action pair, the reward $r$ is bounded by $M$, i.e. $|r|\leq M.$ 
We consider a partially observable scenario in which each agent makes individual observations $\rz \in \gZ$ according to the observation function $O(\rvs,a):\gS \times \gA \rightarrow \gZ$. Each agent has an action-observation history that conditions a stochastic policy $\pi_t$, creating the following joint action value: $Q^{\pi}(\rz_t,\rvu_t) = \mathbb{E}_{s_{t+1:\infty},\mathbf{u}_{t+1:\infty}}\left [R_t  \vert \rvz_t,\mathbf{u}_t\right ]$, where $R_t={\textstyle\sum_{i=0}^\infty}\gamma^i r_{t+i}$ is the discounted return. 

{\bf Centralized training with decentralized execution}
Centralized training with decentralized execution (CTDE) is a commonly used architecture in the MARL context. Each agent is conditioned only on its own action-observation history to make a decision using the learned policy. The centralized value function provides a centralized gradient to update the individual function based on its output. Therefore, a stronger individual value function can benefit the centralized training.

\section{Observation Overlap Percentage Calculation}\label{app:obs overlap}

\subsection{Overlap Percentage Calculation in Games}\label{app:obs overlap smac}
In this part, we demonstrate how to calculate the observation overlap percentage in SMAC \cite{samvelyan19smac}.
As the partial observable area is circular, and the coordinate system is a 2D map with axis X and Y, the observation overlap on one battle scenario can be computed as:
\begin{align}\label{eq:observation overlap}
\begin{split}
l &= \sqrt{(x_{a_0}-x_{a_1})\cdot(x_{a_1}-x_{a_0})+(y_{a_0}-y_{a_1})\cdot(y_{a_1}-y_{a_0})} \\
p &= (1+2\cdot r)/2 \\
s &= 2\cdot \sqrt{p\cdot(p-l)\cdot(p-r)\cdot(p-r)} \\
o &= 2\cdot cos^{-1}(l/(2\cdot r))\cdot r\cdot r-s \\
d_{T}^{a_0,a_1} &= o/(\pi \cdot r^2)
\end{split}
\end{align}

Here $r$ is the vision scope. Notice that if $l<2r$, $d_{T}$ equals zero as no overlap exists. 
We provide the observation overlap curve in Fig.~\ref{fig_spec:traj_role_case} to show how trajectory-based role distance varies in one game.

\subsection{Overlap Percentage Calculation in Real World Scenario (Semantic)}

\begin{figure*}[!h]
\begin{center}
  \includegraphics[width=0.9\linewidth]{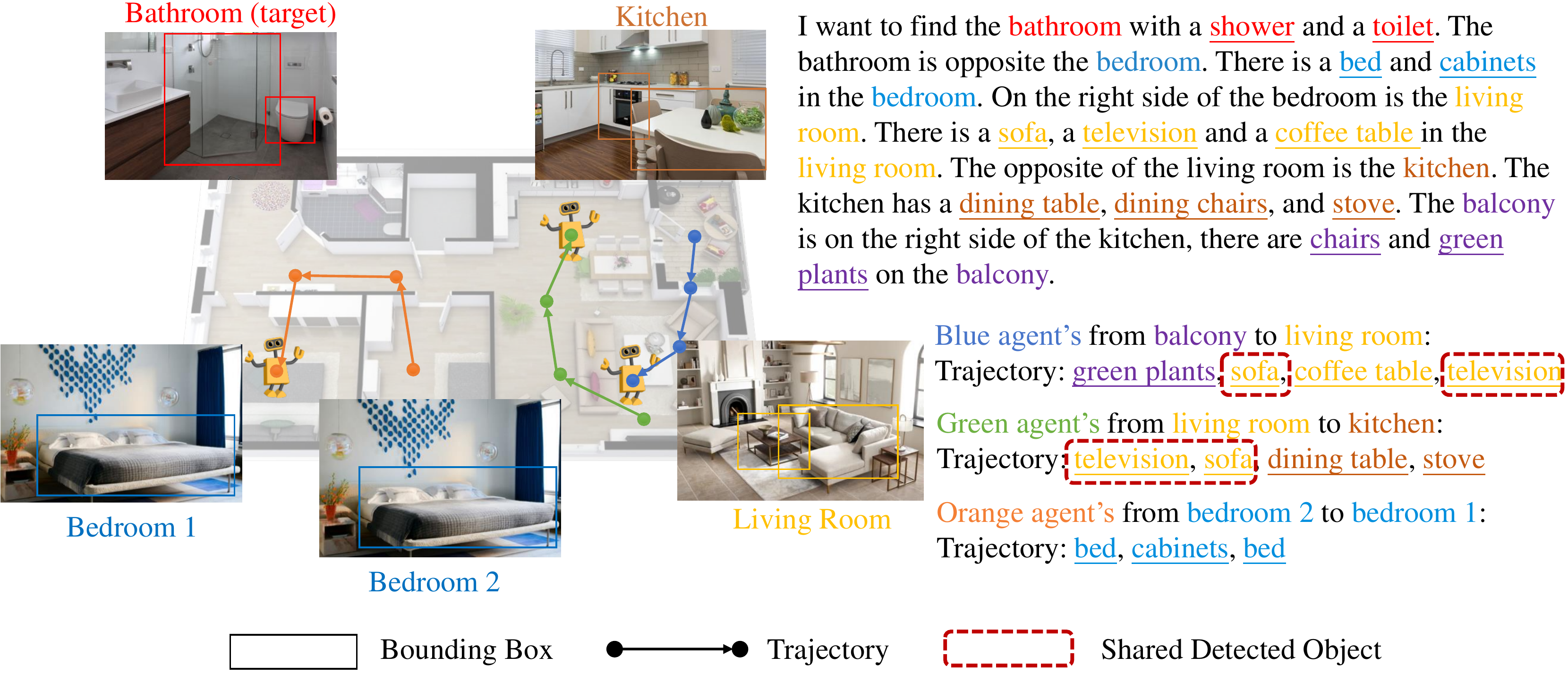}
\end{center}
  \caption{A multi-agent visual language navigation task. Agents are initialized in different locations and the target measurement is given. Agents need to cooperate with each other to find the target location according to the measurement as soon as possible.}
\vspace{-7pt}
\label{fig:vln_framework}
\end{figure*}

In this part, we demonstrate how to apply the observation overlap concept and trajectory-based role diversity calculation (Eq.~\ref{eq:traj based role diversity}) to real-world scenarios. Different from game scenarios like SMAC and MPE, the observation of real-world tasks is usually an image. For example, in the vision language navigation task (VLN\cite{anderson2018vision}), agents take real indoor scene pictures as the input, combine them with language description to locate the target as shown in Fig.~\ref{fig:vln_framework}. Considering the learning efficiency, object detection techniques like YOLO\cite{redmon2016you} and FasterRCNN\cite{ren2015faster} are used in VLN to help extract the objects from the scene pictures as semantic information. The semantic information can be recorded as part of the agents' trajectory, enabling agents to use the past information for future decisions. Under the multi-agent setting, agents are required to cooperate and find the target together. Therefore, trajectory overlap should be avoided, which means that large trajectory-based role diversity is preferred in this task, and policies that cause trajectory overlap should be punished. Directly using scene pictures as input or its feature pattern can bring large noise in observation overlap calculation. Instead, using semantic information from the detected object can significantly reduce the noise and serve as a good observation history representation. As shown in Fig.~\ref{fig:vln_framework}, the red dotted frames indicate that the blue agent and green agent share some similar observation semantic in their trajectories. In this way, the trajectory-based role diversity of multi-agent VLN task can be calculated the same as Eq.~\ref{eq:traj based role diversity}. Only the $d_{T}^{a_0,a_1}$ is replaced by the observation semantic overlap, which is the shared detected object percentage in total detected objects. In this way, without knowing the exact trajectory, we still manage to calculate the trajectory distance. And using overlap to represent the trajectory-based role diversity, we can keep this metric in a fixed range from 0 to 1.

\subsection{Overlap Percentage Calculation in Real World Scenario (Raw)}\label{app:obs raw input}

\begin{figure*}[!h]
\begin{center}
  \includegraphics[width=0.9\linewidth]{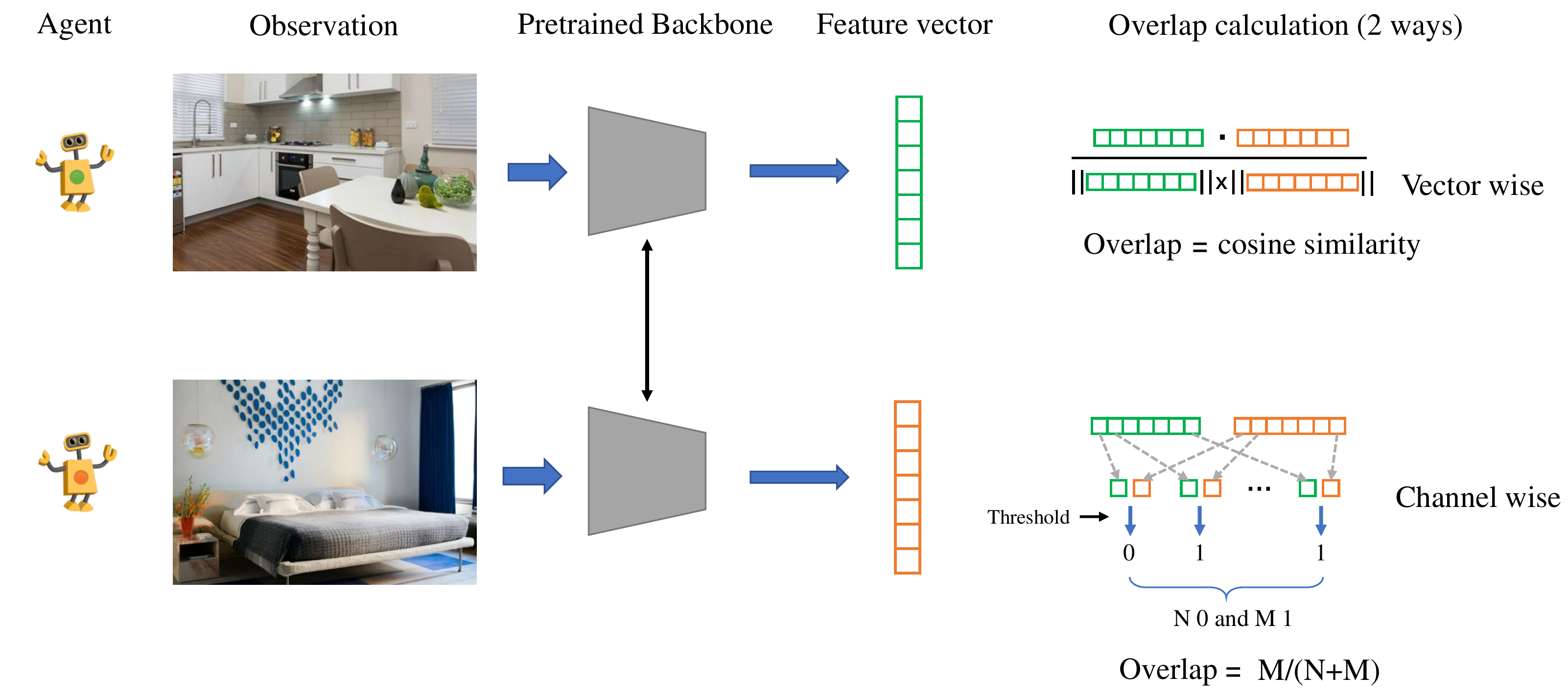}
\end{center}
  \caption{Using real images as observation overlap percentage calculation. Two methods are proposed including vector-wise cosine similarity and channel-wise threshold-based similarity percentage. A detailed discussion can be found in Sec.~\ref{app:obs raw input}}
\vspace{-7pt}
\label{fig:vln_real_input}
\end{figure*}

It is also possible that we can get observation overlap directly based on real image MARL tasks. As showed in Fig.~\ref{fig:vln_real_input}. Passing the input image to pre-trained CNN/Transformer backbone and getting its feature, we can use cosine similarity or channel-wise similarity to compute the overlap between different observation features as $d_{T}^{a_0,a_1}$ in Eq.~\ref{eq:traj based role diversity}. However, these methods can bring large noise to this metric. Moreover, how to stabilize the reinforcement learning with real pictures as input is still under investigation. In addition, it is rare in MARL tasks that the only information provided in the training stage is one single image. Location and communication are necessary auxiliary information to help learn the coordination of agents in most MARL tasks. Therefore, simply using the raw image to calculate the observation overlap can be a choice, but not the best choice.

\section{Types of Role}\label{app:Different Types of Role}

\begin{figure}[!h]
    \centering 
\begin{subfigure}{0.3\textwidth}
  \includegraphics[width=\linewidth]{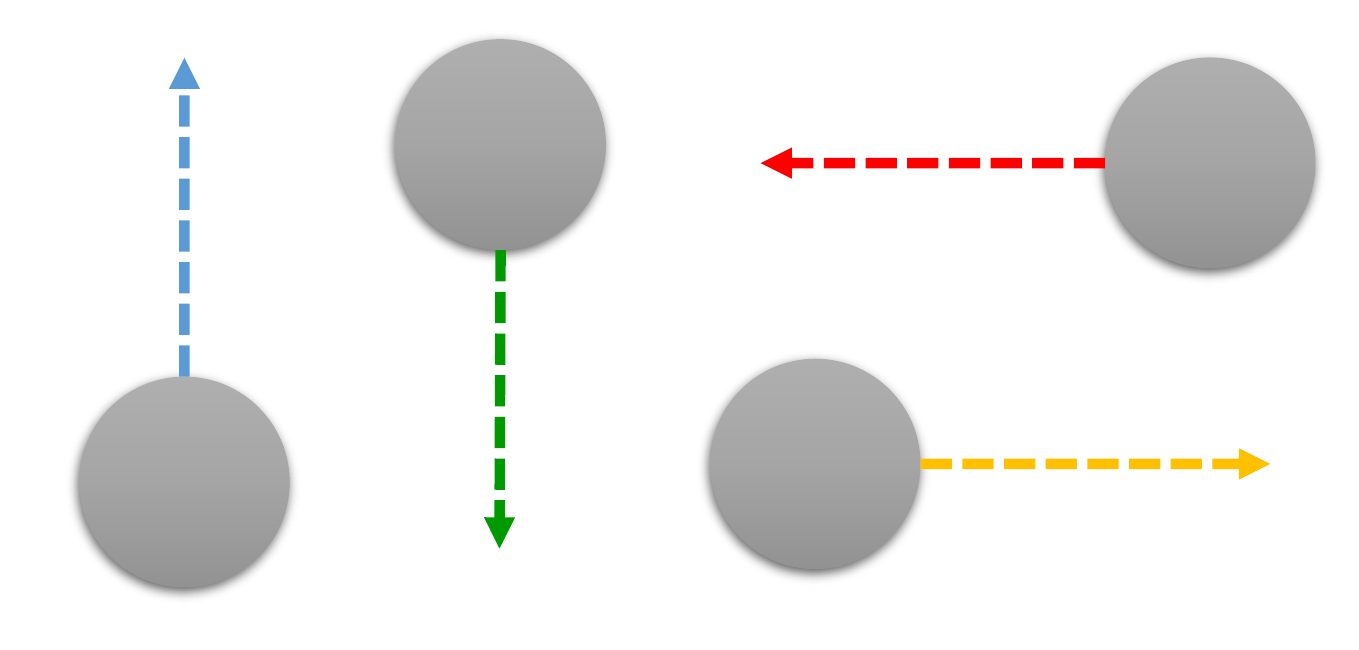}
  \caption{action-based role}
  \label{fig_abs:policy_role_case}
\end{subfigure}\hfil 
\begin{subfigure}{0.3\textwidth}
  \includegraphics[width=\linewidth]{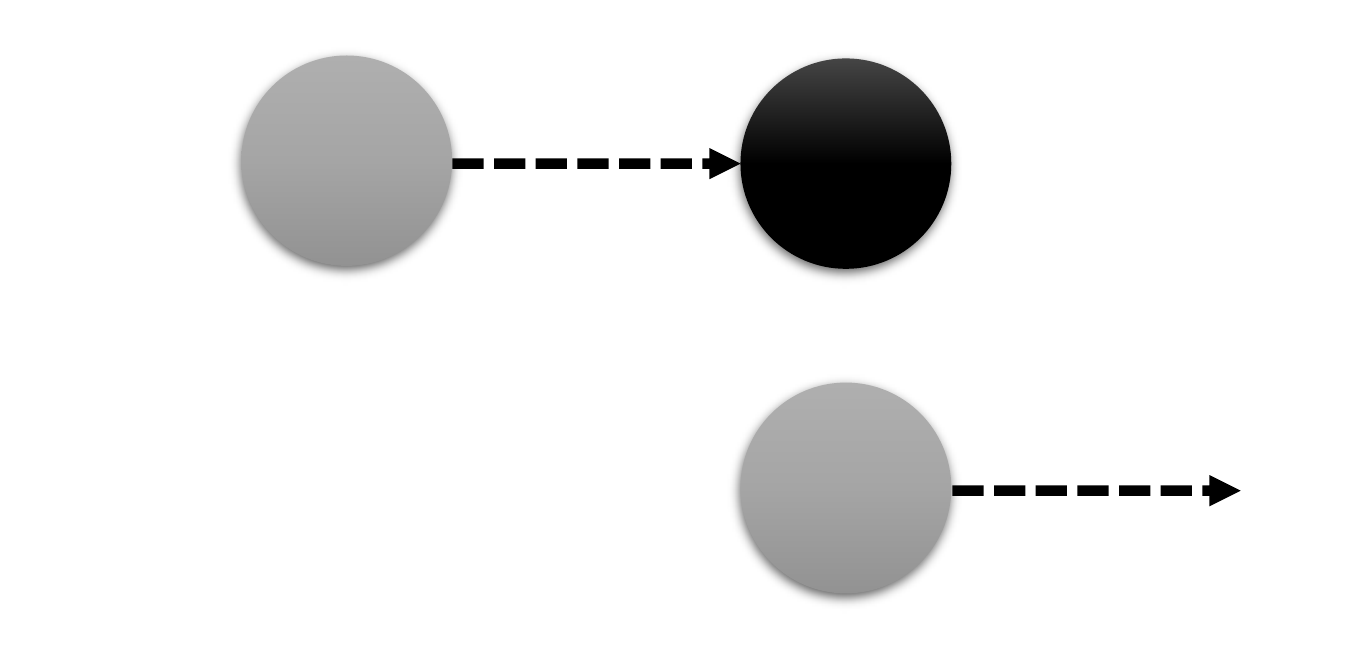}
  \caption{trajectory-based role}
  \label{fig_abs:traj_role_case}
\end{subfigure}\hfil 
\begin{subfigure}{0.3\textwidth}
  \includegraphics[width=\linewidth]{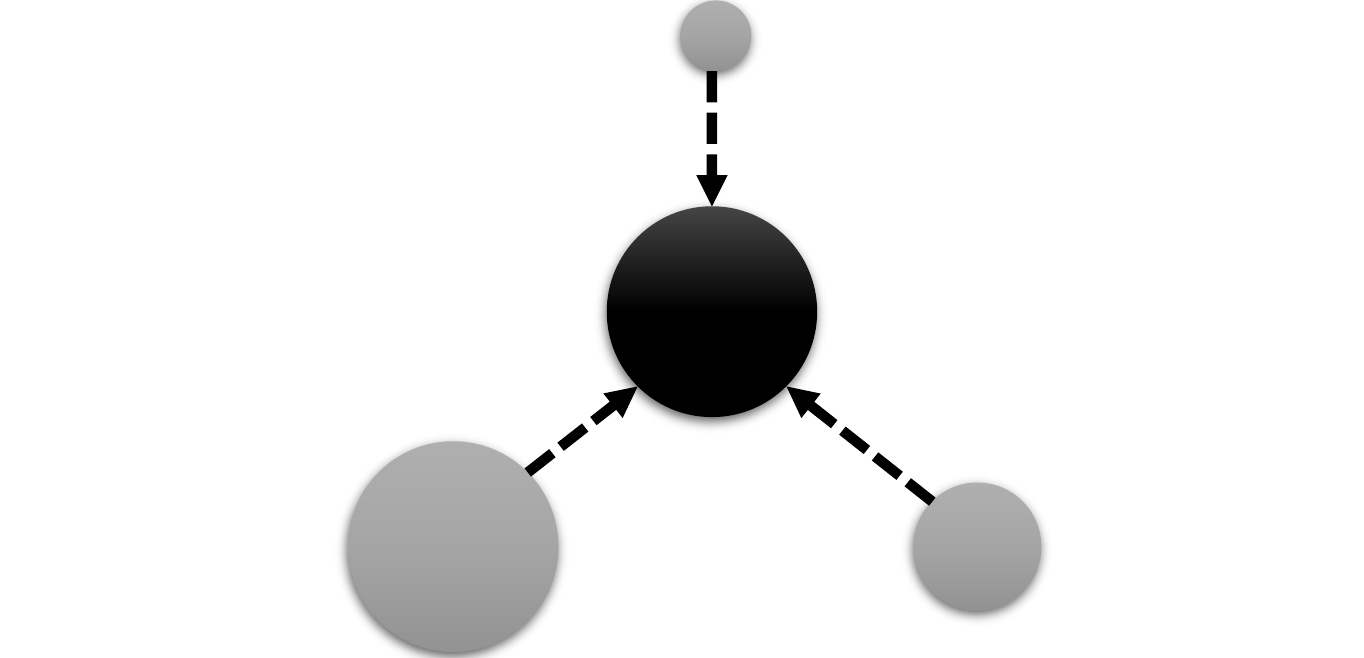}
  \caption{contribution-based role}
  \label{fig_abs:type_role_case}
\end{subfigure}
\caption{An illustration of different role based on MPE.}
\label{fig_abs:role_mpe}
\vspace{-7pt}
\end{figure}

We present two illustration figures for different types of role based on MPE \cite{lowe2017multi} and SMAC \cite{samvelyan19smac}. 
Fig.~\ref{fig_abs:role_mpe} is based on MPE. Grey circles and black circles represent agents and goals respectively. Dashed arrows in different colors represent different actions. Larger circles receive more rewards when they reach the goal.

\begin{figure}[!h]
    \centering 
\begin{subfigure}{0.5\textwidth}
  \includegraphics[width=\linewidth]{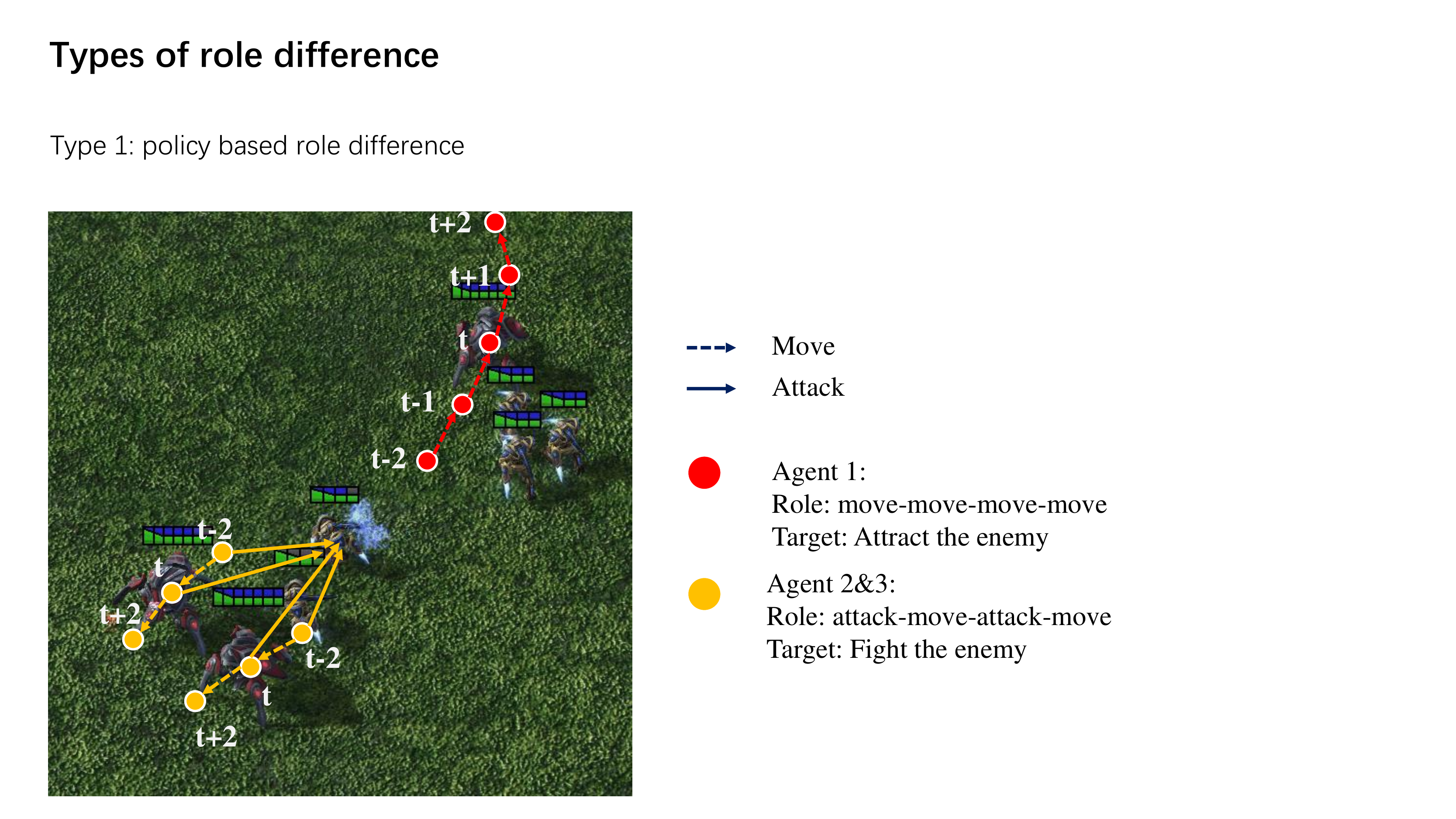}
  \caption{Action-based role difference in a period from t-2 to t+2. Action statistic shows only two different roles among three agents.}
  \label{fig:poli_based_role}
\end{subfigure}\hfil 
\begin{subfigure}{0.45\textwidth}
  \includegraphics[width=\linewidth]{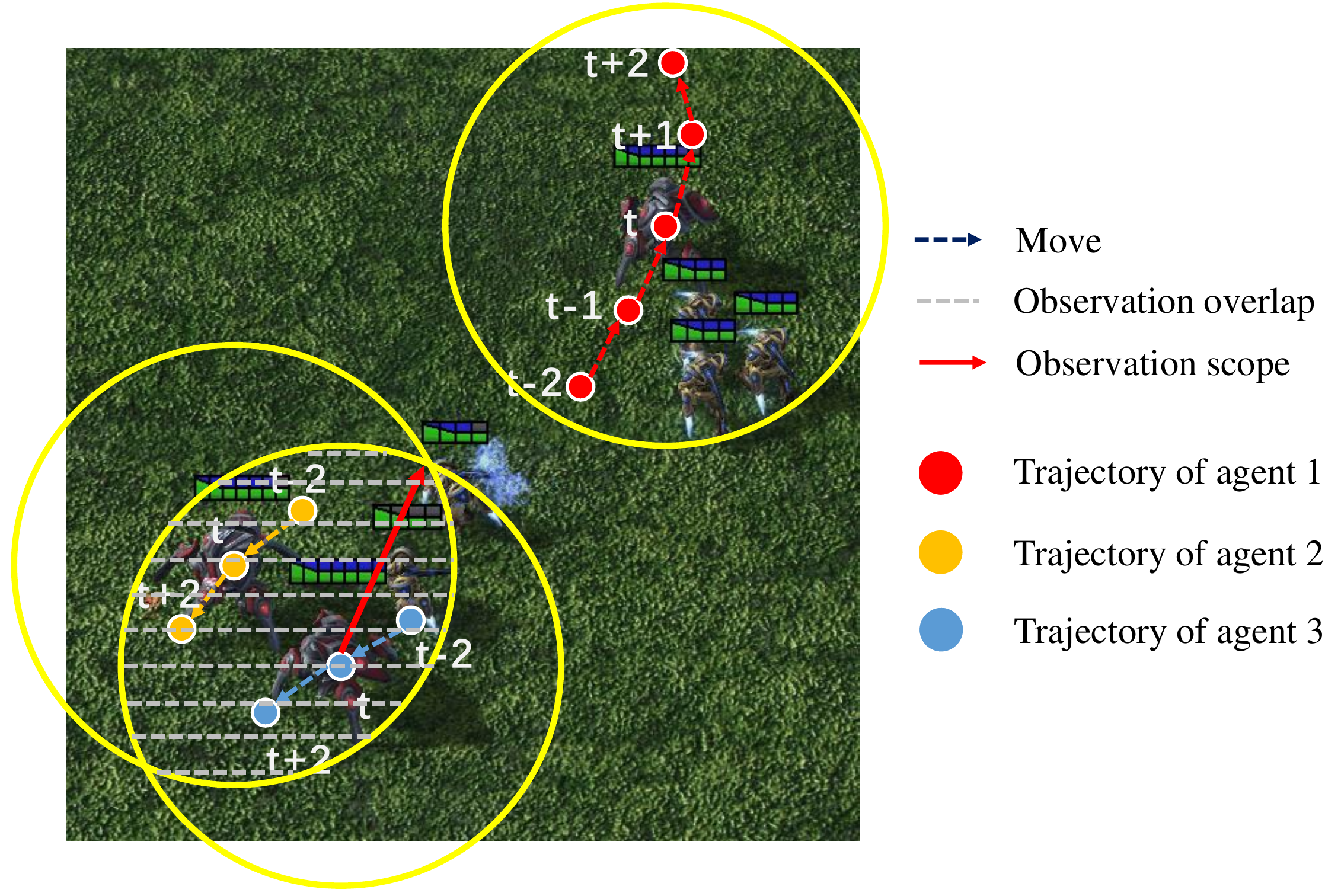}
  \caption{Trajectory-based role difference in a period from t-2 to t+2. The area covered by the grey dotted line is the observation overlap of agents 2 and 3. There is no overlap for agent 1.}
  \label{fig:traj_based_role}
\end{subfigure}\hfil 
\medskip
\vspace{7pt}
\begin{subfigure}{0.9\textwidth}
  \includegraphics[width=\linewidth]{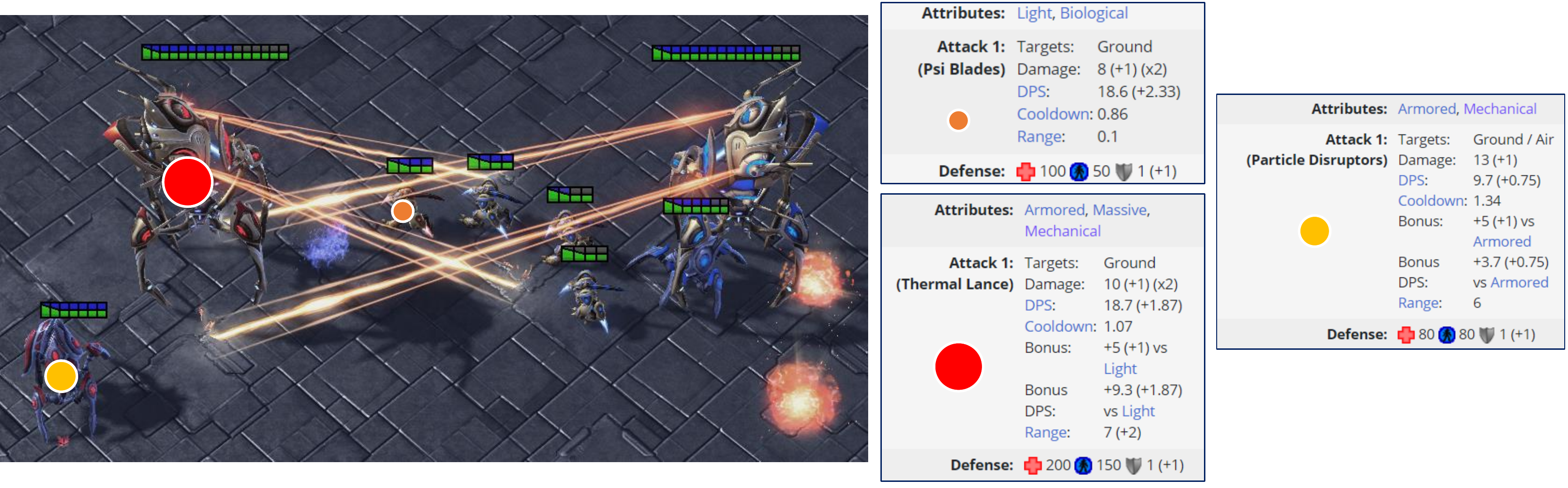}
  \caption{Contribution-based role difference depends on agents' original attributes including attack method and defense.}
  \label{fig:type_based_role}
\end{subfigure}\hfil 
\caption{Illustration of action-based role, trajectory-based role, and contribution-based role on real scenarios from SMAC.}
\label{fig:role_smac}
\end{figure}

Fig.~\ref{fig:role_smac} is based on SMAC. A detailed explanation can be found in the caption.

\section{Connections of Different Roles}\label{app: connection}

\begin{table*}[!h]
\small
 \caption{
Different role diversities on different scenarios from SMAC. The minimum value of one column is labeled in green and the largest value is labeled in red. Detailed analysis can be found in Appx.~\ref{app: connection}.
 }
 
 \begin{center}
 \resizebox{1.0\textwidth}{!}{
 \setlength{\tabcolsep}{1.3em}
 {\renewcommand{\arraystretch}{1.0}
\begin{tabular}{|c|c|c|c|c|}
\hline
{\color[HTML]{000000} Scenario}           & {\color[HTML]{000000} Semantic Action Diversity} & \multicolumn{1}{l|}{{\color[HTML]{000000} Real Action Diversity}} & {\color[HTML]{000000} Trajectory Diversity (overlap)} & {\color[HTML]{000000} Contribution Diversity (max Q)} \\ \hline
{\color[HTML]{000000} 4m\_vs\_5m}         & {\color[HTML]{009901} 1.5}                       & {\color[HTML]{009901} 9.1}                                        & {\color[HTML]{FE0000} 0.47}                           & {\color[HTML]{000000} 0.13}                           \\ \hline
{\color[HTML]{000000} 3s\_vs\_5z}         & {\color[HTML]{000000} 2.7}                       & {\color[HTML]{000000} 18.7}                                       & {\color[HTML]{009901} 0.21}                           & {\color[HTML]{000000} 0.09}                           \\ \hline
{\color[HTML]{000000} 4m\_vs\_4z}         & {\color[HTML]{000000} 3.3}                       & {\color[HTML]{000000} 19.3}                                       & {\color[HTML]{000000} 0.31}                           & {\color[HTML]{009901} 0.06}                           \\ \hline
{\color[HTML]{000000} 4m\_vs\_3z}         & {\color[HTML]{000000} 3.8}                       & {\color[HTML]{009901} 12.1}                                       & {\color[HTML]{000000} 0.35}                           & {\color[HTML]{000000} 0.25}                           \\ \hline
{\color[HTML]{000000} 1c1s1z\_vs\_1c1s3z} & {\color[HTML]{FE0000} 8.7}                       & {\color[HTML]{FE0000} 22.0}                                       & {\color[HTML]{000000} 0.40}                           & {\color[HTML]{009901} 0.03}                           \\ \hline
{\color[HTML]{000000} 1s1m1h1M\_vs\_3z}   & {\color[HTML]{009901} 2.4}                       & {\color[HTML]{000000} 13.2}                                       & {\color[HTML]{FE0000} 0.41}                           & {\color[HTML]{000000} 0.61}                           \\ \hline
{\color[HTML]{000000} 1s1m1h1M\_vs\_4z}   & {\color[HTML]{000000} 2.7}                       & {\color[HTML]{000000} 15.8}                                       & {\color[HTML]{000000} 0.25}                           & {\color[HTML]{FE0000} 0.75}                           \\ \hline
{\color[HTML]{000000} 1s1m1h1M\_vs\_5z}   & {\color[HTML]{FE0000} 6.2}                       & {\color[HTML]{FE0000} 22.5}                                       & {\color[HTML]{009901} 0.18}                           & {\color[HTML]{FE0000} 0.82}                           \\ \hline
\end{tabular}
 }
 }
 \end{center}
  \vspace{-9pt}

 \label{table:connection}
 \end{table*}
 
Is there any redundancy in the definition of different kinds of role diversity in Sec.~\ref{sec:Role Diversity}? Here we discuss the connections of different role diversity.

From the theoretical perspective, the contribution-based role diversity is a compound measurement of role diversity. It corresponds to the variance term in (\ref{eq31}). For the parameter sharing case, we decompose this variance into a sum of two terms: a bias term corresponds to the action-based role diversity and a variance term corresponds to the trajectory-based role diversity. Therefore, under the simple scenario in Sec.~\ref{sec:3}, we can find a clear relationship between different role diversity.  

From the experiment perspective, the decomposition in (\ref{eq32}) may not hold because of the more complicated settings. We have discussed this issue in the remark on Page 18.  Here we collect all different role diversity data in table.~\ref{table:connection}. We find scenarios like 3s\_vs\_5z have relatively small diversity in the action-based role while the observation overlap of trajectory diversity is small. Scenarios like 4m\_vs\_5m have small action-based role diversity while the observation overlap is large. Contribute-based role diversity can not be inferred from the action diversity and trajectory diversity, and is more depending on the agents' behavior difference.

In conclusion, the relationship between different roles exists in MARL training theoretically, while this relationship is not so significant in the experimental perspective due to more complicated settings of the real MARL tasks. Yet the strong relation between different role diversity and the MARL training process still exists with no conflict with the conclusion of this paper.
 
 
 
 

\section{Parameter Sharing}\label{app:Parameter Sharing}

\begin{figure*}[!h]
\begin{center}
  \includegraphics[width=0.95\linewidth]{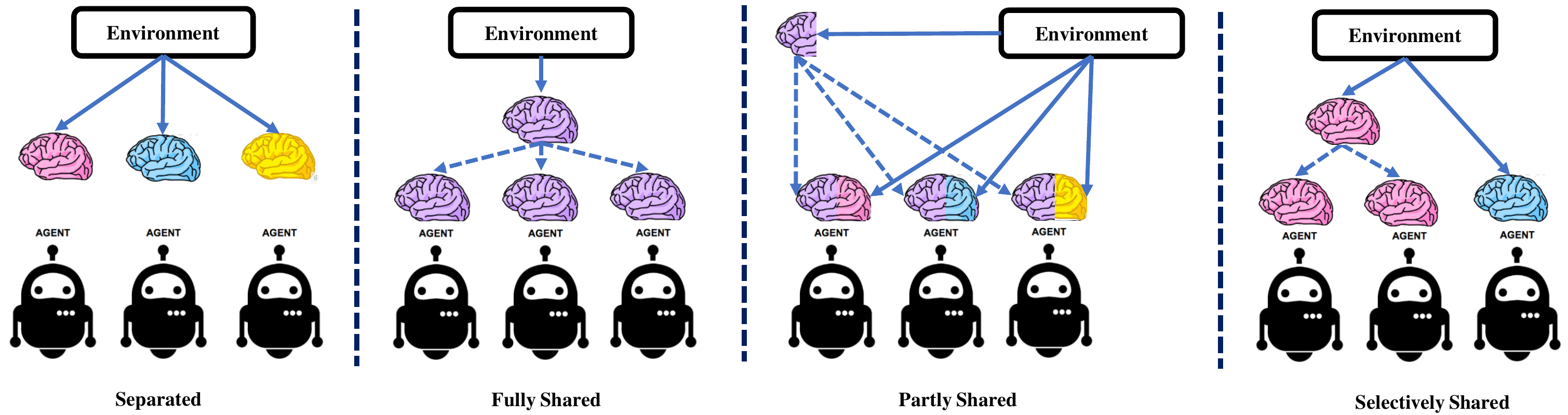}
\end{center}
  \caption{An overview of how knowledge sharing works with the MARL framework. Fully shared, partly shared, no shared (separated), and selectively shared \cite{christianos2021scaling} strategies are shown here. The same color indicates the same policy function part across different agents. Dash line represent only sharing no gradient backpropagation.}
\label{fig:knowledge_sharing}
\end{figure*}

Four different parameter sharing strategies are tested in our experiment including shared, no shared, partly shared, and selectively shared \cite{christianos2021scaling}. For partly shared, we only shared the GRU cell across different agents while keeping the embedding layer of the policy function model separated for each agent. For selectively shared strategy, we reproduce the grouping results following \cite{christianos2021scaling}. An illustration figure can be found in Fig.~\ref{fig:knowledge_sharing}.

\subsection{Selectively Sharing the Parameter}

Here we provide the selective parameter sharing strategy result in table.~\ref{table:selectively sharing} as a supplement for table.~\ref{table:parameter sharing vdn}. The main purpose for doing so is to verify whether this method can serve as a general solution for parameter sharing strategy choosing issue. Selective parameter sharing strategy partitions the agents into the different groups automatically with an encoder-decoder model. However, the partition process is before the MARL training stage, which can not fully catch the policy difference. And according to the grouping result column in table.~\ref{table:selectively sharing}, the selective parameter sharing strategy tends to divide agents by their initial attributes. This works well in scenarios like 1s1m1h1M\_vs\_4z and 1c1s1z\_vs\_1c1s3z but ignores the fact that the same type of agents may evolve to different functions during MARL training, which is the weakness of the selective parameter sharing strategy.

\begin{table*}[!h]
 \caption{
Model performance including selective parameter sharing as a supplement to table.~\ref{table:parameter sharing vdn}. The grouping result is provided in the last column.}
\label{table:selectively sharing}
\vspace{-9pt}
\small
 \begin{center}
 \resizebox{1.0\textwidth}{!}{
 \setlength{\tabcolsep}{1.3em}
 {\renewcommand{\arraystretch}{1.2}
\begin{tabular}{cccccc|cc}
\hline
\hline
Benchmark               & Scenario           & Role Diversity & Warm up & No shared                                     & Shared                                        & Selectively shared                            & Grouping results                   \\ \hline
                        & 4m\_vs\_5m         & 1.5 / 9.1      & 6.5     & {\color[HTML]{000000} +3.6 / +4.4}            & {\color[HTML]{FE0000} \textbf{+5.4 / +6.1}}   & {\color[HTML]{FE0000} \textbf{+5.4 / +6.1}}   & {\color[HTML]{000000} all shared}  \\ \cline{2-8} 
                        & 3s\_vs\_5z         & 2.7 / 18.7     & 5.4     & {\color[HTML]{000000} +7.5 / +11.0}           & {\color[HTML]{FE0000} \textbf{+8.2 / +11.8}}  & {\color[HTML]{FE0000} \textbf{+8.2 / +11.8}}  & {\color[HTML]{000000} all shared}  \\ \cline{2-8} 
                        & 2m                 & 3.1 / 12.2     & 6.0     & +9.2 / +11.1                                  & {\color[HTML]{FE0000} \textbf{+18.1 / +17.6}} & {\color[HTML]{FE0000} \textbf{+18.1 / +17.6}} & {\color[HTML]{000000} all shared}  \\ \cline{2-8} 
                        & 4m\_vs\_4z         & 3.3 / 19.3     & 4.4     & {\color[HTML]{FE0000} \textbf{+ 8.8 / +12.7}} & {\color[HTML]{000000} +5.4 / +8.4}            & {\color[HTML]{000000} + 8.1 / +11.7}          & {\color[HTML]{000000} 2m+2m}       \\ \cline{2-8} 
                        & 4m\_vs\_3z         & 3.8 / 12.1     & 7.2     & +12.4 / +12.1                                 & {\color[HTML]{000000} +11.9 / +12.3}          & {\color[HTML]{FE0000} \textbf{+12.6 / +12.2}} & {\color[HTML]{000000} 2m+2m}       \\ \cline{2-8} 
                        & 3s\_vs\_4z         & 5.2 / 32.5     & 4.8     & {\color[HTML]{FE0000} \textbf{+2.2 / +4.5}}   & {\color[HTML]{000000} +0.9 / +1.2}            & +0.9 / +1.2                                   & {\color[HTML]{000000} all shared}  \\ \cline{2-8} 
                        & 1c1s1z\_vs\_1c1s3z & 8.7 / 22.0     & 11.8    & {\color[HTML]{FE0000} \textbf{+4.1 / +6.1}}   & {\color[HTML]{000000} +2.7 / +5.4}            & {\color[HTML]{FE0000} \textbf{+4.1 / +6.1}}   & {\color[HTML]{000000} 1c+1s+1z}    \\ \cline{2-8} 
                        & 1s1m1h1M\_vs\_3z   & 2.4 / 13.2     & 16.2    & {\color[HTML]{000000} +3.4 / + 3.4}           & {\color[HTML]{FE0000} \textbf{+3.4 / +3.6}}   & +3.4 / + 3.4                                  & {\color[HTML]{000000} 1s+1m+1h+1M} \\
                        & 1s1m1h1M\_vs\_4z   & 2.7 / 15.8     & 8.2     & {\color[HTML]{FE0000} \textbf{+7.8 / +11.6}}  & {\color[HTML]{000000} +5.3 / +10.0}           & {\color[HTML]{FE0000} \textbf{+7.8 / +11.6}}  & {\color[HTML]{000000} 1s+1m+1h+1M} \\
\multirow{-10}{*}{SMAC} & 1s1m1h1M\_vs\_5z   & 6.2 / 22.5     & 6.2     & {\color[HTML]{FE0000} \textbf{+6.4 / +9.1}}   & {\color[HTML]{000000} +3.7 / +6.1}            & {\color[HTML]{FE0000} \textbf{+6.4 / +9.1}}   & {\color[HTML]{000000} 1s+1m+1h+1M} \\ 
\hline
\hline
\end{tabular}
 }
 }
 \end{center}
 \end{table*}

\section{Communication Framework}\label{app:Communication Framework}

\begin{figure}[!h]
    \centering 
\begin{subfigure}{0.45\textwidth}
  \includegraphics[width=\linewidth]{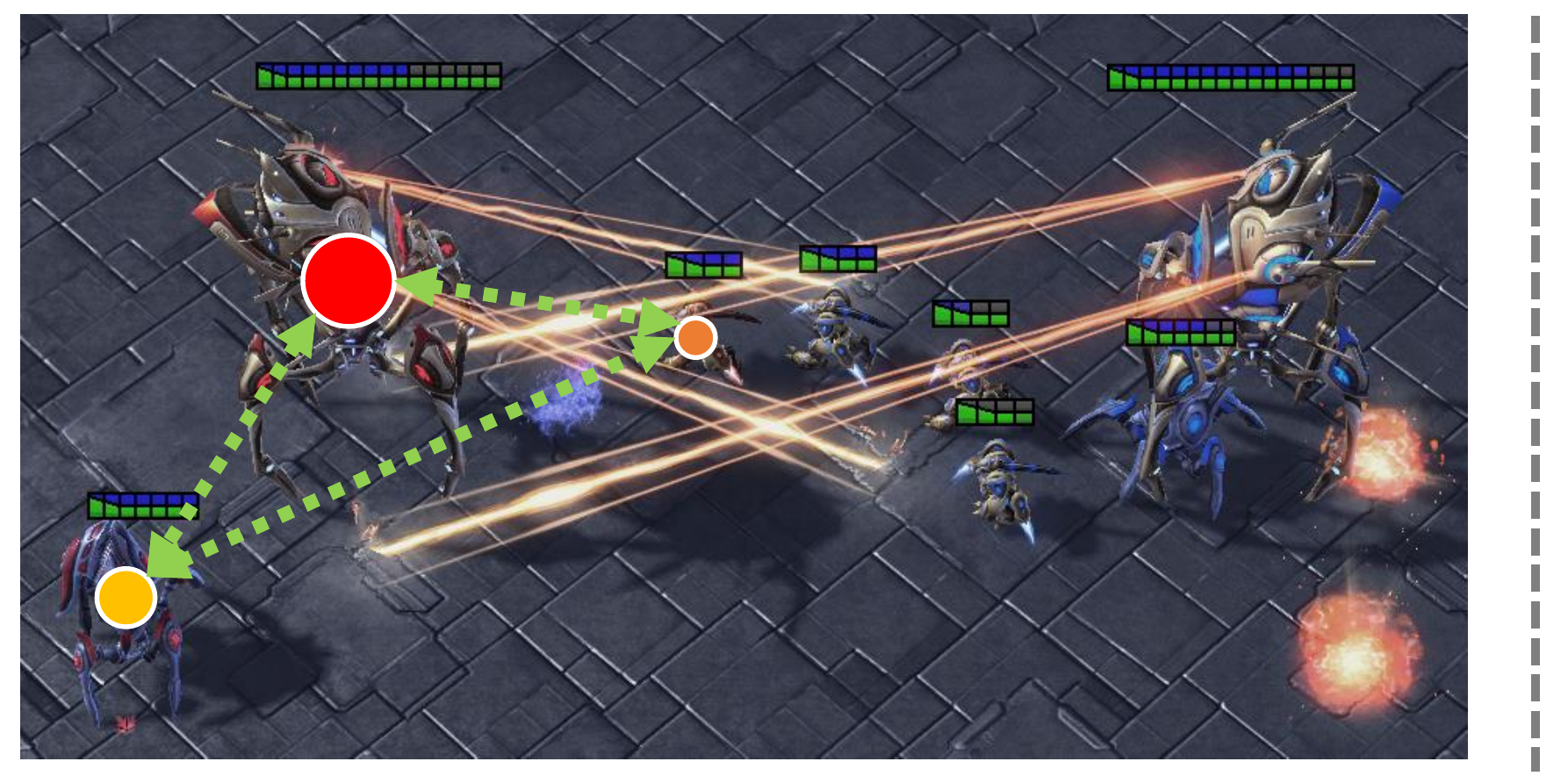}
  \caption{Communicate with others: where I am, what I see and what I will do}
  \label{fig:communicate_example}
\end{subfigure}\hfil 
\begin{subfigure}{0.48\textwidth}
  \includegraphics[width=\linewidth]{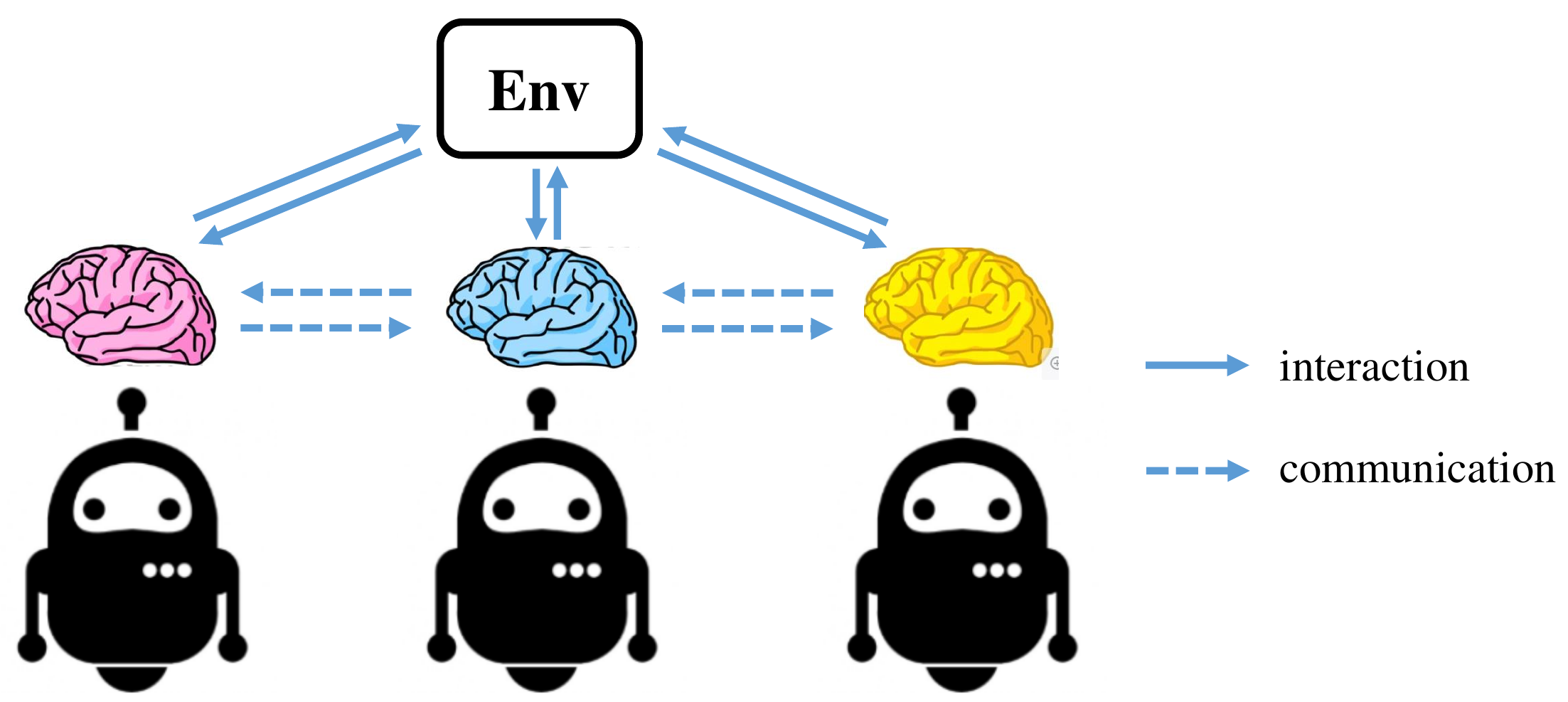}
  \caption{Communication is optional during optimization.}
  \label{fig:communication}
\end{subfigure}\hfil 
\caption{Communication works as a supplement part for MARL under the CTDE framework. (a) The sharing information can be current status or future policy, as the extra information for the decision making. (b) Learning when and how to communicate is critical to help policy learning.}
\label{fig:multi_exp}

\end{figure}

The communication mechanism is important for MARL. The information shared can be location, action, and partial observation as showed in Fig.~\ref{fig:communicate_example}. In many cases, communication is optional where the agent should learn when to communicate and how to ingest the information (dash line in Fig.~\ref{fig:communication}). In our experiment, we only consider the observation sharing method where the support set of policy functions contains both self partial observation and the aggregated observation information from other agents. The aggregated information is obtained by getting the mean value of other agents' observation and concatenate with the self partial observation. This means the support set of policy functions is now much similar to the global state.

\section{Credit Assignment}\label{app:Credit Assignment}

Credit assignment is the key part module for cooperative MARL, especially for the value decomposition-based method as it leverages the reward signal to each agent by approximate the $Q_{tot}$. Then the learned individual policies combine to form a joint policy interacting with the MAS.

\begin{figure*}[!h]
\begin{center}
  \includegraphics[width=0.6\linewidth]{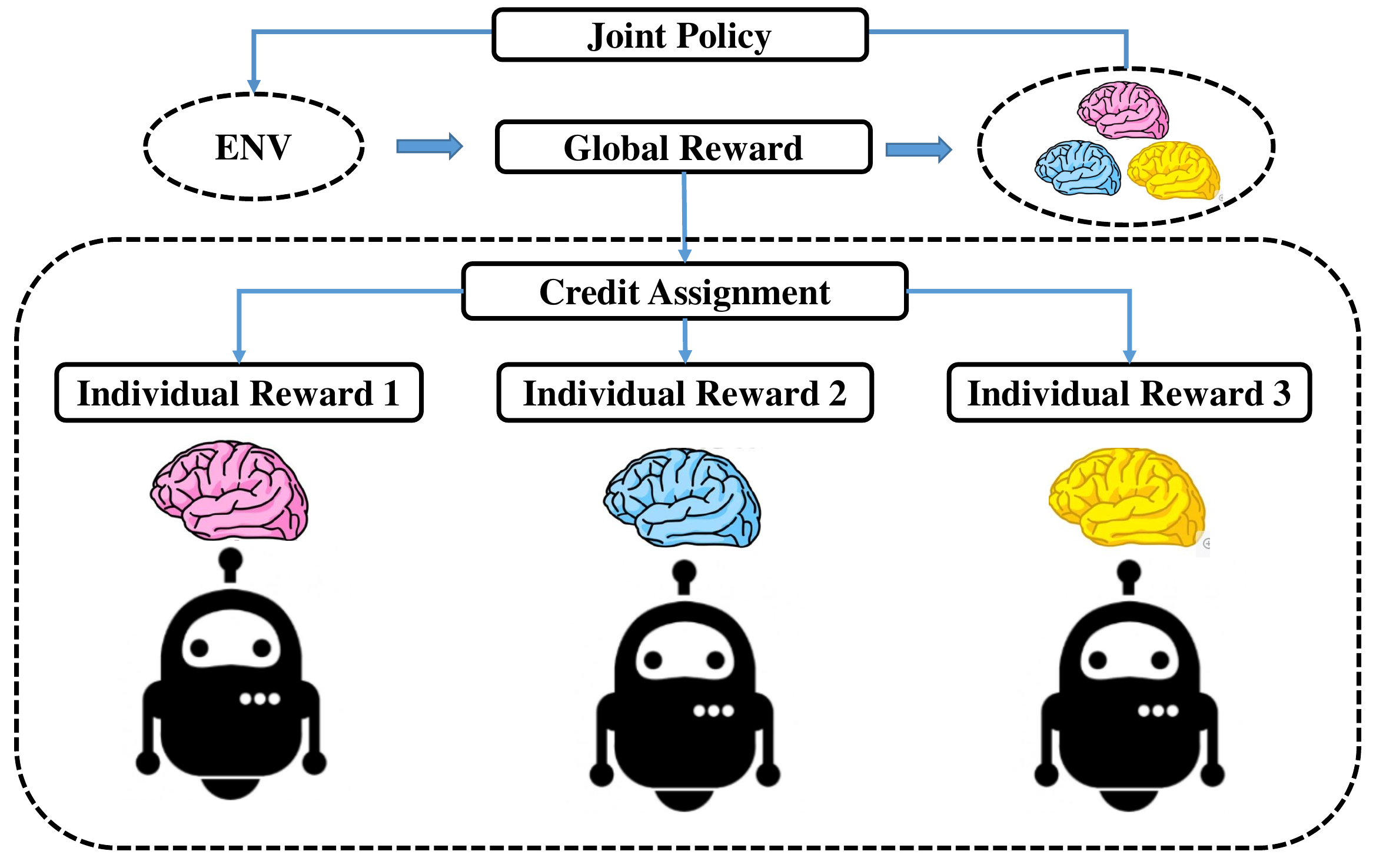}
\end{center}
  \caption{Credit assignment method focuses on assigning the proper individual reward from the total reward to update. }
\label{fig:creidt_assignment}
\end{figure*}

\newpage

\section{Proofs}

In this section, we present more detailed results and the proofs for the theoretical analysis in Sec.~\ref{sec:3}. 
In Sec.~\ref{sec:D1}, we denote more notations and state the concentration property of Markov decision process.
Sec.~\ref{sec:lemmas} presents two useful lemmas about the error propagation and one-step approximation respectively.
In Sec.~\ref{sec:separate}, we consider a simple example of the decentralized and cooperative MARL and provide the finite-sample analysis for the estimation error of the joint action-value function.
We use the value decomposition \cite{sunehag2017value,rashid2018qmix} and the finite-sample results for single-agent RL \cite{fan2020theoretical}. 
For more related results about MARL, please refer to \cite{zhang2021finite} and \cite{wang2020towards}.
Sec.~\ref{sec:share} studies the parameter sharing case that all agents share one deep Q network. 
In Sec.~\ref{sec:rate}, we assume $Q$ is a sparse ReLU network and $T Q$ is a composition of H$\ddot{\text{o}}$lder smooth functions. Then we discuss the convergence rate of the statistical error as the sample size tends to infinity.
According to Sec.~\ref{sec:separate}, \ref{sec:share} and \ref{sec:rate}, one can find that each type of role diversity have different impact to the decomposed estimation error.
Furthermore, we explain the benefits of the training options, e.g. parameter sharing (Sec.~\ref{exp:Parameter Sharing}),  communication (Sec.~\ref{exp:Communicate: When and How}) and credit assignment (Sec.~\ref{exp:Credit Assignment}), and discuss how these options impact the convergence rate of approximation error and statistical error.

We summarize our results as follows:
\begin{itemize}
    \item The parameter sharing strategy introduces a bias term by constraining the diversity of individual action-value functions, which corresponds to the action-based role diversity. At the same time, it speeds up the convergence rate of statistical error by pooling training data.
    \item The communication mechanism reduces the variance caused by the trajectory-based role diversity but slows down the convergence rate of approximation error by introducing more active input variables.
    \item When the contribution-based role diversity is nonnegligible, the credit assignment can significantly reduce the estimation error of the action-value function.
\end{itemize}

\subsection{More Notations and Assumptions}\label{sec:D1}

We denote the joint optimal action-value function by
\benrr
Q^{*}_{tot}(\rvz, \rvu) = Q^{*}_{tot}(\rz_1, \ldots, \rz_n, \ru_1, \ldots, \ru_n),
\eenrr 
where $\rvz$ is the global state of the environment, $\rvu = \{ \ru_1, \ldots, \ru_n\}$ is the action set that collects the action of each agent $\ru_i$ and $\rz_i$ is the observation of the agent $i$ generated from the emission distribution $\rz_i \sim \Lambda(\rz|i, \rvs).$
We further denote the individual optimal action-value function by $Q^*_i$ and write
\benrr
\rmQ^*(\rvz,\rvu) = \big( Q^*_1(\rz_1, \ru_1), \ldots, Q^*_n(\rz_n, \ru_n) \big)
\eenrr
as the vector of all agents' action-value functions. 
According to the value decomposition assumption, the joint optimal Q-function $Q^*_{tot}$ can be approximated with
\benrr
F(\rmQ^*)(\rvz, \rvu) = F(Q^*_1(\rz_1, \ru_1), \ldots, Q^*_n(\rz_n, \ru_n)),
\eenrr
where $F \in \mathcal{F}$ is a credit assignment function. 
The VDN method \cite{sunehag2017value}
approximates the joint 
as a sum of individual action-value functions that condition only on individual observations and
actions. Then a decentralised policy arises simply from each agent selecting actions greedily with respect to its $Q_i.$ Since $n$ is a fixed integer, we write the hypothetical space of credit assign functions that only contains one function as:
\benrr
\gF = \big\{ F(\rmQ) = \frac{1}{n} \sum_{i=1}^n Q_i: \,\,\text{with}\,\, \rmQ=(Q_1,\ldots, Q_n) \big\}.
\eenrr
The QMIX method \cite{rashid2018qmix} generalizes the value decomposition scheme and prove that 
if 
\benrr
\frac{\partial Q_{tot}(\rvz, \rvu)}{\partial Q_i(\rz_i, \ru_i)} \geq 0, \,\, \text{for } \,\,\forall 1\leq i \leq n, \,\, \forall \rvz \in \gZ^n, \,\, \forall \rvu \in \gU^n, 
\eenrr
then the global $\argmax$ performed on joint Q-function yields the same result as a set of individual $\argmax$ operations performed on each agent Q-function, that is
\benrr
\argmax_{\rvu} Q(\rvz, \rvu) =  \begin{pmatrix}
\argmax_{\ru_1} Q_1(\rz_1, \ru_1)  \\
\argmax_{\ru_2} Q_2(\rz_2, \ru_2)  \\
\vdots  \\  
\argmax_{\ru_n} Q_n(\rz_n, \ru_n)  
\end{pmatrix}.
\eenrr
Motivated by VDN and QMIX, we consider a simple case throughout this section: 
\benrr
\gF = \big\{ F(\rmQ) = \rvw^\top \rmQ:\,\, \rvw \in \Delta_n\,\,\text{and}\,\, \rmQ=(Q_1,\ldots, Q_n) \big\},
\eenrr
where $\Delta_n$ is the $n-1$ dimensional probability simplex.

Suppose the individual action-value  function is estimated by the fitted-Q iteration (FQI) algorithm \citep{ernst05a,riedmiller2005neural}. 
At the iteration $0\leq t \leq T$, we write $\tilde Q_{i,t}$ and $\pi_{i,t}$ as the output of FQI algorithm and the corresponding greedy policy respectively.
Let $Q^{\pi_{i,t}}$ be the Q-function corresponding to $\pi_{i,t}$.
Then the joint action-value  function is estimated by $\hat F(\rmQ_t)$, where
\benrr
\rmQ_t(\rvz, \rvu) = \big( Q^{\pi_{1,t}}(\rz_1, \ru_1), Q^{\pi_{2,t}}(\rz_2, \ru_2), \ldots, Q^{\pi_{n,t}}(\rz_n, \ru_n)\big).
\eenrr

To proceed further, we give the following assumption that controls the similarity between
two probability distributions under the Markov decision process.

{\bf Assumption~1}. {\it Let $\mu, \nu \in \gP(\gZ \times \gU)$ be two probability measures that are absolutely continuous with respect to the Lebesgue measure on $\gZ \times \gU.$ 
Let $\{\pi_t\}$ be a sequence of joint policies for all the agents, with $\pi_t: \gZ \rightarrow \gP(\gU)$ for all time $t.$ Suppose the initial state-action pair $(\rz_0, \ru_0)$ has distribution $\mu$, and the action $\ru_t$ is sampled from the joint policy $\pi_t.$ 
For any integer $m$, we denote by $P^{\pi_m} P^{\pi_{m-1}}\cdots P^{\pi_1} \mu$ the distribution of $\{(\rz_t, \ru_t)\}_{t=1}^m$ under the policy sequence $\{ \pi_t \}_{ t=1, \ldots,m}.$ Then, the $m$-th concentration coefficient is defined as
\benrr
\kappa(m; \mu, \nu) = \sup_{\pi_1,\ldots, \pi_m}\left[ \E_{\nu} \Big|\frac{d(P^{\pi_m} P^{\pi_{m-1}} \cdots P^{\pi_1} \mu)}{d \nu}\Big|^2 \right]^{1/2},
\eenrr
where $d(P^{\pi_m} P^{\pi_{m-1}} \cdots P^{\pi_1} \mu) / d \nu$ is the Radon-Nikodym derivative of $P^{\pi_m} P^{\pi_{m-1}} \cdots P^{\pi_1} \mu$ with respect to $\nu$ and the supremum is taken over all possible policies.

Furthermore, let $\nu$ be the stationary distribution of the samples $\{ (\rz_t, \ru_t)\}$ from the  Markov decision process and let $\mu$ be a fixed distribution on $\gS \times \gU.$ We assume that there exists a constant $\phi_{\mu,\nu}$ such that
\benrr
(1-\gamma)^2 \cdot\sum_{m \geq 1} \gamma^{m-1} \cdot m \cdot \kappa(m; \mu, \nu) \leq \phi_{\mu,\nu}.
\eenrr
}

To proceed further, we denote $\gQ$ as the space of individual Q-functions and let 
\benrr
\omega(\gQ) = \sup_{Q \in \gQ} \inf_{Q'\in \gQ} \|Q'-T Q\|_{2,\nu}^2,
\eenrr
where $\|\cdot\|_{2, \nu}$ as the $L_2$ norm with respect to a probability measure $\nu.$ 
In the following, we take $\nu$ as the independent data sampling distribution in the FQI algorithm, e.g. experience replay \cite{lin1992self}.

We say a collection $\{Q^1, \ldots, Q^K \} \subseteq \gQ$ is an $\delta$-cover of $\gQ$ if for each $Q \in \gQ$, there exists $Q^k$ such that $\|Q- Q^k\|\leq \delta $. The $\delta$-covering number of $\gQ$ with respect to $\|\cdot\|$ is
\benrr
N(\gQ, \delta, \|\cdot\|) := \inf \{K \in \sN: \text{there is an $\delta$-cover of $\gQ$ with respect to $\|\cdot\|$} \}.
\eenrr
In the following, we take the $L^{\infty}(\gZ\times\gU)$ norm on $\gQ$ by 
\benrr
\|Q-Q'\|_{L^{\infty}(\gZ \times\gU)} = \sup_{(\rz,\ru)\in \gZ\times\gU} \big| Q(\rz,\ru)-Q'(\rz,\ru) \big|.
\eenrr
For the sake of simplicity, we rewrite $N(\gQ, \delta, \|\cdot\|)$ as $N_\delta.$ 
In Sec.~\ref{sec:separate},
we study the estimation error of the joint action-value function and prove that the statistical error depends on $\ln N_0/N$, where $N_0$ is the $1/N$-cover of $\gQ$ and $N$ is the sample size. 
For the parameter sharing settings, the sample size increases while the cover number also increases due to the smaller $\delta.$   
Thus, we still do not know whether the parameter sharing improves the convergence rate of the statistical error. 
So we present a fine-grain analysis to discuss the convergence rate in Sec~\ref{sec:rate}.

\subsection{Useful Lemmas}
\label{sec:lemmas}

{\bf Lemma~1.} (Theorem 6.1 in \cite{fan2020theoretical}). {\it For each agent $i \in [n]$,  we denote $\{\tilde Q_{i,t}\}_{0\leq t \leq T}$ as the iterates of FQI Algorithm. Let $\pi_{i,t}$ be the one-step greedy policy with respect to $\tilde Q_{i,t}$, and let $Q^{\pi_{i,t}}$ be the action-value function corresponding to $\pi_{i,t}$. Under {\bf Assumption~1}, we have
\benr\label{eq:lemma1}
\| Q^*_i - Q^{\pi_{i,t}} \|_{1,\mu} &\leq& \frac{2 \phi_{\mu,\nu} \gamma}{(1-\gamma)^2} \max_{t \in [T]} \|T  \tilde Q_{i,t-1} - \tilde Q_{i,t}\|_{2, \nu} + \frac{4 \gamma^{T+1}}{(1-\gamma)^2} M.
\eenr
}

{\bf Proof:} Please see Appx~C.1 of \cite{fan2020theoretical} for a complete proof.

\bbox

This lemma quantifies the error propagation procedure of each agent action-value functions through each iteration of FQI Algorithm. The first term on the RHS is the one-step statistical error and will not vanish even when the iteration goes to infinity ($T\to \infty$).
For more related error propagation results, please refer to 
\cite{munos2008finite,farahmand2010error,scherrer2015approximate,farahmand2016regularized,lazaric2016analysis}.

{\bf Lemma~2.} (Theorem 6.2 in \cite{fan2020theoretical}).
Let $\{(\rz_{ij},\ru_{ij})\}_{j \in [N]}$ be $N$ i.i.d. random variables. For each $j \in [N]$, let $\rr_{ij}$ and $\rz'_{ij}$ be the reward and the next state corresponding to $(\rz_{ij}, \ru_{ij})$. In addition, for any fixed $\tilde Q_{i,t-1} \in \gQ$, we define $\ry_{ij} = \rr_{ij} + \gamma \cdot \max_\ru \tilde Q_{i,t-1}(\rz'_{ij}, \ru)$. Based on $\{( \rz_{ij}, \ru_{ij}, \ry_{ij})\}_{j\in [N]}$, we define $\tilde Q_{i,t}$ as 
\benrr
\tilde Q_{i,t} = \argmin_{Q\in \gQ} \frac{1}{N} \sum_{j=1}^N \big(Q(\rz_{ij}, \ru_{ij}) - \ry_{ij}\big)^2.
\eenrr
Then for any $\delta > 0$, we have
\benr\label{eq:lemma2}
\|T  \tilde Q_{i,t-1} - \tilde Q_{i,t}\|_{2, \nu}^2 &\leq& 4 \omega(\gQ) + C \frac{ M^2 }{ (1-\gamma)^2 } \frac{\ln N_\delta}{N} + C \frac{M\delta}{1-\gamma} ,
\eenr
where 
$
\omega(\gQ) = \sup_{Q\in \gQ} \inf_{Q'\in \gQ} \| Q' - T Q\|^2_{\nu}
$
and $N_\delta$ is the $\delta$-covering number of $\gQ$ with respect to the norm $\|\cdot\|_{\infty}.$

{\bf Proof:} Please see Appx~C.2 of \cite{fan2020theoretical} for a complete proof.

\subsection{Individual Q-Function} \label{sec:separate}

{\bf Theorem~1}. {\it We consider the separated strategy that each agent has its own action-value function and reward.
All agents' learning process is independent. Suppose $\{\tilde Q_{i,t}\}_{0\leq t \leq T}$ are the output of FQI Algorithm for the agent $i.$
Let $\pi_{i,t}$ be the one-step greedy policy with respect to $\tilde Q_{i,t}$, and let $Q^{\pi_{i,t}}$ be the action-value function corresponding to $\pi_{i,t}$.
We rewrite 
\benrr
\rmQ_t = \big(Q^{\pi_{1,t}}, Q^{\pi_{2,t}}, \ldots, Q^{\pi_{n,t}}\big), \quad \text{and} \quad \rmQ^* = \big(Q^*_1, Q^*_2, \ldots, Q^*_n\big).
\eenrr
Recall that $0\leq \gamma < 1$ is the discount factor, the reward function is bounded, i.e., $|r(\rs, \ru)| \leq M$, $\gQ$ is the space of individual Q-functions and $\omega(\gQ) = \sup_{Q \in \gQ} \inf_{Q'\in \gQ} \|Q'-T Q\|_{2,\nu}^2.$
Then, under {\bf Assumption~1}, we have
\benrr
\|Q_{tot}^* - \hat F (\rmQ_t)\|_{1,\mu} &\leq & \|Q_{tot}^* - (\rvw^*)^\top \rmQ^*\|_{1,\mu} \\
&& + \sqrt{n} \times \|\rvw^* - \hat \rvw\| \times \left\|\sqrt{\Var_n \big(\rmQ^*(\rvz,\rvu) \big)}\right\|_{1,\mu}\\
&& + \frac{4 \phi_{\mu,\nu} \gamma}{(1-\gamma)^2} \sqrt{\omega(\gQ)} + O(\sqrt{\frac{\ln N_0}{N}}) +  \frac{4 \gamma^{T+1}}{(1-\gamma)^2} M,
\eenrr
where $N_0$ is the $1/N$-covering number of $\gQ$ with respect to the norm $\|\cdot\|_{\infty}$ and 
\benrr
\Var_n \big(\rmQ^*(\rvz,\rvu) \big) = \frac{1}{n} \sum_{i=1}^n  \left(Q_i^*(\rz_i, \ru_i) - \frac{1}{n}\sum_{i=1}^n Q_i^*(\rz_i, \ru_i) \right)^2.
\eenrr  
}

{\bf Proof}: It is easy to see that
\benr\label{eq:A1}
&& \|Q_{tot}^* - \hat F (\rmQ_t)\|_{1,\mu} \nonumber\\
&=& \|Q_{tot}^* - F^* (\rmQ^*) + F^* (\rmQ^*) - \hat F (\rmQ^*) +  \hat F (\rmQ^*) - \hat F (\rmQ_t)\|_{1,\mu} \nonumber\\
&\leq& \|Q_{tot}^* - F^* (\rmQ^*)\|_{1,\mu} + \| F^* (\rmQ^*) - \hat F (\rmQ^*) \|_{1,\mu} + \|\hat F (\rmQ^*) - \hat F (\rmQ_t)\|_{1,\mu}.
\eenr
Here the first term at the RHS of (\ref{eq:A1}):
\benr\label{eq:A2}
\|Q_{tot}^* - F^* (\rmQ^*)\|_{1,\mu} = \|Q_{tot}^* - (\rvw^*)^\top \rmQ^*\|_{1,\mu}
\eenr
represents the best achievable estimation error under the value decomposition assumption.
Next we consider the second term in the inequality (\ref{eq:A1}):
\benrr
\| F^* (\rmQ^*) - \hat F (\rmQ^*) \|_{1,\mu} &=& \| (\rvw^*)^\top \rmQ^* - \hat \rvw^\top \rmQ^* \|_{1,\mu} \\
&=& \| (\rvw^* - \hat \rvw)^\top \rmQ^*\|_{1,\mu}
\eenrr
For any given $\rvz = (\rz_1, \ldots, \rz_n)$ and $\rvu=(\ru_1, \ldots, \ru_n)$,
\benrr
(\rvw^* - \hat \rvw)^\top \rmQ^*(\rvz, \rvu) &=& \sum_{i=1}^n (\rw^*_i - \hat \rw_i) Q_i^*(\rz_i, \ru_i) \\
& = &  \sum_{i=1}^n (\rw^*_i - \hat \rw_i) \left(Q_i^*(\rz_i, \ru_i) - \frac{1}{n}\sum_{i=1}^n Q_i^*(\rz_i, \ru_i) \right).
\eenrr
The second equality holds since $\sum_{i=1}^n (\rw^*_i - \hat \rw_i) \times c =0$ for any constant $c.$
By the Cauchy–Schwarz inequality, we have
\benrr
(\rvw^* - \hat \rvw)^\top \rmQ^*(\rvz, \rvu) &\leq& \sqrt{\sum_{i=1}^n (\rw^*_i - \hat \rw_i)^2} \times \sqrt{\sum_{i=1}^n  \left(Q_i^*(\rz_i, \ru_i) - \frac{1}{n}\sum_{i=1}^n Q_i^*(\rz_i, \ru_i) \right)^2} \\
&=& \|\rvw^* - \hat \rvw\| \times \sqrt{n \Var_n \big(\rmQ^*(\rvz,\rvu) \big) }
\eenrr
where 
\benrr
\Var_n \big(\rmQ^*(\rvz,\rvu) \big) = \frac{1}{n} \sum_{i=1}^n  \left(Q_i^*(\rz_i, \ru_i) - \frac{1}{n}\sum_{i=1}^n Q_i^*(\rz_i, \ru_i) \right)^2
\eenrr
is the variance of the output vector of $\rmQ^*$ given $\rvz$ and $\rvu.$ Plugging the positive upper boundary of $(\rvw^* - \hat \rvw)^\top \rmQ^*$ into the expression of $\| F^* (\rmQ^*) - \hat F (\rmQ^*) \|_{1,\mu}$, we obtain that
\benr\label{eq:A3}
\| F^* (\rmQ^*) - \hat F (\rmQ^*) \|_{1,\mu} &\leq& \left\| \|\rvw^* - \hat \rvw\| \cdot \sqrt{n \cdot \Var_n \big(\rmQ^*(\rvz,\rvu) \big) } \right \|_{1,\mu} \nonumber\\
&=& \sqrt{n} \times \|\rvw^* - \hat \rvw\| \times \left\|\sqrt{\Var_n \big(\rmQ^*(\rvz,\rvu) \big)}\right\|_{1,\mu}.
\eenr
Here the term $\|\sqrt{\Var_n \big(\rmQ^*(\rvz,\rvu) \big)}\|_{1,\mu}$ stands for the diversity of the agents. 

Finally, we deal with the third term on the RHS of (\ref{eq:A1}). Notice that 
\benrr
\|\hat F (\rmQ^*) - \hat F (\rmQ_t)\|_{1,\mu} &=& \| \sum_{i=1}^n \hat w_i (Q^*_i - Q^{\pi_t}_i) \|_{1,\mu} \\
&\leq& \sum_{i=1}^n  \hat w_i \| Q^*_i - Q^{\pi_t}_i \|_{1,\mu}.
\eenrr
Therefore, it suffices to consider the deep Q-learning procedure of each agent separately and to give upper bound of $\| Q^*_i - Q^{\pi_t}_i \|_{1,\mu}$ for each $i \in [n].$
According to (\ref{eq:lemma1}) and (\ref{eq:lemma2}),
\benrr
\| Q^*_i - Q^{\pi_t}_i \|_{1,\mu} &\leq&  \frac{2 \phi_{\mu,\nu} \gamma}{(1-\gamma)^2} \max_{t \in [T]} \|T  \tilde Q_{i,t-1} - \tilde Q_{i,t}\|_{2, \nu} + \frac{4 \gamma^{T+1}}{(1-\gamma)^2} M \nonumber\\
&\leq& \frac{2 \phi_{\mu,\nu} \gamma}{(1-\gamma)^2} \sqrt{ 4 \omega(\gQ) + C \frac{ M^2 }{ (1-\gamma)^2 } \frac{\ln N_\delta}{n} + C' \frac{M\delta}{1-\gamma}}  + \frac{4 \gamma^{T+1}}{(1-\gamma)^2} M \nonumber \\
&\leq& \frac{4 \phi_{\mu,\nu} \gamma}{(1-\gamma)^2} \sqrt{\omega(\gQ)} + C \frac{2 M \phi_{\mu,\nu} \gamma}{(1-\gamma)^3} \sqrt{\frac{\ln N_\delta}{N}} + C' \frac{2 \sqrt{M} \phi_{\mu,\nu} \gamma}{(1-\gamma)^{5/2}} \sqrt{\delta} + \frac{4 \gamma^{T+1}}{(1-\gamma)^2} M.
\eenrr
We take $\delta = 1/N$ and write $N_0$ as the $1/N$-covering number of $\gQ$. Then, we have
\benrr
\| Q^*_i - Q^{\pi_t}_i \|_{1,\mu} &\leq&  \frac{4 \phi_{\mu,\nu} \gamma}{(1-\gamma)^2} \sqrt{\omega(\gQ)} + O(\sqrt{\frac{\ln N_0}{N}}) +  \frac{4 \gamma^{T+1}}{(1-\gamma)^2} M.
\eenrr
Furthermore,
\benr\label{eq:A4}
\|\hat F (\rmQ^*) - \hat F (\rmQ_t)\|_{1,\mu} &\leq&  \frac{4 \phi_{\mu,\nu} \gamma}{(1-\gamma)^2} \sqrt{\omega(\gQ)} + O(\sqrt{\frac{\ln N_0}{N}}) +  \frac{4 \gamma^{T+1}}{(1-\gamma)^2} M.
\eenr
Combining the results of (\ref{eq:A2}), (\ref{eq:A3}) and (\ref{eq:A4}), we know 
\benrr
\|Q_{tot}^* - \hat F (\rmQ_t)\|_{1,\mu} &\leq & \|Q_{tot}^* - (\rvw^*)^\top \rmQ^*\|_{1,\mu} \\
&& + \sqrt{n} \times \|\rvw^* - \hat \rvw\| \times \left\|\sqrt{\Var_n \big(\rmQ^*(\rvz,\rvu) \big)}\right\|_{1,\mu}\\
&& + \frac{4 \phi_{\mu,\nu} \gamma}{(1-\gamma)^2} \sqrt{\omega(\gQ)} + O(\sqrt{\frac{\ln N_0}{N}}) +  \frac{4 \gamma^{T+1}}{(1-\gamma)^2} M.
\eenrr

\bbox

{\bf Remark.} The term
\benrr
\sqrt{n} \times \|\rvw^* - \hat \rvw\| \times \left\|\sqrt{\Var_n \big(\rmQ^*(\rvz,\rvu) \big)}\right\|_{1,\mu}
\eenrr
shows the benefits of learning credit assignment, where $\rvw^*$ stands for the best  credit assignment scheme. Here we assume $\hat \rvw$ is given and do not take the modelling and learning of $\hat \rvw$ into consideration. In practice, $\hat \rvw$ is the output of a credit distribution network and its learning procedure also influence the convergence properties of individual Q-functions.  
On the other hand, $\Var_n (\rmQ)$ corresponds to the contribution-based role diversity in Sec.~\ref{sec:Contribution Based Role}.
Therefore, when the variance is nonzero, a good credit assignment $\hat \rvw$ can the estimation error. 
For the parameter sharing case in the next section, we decompose the variance term  into the sum of a bias and a variance caused by action-based role diversity and the trajectory-based role diversity respectively.
This decomposition does not always hold. 
Here, we assume that all agents' learning processes are independent and that each agent has its own reward function. In practice, these assumptions are idealistic and limited.

\subsection{Shared Q-Function} \label{sec:share}

{\bf Theorem~2}. {\it We consider the parameter sharing strategy that all individual agents shares one action-value function. Suppose $\{\tilde Q_{t}\}_{0\leq t \leq T}$ are the output of FQI Algorithm.
Let $\pi_{t}$ be the one-step greedy policy with respect to $\tilde Q_{t}$, and let $Q^{\pi_{t}}$ be the action-value function corresponding to $\pi_{t}$.
We further denote $\bar Q^*$ as the optimal shared action-value function and write
\benrr
\bar \rmQ_t(\rvz, \rvu) &=& \big(Q^{\pi_{t}}(\rz_1, \ru_1), Q^{\pi_{t}}(\rz_2, \ru_2), \ldots, Q^{\pi_{t}}(\rz_n, \ru_n)\big), \\
\bar \rmQ^*(\rvz, \rvu) &=& \big(\bar Q^*(\rz_1, \ru_1), \bar Q^*(\rz_2, \ru_2), \ldots, \bar Q^*(\rz_n, \ru_n)\big).
\eenrr
Recall that $0\leq \gamma < 1$ is the discount factor, the reward function is bounded, i.e., $|r(\rs, \ru)| \leq M$, $\gQ$ is the space of individual Q-functions and $\omega(\gQ) = \sup_{Q \in \gQ} \inf_{Q'\in \gQ} \|Q'-T Q\|_{2,\nu}^2.$
Then, under {\bf Assumption~1}, we have
\benrr
\|Q_{tot}^* - \hat F (\bar \rmQ_t)\|_{1,\mu} &\leq& \|Q_{tot}^* - (\rvw^*)^\top \rmQ^*\|_{1,\mu} + \left\| \sum_{i=1}^n \rw_i^* (Q^*_i - \bar Q^*) \right\|_{1,\mu} \\
&& + \sqrt{n} \times \|\rvw^* - \hat \rvw\| \times \left\|\sqrt{ \Var_n \big(\bar \rmQ^*(\rvz,\rvu) \big)}\right\|_{1,\mu}\\
&& + \frac{4 \phi_{\mu,\nu} \gamma}{(1-\gamma)^2} \sqrt{\omega(\gQ)} + O(\sqrt{\frac{\ln N'_0}{nN}}) +  \frac{4 \gamma^{T+1}}{(1-\gamma)^2} M.
\eenrr
where $N'_0$ is the $1/(nN)$-covering number of $\gQ$ with respect to the norm $\|\cdot\|_{\infty}$ and 
\benrr
\Var_n \big(\bar \rmQ^*(\rvz,\rvu) \big) = \frac{1}{n} \sum_{i=1}^n  \left(\bar Q^*(\rz_i, \ru_i) - \frac{1}{n}\sum_{i=1}^n \bar Q^*(\rz_i, \ru_i) \right)^2.
\eenrr 
}

{\bf Proof}: Similar to the arguments in (\ref{eq:A1}), we have 
\benr\label{eq:A5}
\|Q_{tot}^* - \hat F (\bar \rmQ_t)\|_{1,\mu} 
&\leq& \|Q_{tot}^* - F^* (\rmQ^*)\|_{1,\mu}
+ \| F^* (\rmQ^*)  - F^* (\bar \rmQ^*)\|_{1,\mu}  \nonumber \\
&& + \| F^* (\bar\rmQ^*) - \hat F (\bar\rmQ^*) \|_{1,\mu} + \|\hat F (\bar\rmQ^*) - \hat F (\bar \rmQ_t)\|_{1,\mu}.
\eenr
The term $\|Q_{tot}^* - F^* (\rmQ^*)\|_{1,\mu}$ caused by the value decomposition is the same to that in (\ref{eq:A1}). So (\ref{eq:A2}) still holds. For the second term on the RHS of (\ref{eq:A5}),
\benrr
\| F^* (\rmQ^*)  - F^* (\bar \rmQ^*)\|_{1,\mu} &=& \| (\rvw^*)^\top (\rmQ^*  - \bar \rmQ^*) \|_{1,\mu} \\
&=& \left\| \sum_{i=1}^n \rw_i^* (Q^*_i - \bar Q^*) \right\|_{1,\mu},
\eenrr
which is the bias term caused by the parameter sharing. 
Next, we turn to a turn that is related to the trajectory-based role diversity.
Similar to (\ref{eq:A3}), we know
\benrr
\big(F^* (\bar\rmQ^*) - \hat F (\bar\rmQ^*)\big)(\rvz, \rvu) &=& (\rvw^* - \hat \rvw)^\top \bar\rmQ^* (\rvz, \rvu) \\
&=& \sum_{i=1}^n (\rw^*_i - \hat \rw_i) \bar Q^*(\rz_i, \ru_i) \\
&=& \sum_{i=1}^n (\rw^*_i - \hat \rw_i) \left(\bar Q^*(\rz_i, \ru_i) - \frac{1}{n}\sum_{i=1}^n \bar Q^*(\rz_i, \ru_i) \right) \\
&=& \sqrt{n} \times \|\rvw^* - \hat \rvw\| \times \sqrt{ \Var_n \big(\bar \rmQ^*(\rvz,\rvu) \big)}.
\eenrr
On the other hand, by (\ref{eq:lemma1}) and (\ref{eq:lemma2}), 
\benrr
\| Q^* - Q^{\pi_t} \|_{1,\mu} &\leq&  \frac{2 \phi_{\mu,\nu} \gamma}{(1-\gamma)^2} \max_{t \in [T]} \|T  \tilde Q_{t-1} - \tilde Q_{t}\|_{2, \nu} + \frac{4 \gamma^{T+1}}{(1-\gamma)^2} M \nonumber\\
&\leq& \frac{2 \phi_{\mu,\nu} \gamma}{(1-\gamma)^2} \sqrt{ 4 \omega(\gQ) + C \frac{ M^2 }{ (1-\gamma)^2 } \frac{\ln N_\delta}{n N } + C' \frac{M\delta}{1-\gamma}}  + \frac{4 \gamma^{T+1}}{(1-\gamma)^2} M \nonumber \\
&\leq& \frac{4 \phi_{\mu,\nu} \gamma}{(1-\gamma)^2} \sqrt{\omega(\gQ)} + C \frac{2 M \phi_{\mu,\nu} \gamma}{(1-\gamma)^3} \sqrt{\frac{\ln N_\delta}{nN}} + C' \frac{2 \sqrt{M} \phi_{\mu,\nu} \gamma}{(1-\gamma)^{5/2}} \sqrt{\delta} + \frac{4 \gamma^{T+1}}{(1-\gamma)^2} M.
\eenrr
We take $\delta = 1/(n N)$ and write $N'_0$ as the $1/(nN)$-covering number of $\gQ$. Then, we have
\benrr
\| Q^* - Q^{\pi_t} \|_{1,\mu} &\leq&  \frac{4 \phi_{\mu,\nu} \gamma}{(1-\gamma)^2} \sqrt{\omega(\gQ)} + O(\sqrt{\frac{\ln N'_0}{n N}}) +  \frac{4 \gamma^{T+1}}{(1-\gamma)^2} M.
\eenrr
Therefore,
\benrr
\|\hat F (\bar \rmQ^*) - \hat F (\bar \rmQ_t)\|_{1,\mu} &\leq&  \frac{4 \phi_{\mu,\nu} \gamma}{(1-\gamma)^2} \sqrt{\omega(\gQ)} + O(\sqrt{\frac{\ln N'_0}{nN}}) +  \frac{4 \gamma^{T+1}}{(1-\gamma)^2} M.
\eenrr
Summarizing the above results, we have
\benrr
\|Q_{tot}^* - \hat F (\bar \rmQ_t)\|_{1,\mu} &\leq& \|Q_{tot}^* - (\rvw^*)^\top \rmQ^*\|_{1,\mu} + \left\| \sum_{i=1}^n \rw_i^* (Q^*_i - \bar Q^*) \right\|_{1,\mu} \\
&& + \sqrt{n} \times \|\rvw^* - \hat \rvw\| \times \left\|\sqrt{ \Var_n \big(\bar \rmQ^*(\rvz,\rvu) \big)}\right\|_{1,\mu}\\
&& + \frac{4 \phi_{\mu,\nu} \gamma}{(1-\gamma)^2} \sqrt{\omega(\gQ)} + O(\sqrt{\frac{\ln N'_0}{nN}}) +  \frac{4 \gamma^{T+1}}{(1-\gamma)^2} M.
\eenrr

\bbox


\subsection{Convergence Rate} \label{sec:rate}

Similar to the Theorem~4.4 in \cite{fan2020theoretical}, we assume that $Q$ belongs to a family of sparse ReLU networks and $T Q$ can be written as compositions of H$\ddot{\text{o}}$lder smooth functions. 
Here $T$ is the optimal Bellman operator. We start with the definition of a $(L+1)$ layers and $\{d_j\}_{j=1}^{L+1}$ width ReLU networks:
\benrr
f(x) = W_{L+1} \sigma( W_L \sigma(W_{L-1} \ldots \sigma( W_2 \sigma(W_1 x + v_1) + v_2) \ldots v_{L-1}) + v_L),
\eenrr 
where $\sigma$ is the ReLU activation function, and $W_l$ and $v_l$ are the weight matrix and the bias in the $l$-th layer, respectively. 
The family of sparse ReLU networks is defined as
\benrr
\gF(L, \{ d_j \}^{L+1}_{i=0}, s) = \left\{ f: \max_{l \in [L+1]} \|\tilde W_l\|_\infty \leq 1,  \sum_{l=1}^{L+1}\| \tilde W_l\|_0 \leq s,\,\, \max_{j \in [d_{L+1}]}\|f_j\|_\infty \leq \frac{M}{1-\gamma} \right\},
\eenrr
where $\tilde W_l$ is the parameter matrix that contains $W_l$ and $v_l$ and $f_j$ is the $j$-th output of $f.$
On the other hand, the set of H$\ddot{\text{o}}$lder smooth functions is 
\benrr
\gC_r(\gD, \beta, H) = \left\{f: \gD \to \R: \sum_{\bm{\alpha}: |\bm{\alpha}| < \beta} \|\partial^{\bm{\alpha}} f\|_{\infty} + \sum_{\bm{\alpha}: \|\bm{\alpha}\|_1 =\lfloor \beta \rfloor} \sup_{\substack{x,y \in \gD,\\ x  \neq y}} \frac{|\partial^{\bm{\alpha}}f(x) - \partial^{\bm{\alpha}} f(y)|}{\|x- y\|_{\infty}^{\beta-\lfloor \beta \rfloor}} \leq H \right\}, 
\eenrr
where $\gD$ is a a compact subset of $\R^r$, $\lfloor \cdot \rfloor$ stands for the floor function and $\partial^{\bm{\alpha}} = \partial^{\alpha_1}\partial^{\alpha_2}\cdots \partial^{\alpha_r}$ with $\bm{\alpha} = (\alpha_1,\alpha_2,\ldots, \alpha_n)^\top.$
Furthermore, we write $\gG(\{p_j, t_j, \beta_j, H_j\}_{j \in [q]})$ as the family of functions
that can be decomposed into the composition of a sequence of H$\ddot{\text{o}}$lder smooth functions $\{g_j\}_{j \in [q]}.$ That is, for any function $f \in \gG(\{p_j, t_j, \beta_j, H_j\}_{j \in [q]})$,
\benrr
f = g_q \circ g_{q-1} \circ \cdots \circ g_2 \circ g_1,
\eenrr
where for any $k \in [p_{j+1}]$ and $j \in [q]$, the $k$-th component of the function $g_j$ is a H$\ddot{\text{o}}$lder smooth function, i.e., $g_{jk} \in \gC_{t_j}([a_j, b_j]^{t_j}, \beta_j, H_j).$ For simplicity, we take $p_{j+1}=1.$
Here we assume that the input of $g_{jk}$ is $t_j$-dimensional, where $t_j$ can be much smaller than $p_j.$
More specific, the deep Q network we used is a sparse ReLU network for any given action $\ru.$ Therefore, we rewrite the space of individual Q-functions $\gQ$ as
\benrr
\gF_0 = \{f: S \times U \to \R:\,\,  f(\cdot, \ru) \in  \gF( L, \{d_j\}_{j=0}^{L+1}, s)\,\, \text{for any} \,\, \ru \in U\}.
\eenrr 
Furthermore, for any $Q \in \gF_0$, we assume $T Q$ is a composition of H$\ddot{\text{o}}$lder smooth functions and belongs to the following family:
\benrr
\gG_0 = \{f: S \times U \to \R:\,\,  f(\cdot, \ru) \in  \gG( \{p_j, t_j, \beta_j, H_j\}_{j \in [q]} )\,\, \text{for any} \,\, \ru \in U\}.
\eenrr 
To proceed further, we denote
\benrr
\alpha^* = \max_{j\in[q]} \frac{t_j}{2 \beta^*_j + t_j}, \quad \beta^*_j = \beta_j \times \prod_{l=j+1}\min(\beta_l, 1), \quad \text{and} \quad \beta^*_q = 1.
\eenrr 
Now we are ready to state the following result.

{\bf Theorem~3}. {\it Suppose the assumptions of Theorem~1 hold and for any $Q \in \gF_0$, $T Q \in \gG_0$, where $T$ is the optimal Bellman operator. 
The sample size $N$ is sufficiently large such that there exists a constant $\xi>0$ satisfies
\benrr
\max \left\{ \sum_{j=1}^q (t_j + \beta_j + 1)^{3+t_j},\,\, \sum_{j\in[q]} \ln (t_j + \beta_j), \,\, \max_{j\in[q]} p_j \right\}  \lesssim   (\ln N)^\xi.
\eenrr
The network architecture of the Q-function is well designed such that
\benrr
L \lesssim (\ln N)^{\xi^*}, \quad r \leq \min_{j\in [L]} d_j \leq \max_{j\in [L]} d_j \lesssim N^{\xi^*}, \quad \text{and} \quad s \asymp N^{\alpha^*} (\ln N)^{\xi^*}
\eenrr
for some constant $\xi^* > 1 + 2 \xi.$
The number of iterations $T$ is sufficiently large, such that 
\benrr
T \geq C' (1-\alpha^*) \ln N, 
\eenrr 
where $C'$ is a constant.
Then, under {\bf Assumption~1}, we have
\benrr
\|Q_{tot}^* - \hat F (\rmQ_t)\|_{1,\mu} &\leq & \|Q_{tot}^* - (\rvw^*)^\top \rmQ^*\|_{1,\mu} \\
&& + \sqrt{n} \times \|\rvw^* - \hat \rvw\| \times \left\|\sqrt{\Var_n \big(\rmQ^*(\rvz,\rvu) \big)}\right\|_{1,\mu}\\
&& + O\left( (\ln N)^{1+2\xi^*}  N^{-\min_{j\in [q]}\frac{\beta^*_j}{2 \beta^*_j + t_j}} \right).
\eenrr
}

{\bf Proof:} This is a direct conclusion reached by Theorem 4.4 in \cite{fan2020theoretical}. That is, 
for any agent $i \in [n]$, 
\benrr
\| Q^*_i - Q^{\pi_{i,t}} \|_{1,\mu} &\leq& O\left( (\ln N)^{1+2 \xi^*}N ^{(\alpha^*-1)/2} \right) + \frac{4 \gamma^{T+1}}{(1-\gamma)^2} M.
\eenrr
The approximation error in Theorem~1 that involves $\omega(\gQ)$ is bounded above via Theorem~5 in \cite{schmidt2020nonparametric}. The upper bound for the cover number $N_0$ is derived from Theorem 14.5 in \cite{anthony2009neural}. Please refer to Section~6 in \cite{fan2020theoretical} and Sec.~\ref{sec:separate} for a complete proof.

\bbox

{\bf Remark:} Note that
\benrr
\frac{4 \gamma^{T+1}}{(1-\gamma)^2} M \to 0 \quad \text{as} \quad T \to \infty,
\eenrr
which is the algorithmic error that converges to zero at a linear rate of $T.$
In Theorem~3, we assume $T$ is sufficiently large such that this error is negligible comparing to the statistical error. If we ignore the logarithmic term, the convergence rate of the statistical error is about
\benrr
\max_{j\in [q]} N^{-\frac{\beta^*_j}{2 \beta^*_j + t_j}}.
\eenrr
Here $\beta_j^*$ and $t_j$ are parameters of the functional space of $T Q$.  
Therefore, the parameter sharing (Sec.~\ref{exp:Parameter Sharing}) keeps $\beta_j^*$ and $t_j$ unchanged and increases the sample size $N$ to $nN$ by pooling training data.
In addition, $t_j$ is the number of active input variables of $g_j.$ Thus, the communication mechanism (Sec.~\ref{exp:Communicate: When and How}) slows down the convergence rate by enlarging $t_j.$



\newpage

\section{Experiment Table \& Curve}

\begin{table*}[!h]
\caption{
 Action-based role diversity influence the performance of different parameter sharing strategies on the MPE \cite{lowe2017multi} and SMAC \cite{samvelyan19smac} benchmarks. 
 }
\small
 \begin{center}
 \resizebox{1.0\textwidth}{!}{
 \setlength{\tabcolsep}{1.3em}
 {\renewcommand{\arraystretch}{1.0}
\begin{tabular}{ccccccccc}
\hline\hline
{\color[HTML]{000000} Benchmark}              & {\color[HTML]{000000} Scenario}                             & {\color[HTML]{000000} Sharing}   & {\color[HTML]{000000} IQL}                  & {\color[HTML]{000000} IA2C}                 & {\color[HTML]{000000} MADDPG}              & {\color[HTML]{000000} MAPPO}                & {\color[HTML]{000000} MAA2C}               & {\color[HTML]{000000} QMIX}               \\ \hline
{\color[HTML]{000000} }                       & {\color[HTML]{000000} }                                     & {\color[HTML]{000000} no shared} & {\color[HTML]{FE0000} 19.4 / 53.0 / 52.6}   & {\color[HTML]{000000} 1.4 / 13.1 / 14.7}    & {\color[HTML]{000000} 3.3 / 2.5 / 2.3}     & {\color[HTML]{000000} 1.1 / 55.6 / 47.2}    & {\color[HTML]{000000} 0.6 / 11.3 / 47.9}   & {\color[HTML]{000000} 2.4 / 15.2 / 22.5}  \\ \cline{3-9} 
{\color[HTML]{000000} }                       & \multirow{-2}{*}{{\color[HTML]{000000} Tag}}                & {\color[HTML]{000000} shared \checkmark}    & {\color[HTML]{000000} 16.8 / 50.3 / 47.9}   & {\color[HTML]{FE0000} 1.0 / 16.6 / 27.5}    & {\color[HTML]{FE0000} 3.1 / 5.9 / 32.8}    & {\color[HTML]{FE0000} 1.4 / 40.0 / 45.9}    & {\color[HTML]{FE0000} 0.8 / 42.1 / 60.9}   & {\color[HTML]{FE0000} 2.9 / 23.3 / 36.0}  \\ \cline{2-9} 
{\color[HTML]{000000} }                       & {\color[HTML]{000000} }                                     & {\color[HTML]{000000} no shared} & {\color[HTML]{FE0000} 15.8 / 16.3 / 16.7}   & {\color[HTML]{000000} 17.1 / 19.7 / 19.9*}  & {\color[HTML]{000000} 16.8 / 19.0 / 16.0*} & {\color[HTML]{000000} 18.8 / 20.1 / 20.8*}  & {\color[HTML]{000000} 15.3 / 19.6 / 20.4*} & {\color[HTML]{000000} 13.3 / 16.1 / 16.5} \\ \cline{3-9} 
{\color[HTML]{000000} }                       & \multirow{-2}{*}{{\color[HTML]{000000} Adversary}}          & {\color[HTML]{000000} shared}    & {\color[HTML]{000000} 15.3 / 15.8 / 15.5}   & {\color[HTML]{000000} 16.7 / 19.9 / 20.3*}  & {\color[HTML]{000000} 16.5 / 18.4 / 16.4*} & {\color[HTML]{000000} 19.8 / 19.9 / 20.5*}  & {\color[HTML]{000000} 17.9 / 19.8 / 20.4*} & {\color[HTML]{FE0000} 14.8 / 17.3 / 17.3} \\ \cline{2-9} 
{\color[HTML]{000000} }                       & {\color[HTML]{000000} }                                     & {\color[HTML]{000000} no shared \checkmark} & {\color[HTML]{FE0000} 7.1 / 59.4 / 59.8}    & {\color[HTML]{FE0000} 0.3 / 59.9 / 64.1}    & {\color[HTML]{000000} 5.3 / 10.5 / 11.6*}  & {\color[HTML]{FE0000} 0.6 / 63.0 / 63.7}    & {\color[HTML]{FE0000} 0.2 / 41.6 / 63.8}   & {\color[HTML]{FE0000} 0.5 / 0.9 / 9.5}    \\ \cline{3-9} 
{\color[HTML]{000000} }                       & \multirow{-2}{*}{{\color[HTML]{000000} DoubleSpread-2}}     & {\color[HTML]{000000} shared}    & {\color[HTML]{000000} 4.3 / 5.4 / 8.0}      & {\color[HTML]{000000} 0.2 / 7.9 / 11.1}     & {\color[HTML]{000000} 5.3 / 10.5 / 11.6*}  & {\color[HTML]{000000} 3.1 / 25.6 / 56.5}    & {\color[HTML]{000000} 0.3 / 10.1 / 19.2}   & {\color[HTML]{000000} 0.6 / 4.3 / 6.0}    \\ \cline{2-9} 
{\color[HTML]{000000} }                       & {\color[HTML]{000000} }                                     & {\color[HTML]{000000} no shared \checkmark} & {\color[HTML]{FE0000} 32.2 / 144.3 / 212.2} & {\color[HTML]{FE0000} 12.3 / 436.4 / 480.8} & {\color[HTML]{000000} 1.1 / 1.2 / 1.3*}    & {\color[HTML]{000000} 47.0 / 261.4 / 261.1} & {\color[HTML]{FE0000} 4.7 / 343.6 / 390.5} & {\color[HTML]{000000} 3.2 / 3.2 / 2.9*}   \\ \cline{3-9} 
\multirow{-8}{*}{{\color[HTML]{000000} MPE}}  & \multirow{-2}{*}{{\color[HTML]{000000} DoubleSpread-4}}     & {\color[HTML]{000000} shared}    & {\color[HTML]{000000} 31.3 / 29.1 / 20.1}   & {\color[HTML]{000000} 18.6 / 83.4 / 106.3}  & {\color[HTML]{000000} 4.9 / 1.2 / 1.3*}    & {\color[HTML]{FE0000} 61.9 / 291.7 / 504.8} & {\color[HTML]{000000} 32.7 / 94.4 / 231.0} & {\color[HTML]{000000} 3.8 / 3.5 / 3.2*}   \\ \hline
{\color[HTML]{000000} }                       & {\color[HTML]{000000} }                                     & {\color[HTML]{000000} no shared} & {\color[HTML]{000000} 4.8 / 7.6 / 8.1*}     & {\color[HTML]{000000} 6.4 / 6.6 / 6.7}      & {\color[HTML]{000000} 2.5 / 2.4 / 1.1}     & {\color[HTML]{000000} 6.9 / 7.1 / 7.2*}     & {\color[HTML]{000000} 6.6 / 7.0 / 7.1*}    & {\color[HTML]{000000} 7.0 / 9.9 / 10.9}   \\ \cline{3-9} 
{\color[HTML]{000000} }                       & \multirow{-2}{*}{{\color[HTML]{000000} 4m\_vs\_5m}}         & {\color[HTML]{000000} shared \checkmark}    & {\color[HTML]{000000} 5.1 / 8.1 / 8.5*}     & {\color[HTML]{FE0000} 5.8 / 7.1 / 7.9}      & {\color[HTML]{FE0000} 4.7 / 4.0 / 3.1}     & {\color[HTML]{000000} 7.0 / 7.1 / 7.2*}     & {\color[HTML]{000000} 6.9 / 7.0 / 7.1*}    & {\color[HTML]{FE0000} 6.9 / 12.4 / 13.3}  \\ \cline{2-9} 
{\color[HTML]{000000} }                       & {\color[HTML]{000000} }                                     & {\color[HTML]{000000} no shared} & {\color[HTML]{000000} 4.6 / 5.1 / 7.9*}     & {\color[HTML]{000000} 4.2 / 4.3 / 4.4*}     & {\color[HTML]{000000} 2.8 / 4.1 / 4.5}     & {\color[HTML]{000000} 4.3 / 6.0 / 6.1}      & {\color[HTML]{000000} 4.1 / 4.4 / 5.3}     & {\color[HTML]{000000} 4.6 / 13.5 / 17.0}  \\ \cline{3-9} 
{\color[HTML]{000000} }                       & \multirow{-2}{*}{{\color[HTML]{000000} 3s\_vs\_5z}}         & {\color[HTML]{000000} shared \checkmark}    & {\color[HTML]{000000} 4.3 / 5.3 / 7.8*}     & {\color[HTML]{000000} 4.1 / 4.4 / 4.4*}     & {\color[HTML]{FE0000} 3.3 / 4.5 / 5.1}     & {\color[HTML]{000000} 5.7 / 6.9 / 6.5*}     & {\color[HTML]{FE0000} 4.1 / 5.8 / 6.0}     & {\color[HTML]{FE0000} 4.2 / 12.9 / 20.0}  \\ \cline{2-9} 
{\color[HTML]{000000} }                       & {\color[HTML]{000000} }                                     & {\color[HTML]{000000} no shared \checkmark} & {\color[HTML]{FE0000} 10.8 / 12.3 / 12.2}   & {\color[HTML]{FE0000} 11.0 / 11.4 / 11.5}   & {\color[HTML]{000000} 9.5 / 13.4 / 13.5*}  & {\color[HTML]{FE0000} 11.1 / 12.9 / 13.5}   & {\color[HTML]{FE0000} 10.2 / 11.6 / 12.2}  & {\color[HTML]{FE0000} 12.9 / 17.8 / 19.4} \\ \cline{3-9} 
{\color[HTML]{000000} }                       & \multirow{-2}{*}{{\color[HTML]{000000} 1c1s1z\_vs\_1c1s3z}} & {\color[HTML]{000000} shared}    & {\color[HTML]{000000} 9.8 / 11.2 / 11.9}    & {\color[HTML]{000000} 9.0 / 9.2 / 9.4}      & {\color[HTML]{000000} 9.7 / 12.8 / 13.4}   & {\color[HTML]{000000} 10.7 / 12.2 / 12.6}   & {\color[HTML]{000000} 9.7 / 10.6 / 11.0}   & {\color[HTML]{000000} 12.5 / 15.8 / 18.4} \\ \cline{2-9} 
{\color[HTML]{000000} }                       & {\color[HTML]{000000} }                                     & {\color[HTML]{000000} no shared \checkmark} & {\color[HTML]{FE0000} 3.3 / 3.2 / 3.7}      & {\color[HTML]{FE0000} 2.6 / 4.0 / 5.4}      & {\color[HTML]{FE0000} 2.4 / 2.0 / 3.0}     & {\color[HTML]{000000} 4.3 / 4.5 / 4.5*}     & {\color[HTML]{000000} 4.0 / 4.6 / 4.5*}    & {\color[HTML]{FE0000} 4.3 / 18.3 / 18.8}  \\ \cline{3-9} 
\multirow{-8}{*}{{\color[HTML]{000000} SMAC}} & \multirow{-2}{*}{{\color[HTML]{000000} 4m\_vs\_4z}}         & {\color[HTML]{000000} shared}    & {\color[HTML]{000000} 2.6 / 3.2 / 3.2}      & {\color[HTML]{000000} 3.7 / 3.9 / 3.9}      & {\color[HTML]{000000} 3.2 / 2.9 / 1.4}     & {\color[HTML]{000000} 4.1 / 4.2 / 4.3*}     & {\color[HTML]{000000} 3.8 / 4.1 / 4.2*}    & {\color[HTML]{000000} 4.3 / 14.8 / 16.5}  \\ \hline\hline
\end{tabular}
 }
 }
\end{center}
\label{table:policy vdn}
\end{table*}
 
\begin{table*}[!h]
\caption{
Model performance with and without communication. The performance is recorded in the 'warmup performance / half steps performance / final performance' pattern. Detailed analysis can be found in Sec.~\ref{exp:Communicate: When and How}.
 }
\label{table:communication}
\small
 \begin{center}
 \resizebox{1.0\textwidth}{!}{
 \setlength{\tabcolsep}{1.3em}
 {\renewcommand{\arraystretch}{1.0}
\begin{tabular}{cccccccc}
\hline\hline
{\color[HTML]{000000} scenario}                             & {\color[HTML]{000000} obs overlap}            & {\color[HTML]{000000} baseline}                             & {\color[HTML]{000000} communication}                           & {\color[HTML]{000000} scenario}                           & {\color[HTML]{000000} obs overlap}            & {\color[HTML]{000000} baseline}                            & {\color[HTML]{000000} communication}     \\ \hline
{\color[HTML]{000000} }                                     & {\color[HTML]{000000} }                       & {\color[HTML]{000000} }                                     & \multicolumn{1}{c|}{{\color[HTML]{000000} 6.6 / 11.5 / 11.4}}  & {\color[HTML]{000000} }                                   & {\color[HTML]{000000} }                       & {\color[HTML]{000000} }                                    & {\color[HTML]{000000} 7.7 / 19.6 / 19.4} \\
\multirow{-2}{*}{{\color[HTML]{000000} 4m\_vs\_5m}}         & \multirow{-2}{*}{{\color[HTML]{000000} 0.47}} & \multirow{-2}{*}{{\color[HTML]{000000} 6.5 / 10.1 / 10.9}}  & \multicolumn{1}{c|}{{\color[HTML]{FE0000} +1.4 +0.5}}          & \multirow{-2}{*}{{\color[HTML]{000000} 4m\_vs\_3z}}       & \multirow{-2}{*}{{\color[HTML]{000000} 0.37}} & \multirow{-2}{*}{{\color[HTML]{000000} 7.5 / 19.6 / 19.3}} & {\color[HTML]{000000} 0.0 +0.1}          \\ \hline
{\color[HTML]{000000} }                                     & {\color[HTML]{000000} }                       & {\color[HTML]{000000} }                                     & \multicolumn{1}{c|}{{\color[HTML]{000000} 12.4 / 15.7 / 18.1}} & {\color[HTML]{000000} }                                   & {\color[HTML]{000000} }                       & {\color[HTML]{000000} }                                    & {\color[HTML]{000000} 5.4 / 12.4 / 15.5} \\
\multirow{-2}{*}{{\color[HTML]{000000} 1c1s1z\_vs\_1c1s3z}} & \multirow{-2}{*}{{\color[HTML]{000000} 0.40}} & \multirow{-2}{*}{{\color[HTML]{000000} 12.3 / 15.9 / 17.9}} & \multicolumn{1}{c|}{{\color[HTML]{000000} -0.2 +0.2}}          & \multirow{-2}{*}{{\color[HTML]{000000} 3s\_vs\_5z}}       & \multirow{-2}{*}{{\color[HTML]{000000} 0.21}} & \multirow{-2}{*}{{\color[HTML]{000000} 5.4 / 12.9 / 16.4}} & {\color[HTML]{009901} -0.5 -0.9}         \\ \hline
{\color[HTML]{000000} }                                     & {\color[HTML]{000000} }                       & {\color[HTML]{000000} }                                     & \multicolumn{1}{c|}{{\color[HTML]{000000} 4.1 / 15.9 / 18.3}}  & {\color[HTML]{000000} }                                   & {\color[HTML]{000000} }                       & {\color[HTML]{000000} }                                    & {\color[HTML]{000000} 7.9 / 13.2 / 19.0} \\
\multirow{-2}{*}{{\color[HTML]{000000} 4m\_vs\_4z}}         & \multirow{-2}{*}{{\color[HTML]{000000} 0.32}} & \multirow{-2}{*}{{\color[HTML]{000000} 4.3 / 13.2 / 17.1}}  & \multicolumn{1}{c|}{{\color[HTML]{FE0000} +2.7 +1.2}}          & \multirow{-2}{*}{{\color[HTML]{000000} 1s1m1h1M\_vs\_4z}} & \multirow{-2}{*}{{\color[HTML]{000000} 0.25}} & \multirow{-2}{*}{{\color[HTML]{000000} 8.4 / 16.0 / 19.8}} & {\color[HTML]{009901} -2.8 -0.8}         \\ \hline\hline
\end{tabular}
 }
 }
 \end{center}
 \end{table*}

\begin{figure*}[!h]
\begin{center}
  \includegraphics[width=1.0\linewidth]{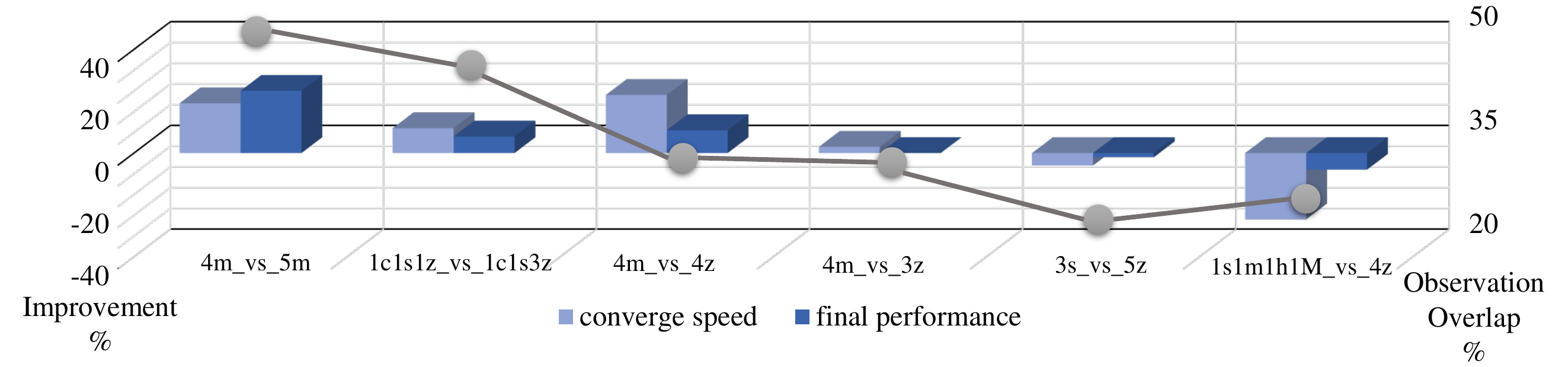}
\end{center}
  \caption{Histogram of performance improvement when adopting communication mechanism compared to baseline (w/o communication). Grey dots represent for the observation overlap. Larger the overlap, smaller  trajectory-based role diversity. }
\label{fig:variants}
\end{figure*}

\begin{figure*}[!h]
    \centering 
\begin{subfigure}{0.45\textwidth}
  \includegraphics[width=\linewidth]{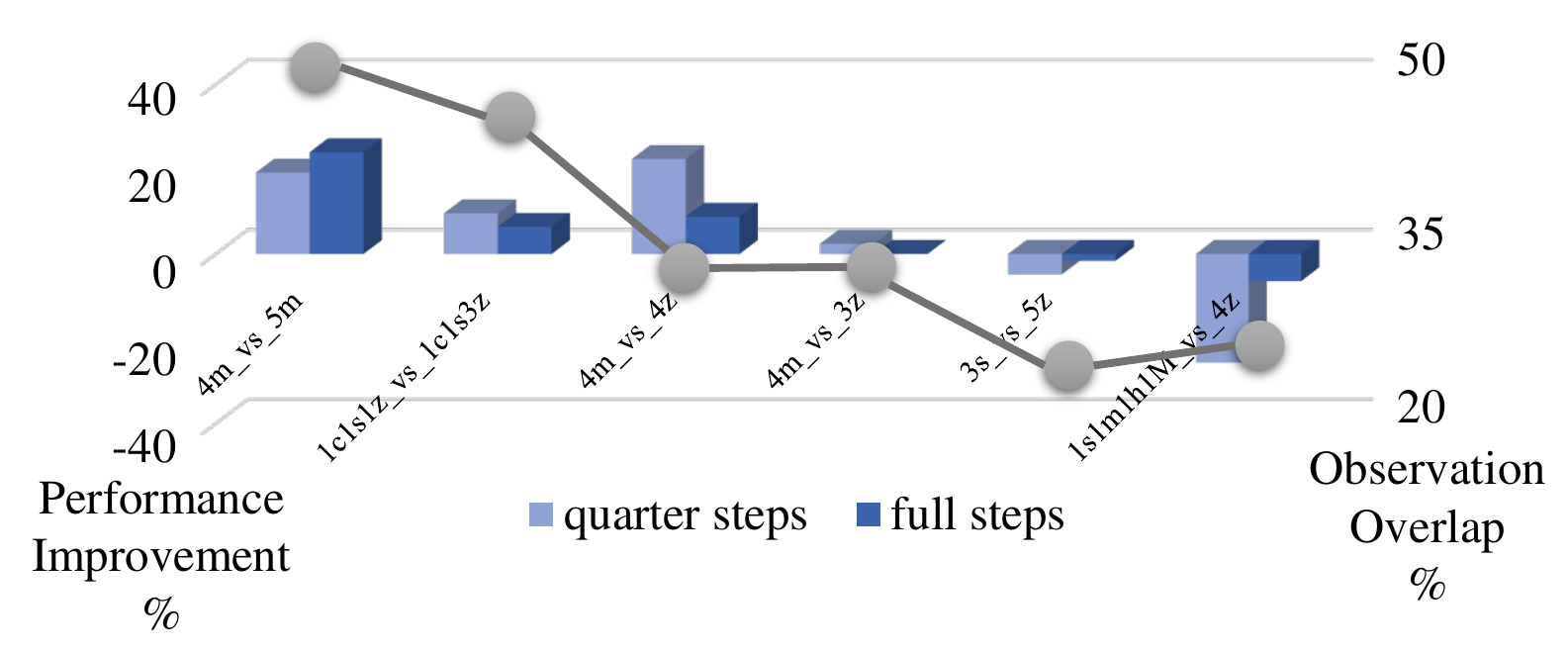}
  \caption{Model performance with or without communication}
  \label{fig:exp_comm}
\end{subfigure}\hfil 
\begin{subfigure}{0.45\textwidth}
  \includegraphics[width=\linewidth]{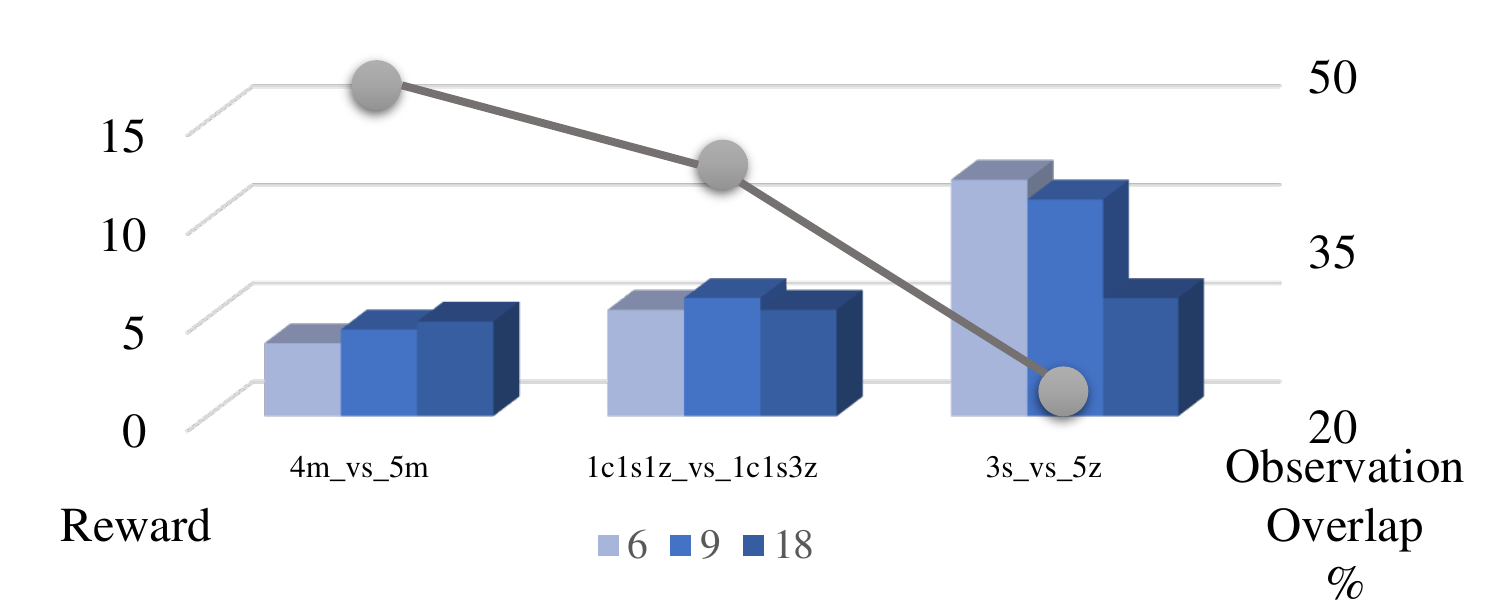}
  \caption{Model performance with various vision scope}
  \label{fig:exp_scope}
\end{subfigure}\hfil 
\caption{Histogram of (a) model performance when adopting communication mechanism compared to baseline (w/o communication). (b) model performance with different vision scope (6-9-18), where scope 9 is the standard setting in SMAC. Grey dots represent the observation overlap. Larger the overlap, the smaller the trajectory-based role diversity. Detailed analysis can be found in Sec.~\ref{exp:Communicate: When and How}}
\label{fig:Histogram}
\end{figure*}

\begin{figure}[!h]
    \centering 
\begin{subfigure}{0.25\textwidth}
  \includegraphics[width=\linewidth]{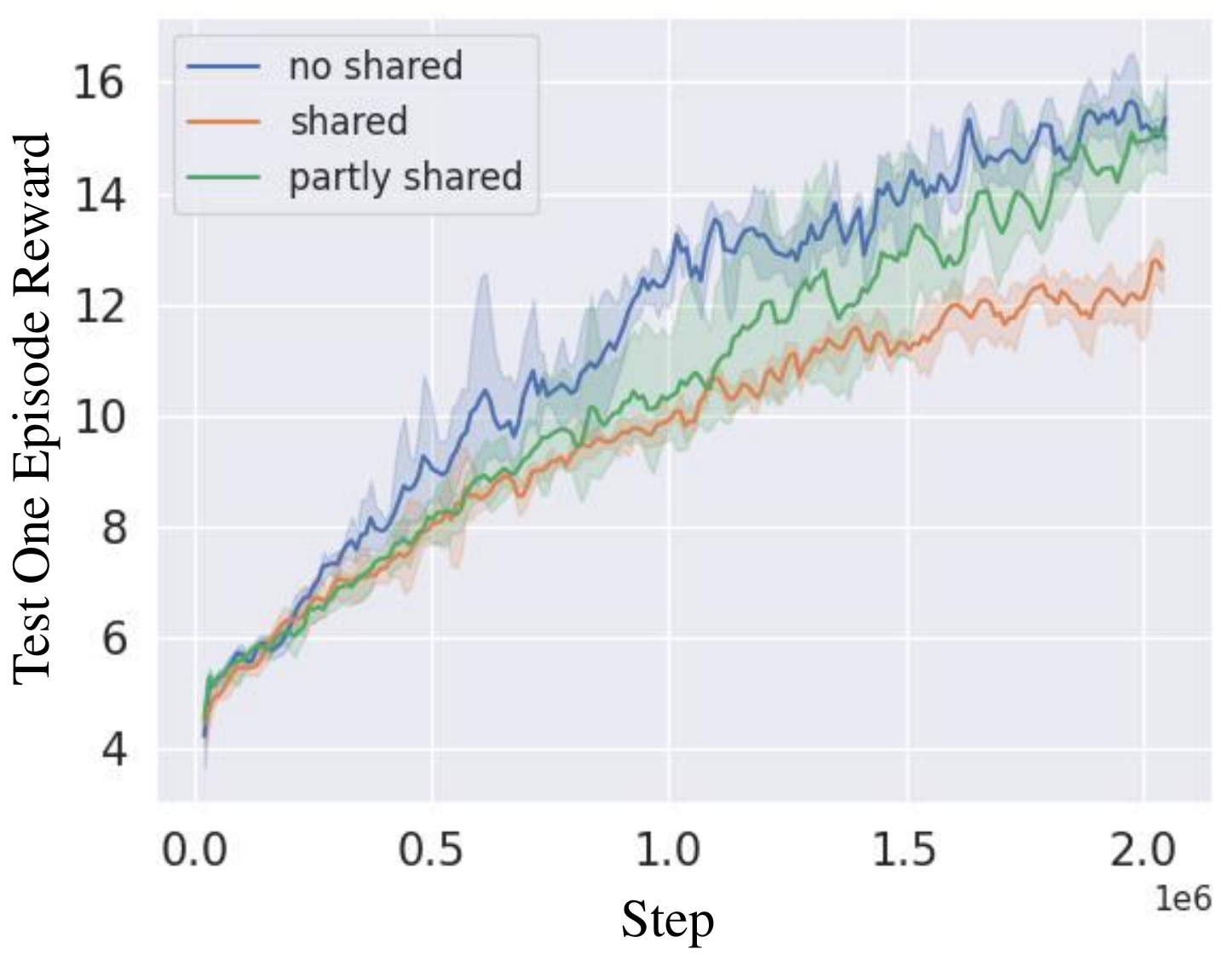}
  \caption{3s\_vs\_5z}
  \label{chart1:3s_vs_5z}
\end{subfigure}\hfil 
\begin{subfigure}{0.25\textwidth}
  \includegraphics[width=\linewidth]{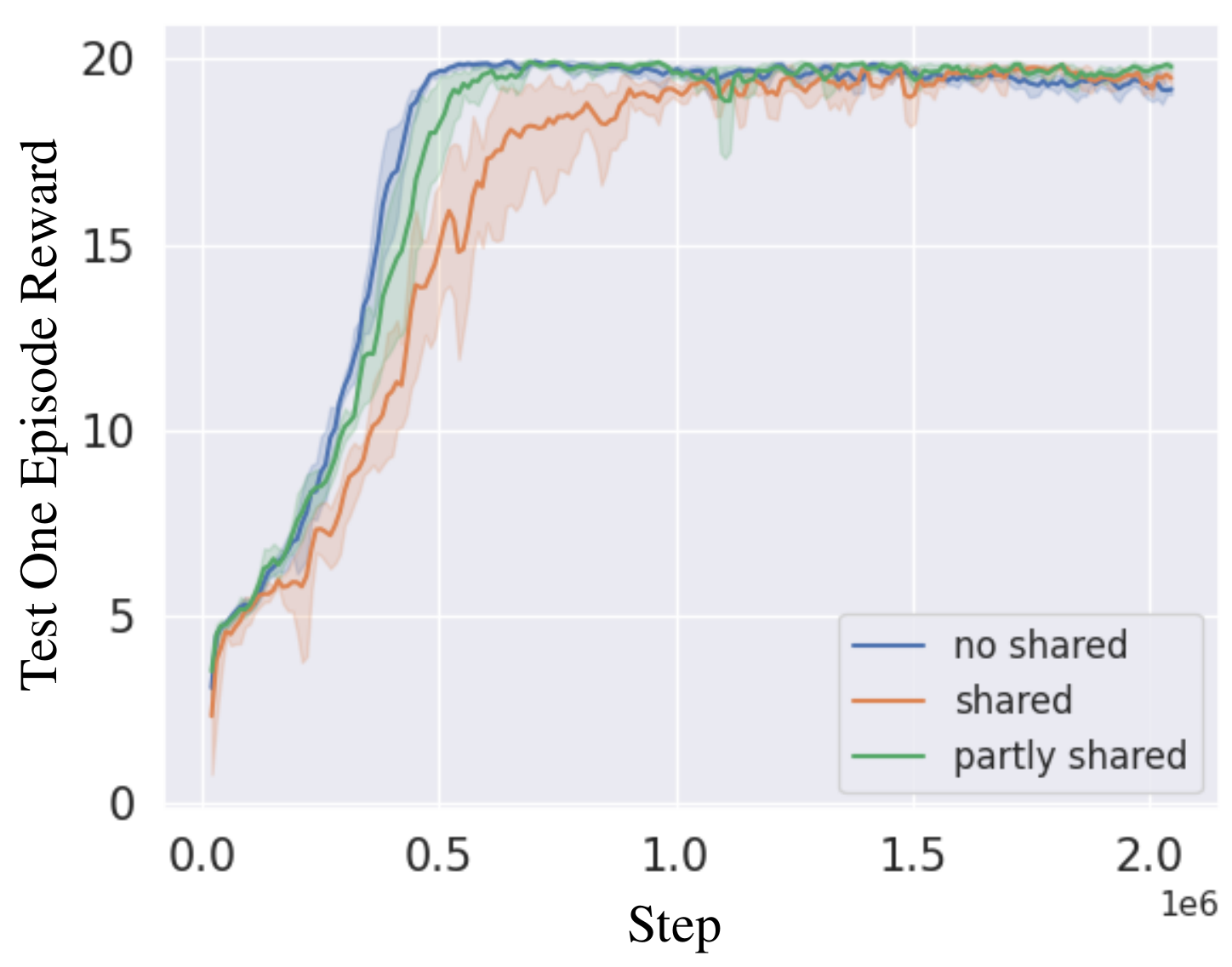}
  \caption{4m\_vs\_3z}
  \label{chart1:4m_vs_3z}
\end{subfigure}\hfil 
\begin{subfigure}{0.25\textwidth}
  \includegraphics[width=\linewidth]{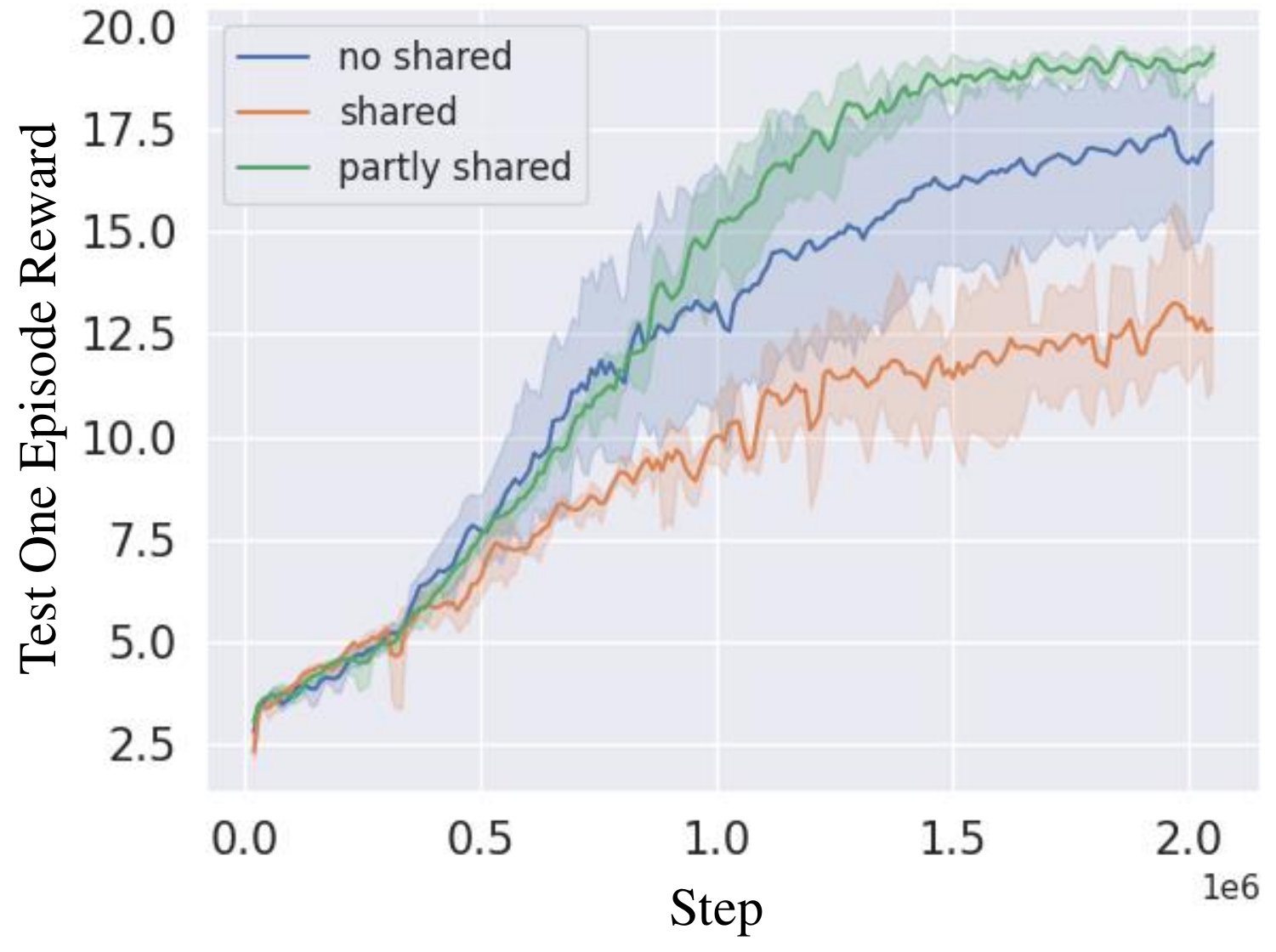}
  \caption{4m\_vs\_4z}
  \label{chart1:4m_vs_4z}
\end{subfigure}\hfil 
\begin{subfigure}{0.25\textwidth}
  \includegraphics[width=\linewidth]{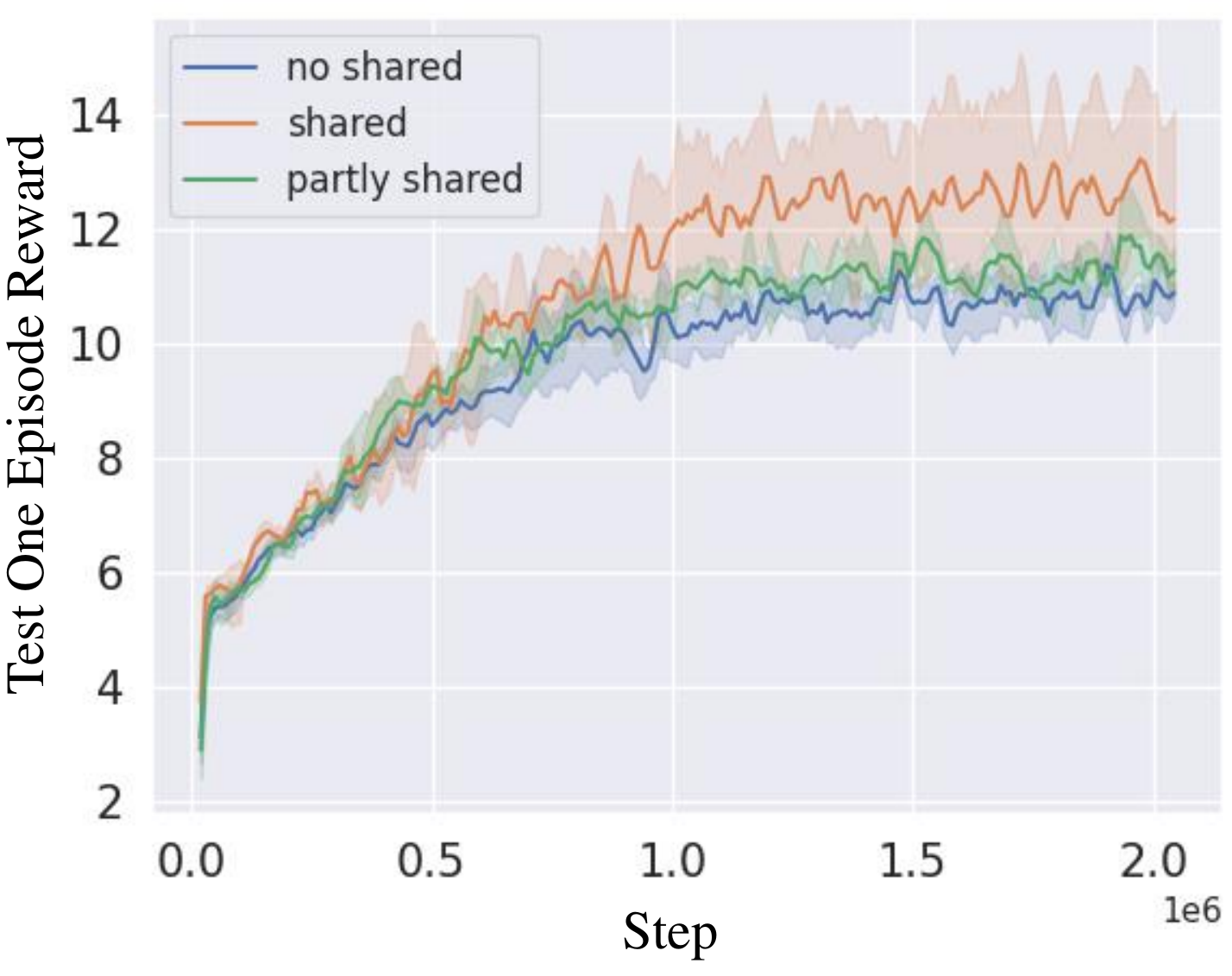}
  \caption{4m\_vs\_5m}
  \label{chart1:4m_vs_5m}
\end{subfigure}\hfil 
\medskip
\begin{subfigure}{0.25\textwidth}
  \includegraphics[width=\linewidth]{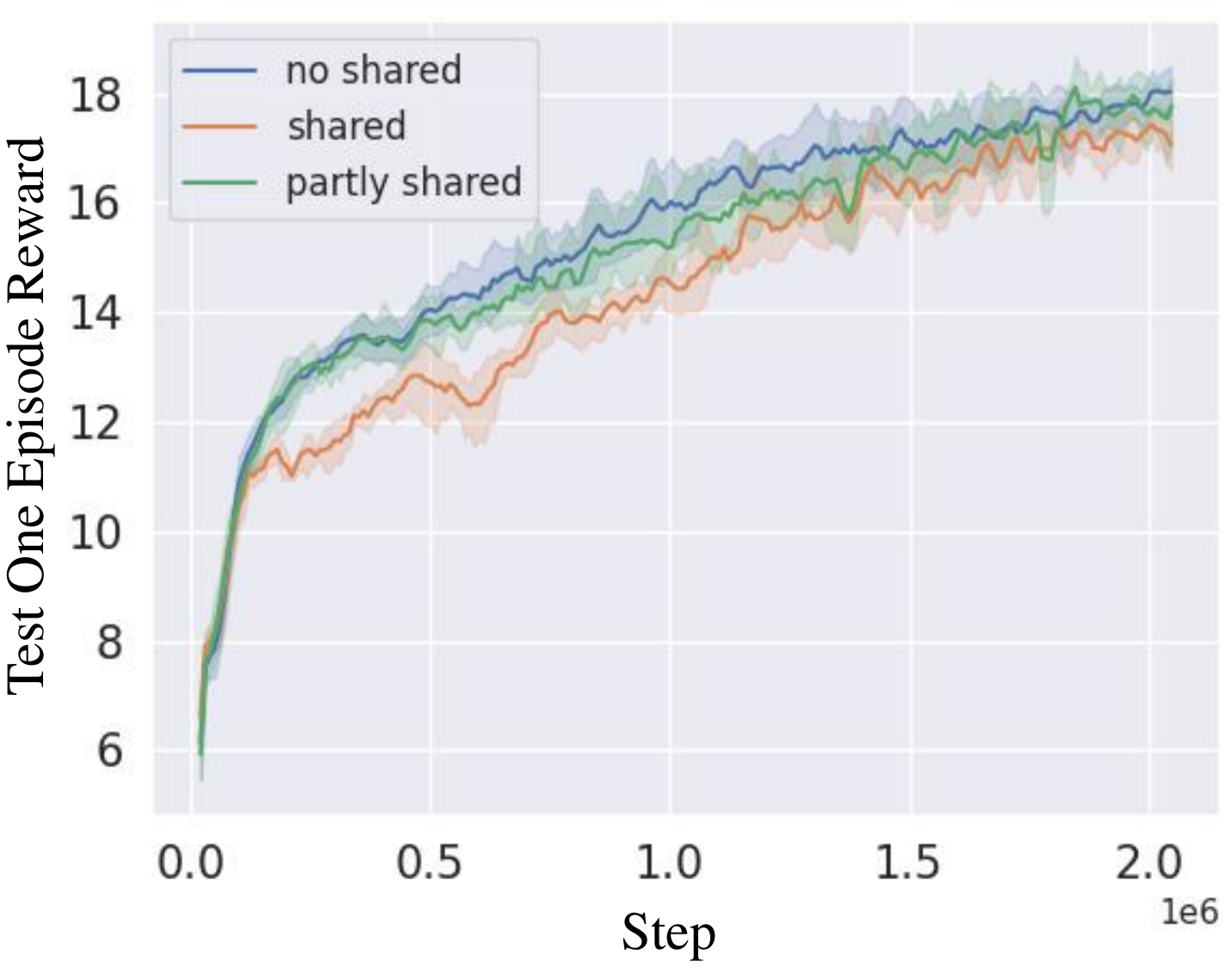}
  \caption{1c1s1z\_vs\_1c1s3z}
  \label{chart1:1c1s1z_vs_1c1s3z}
\end{subfigure}\hfil 
\begin{subfigure}{0.25\textwidth}
  \includegraphics[width=\linewidth]{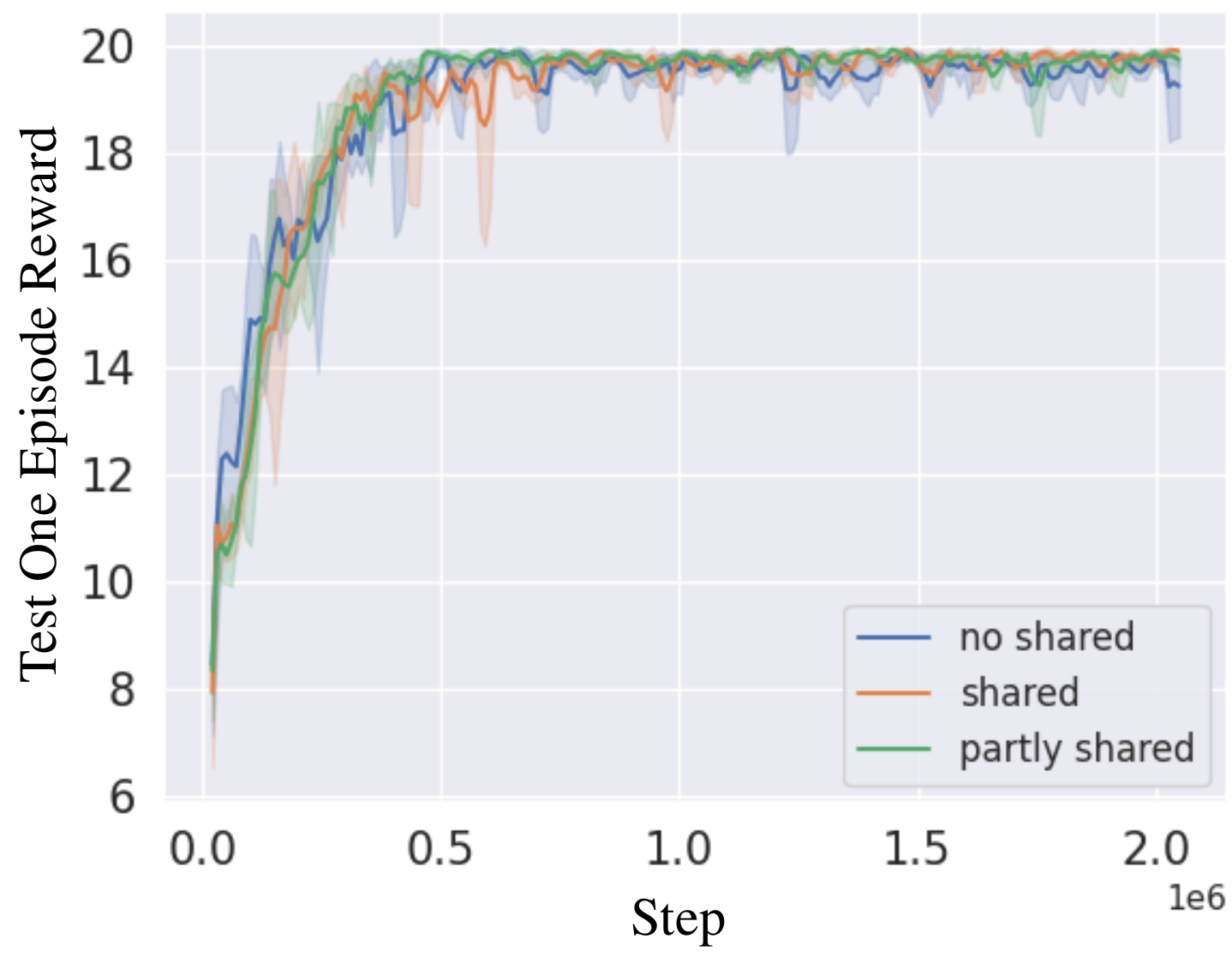}
  \caption{1s1m1h1M\_vs\_3z}
  \label{chart1:1s1m1h1M_vs_3z}
\end{subfigure}\hfil 
\begin{subfigure}{0.25\textwidth}
  \includegraphics[width=\linewidth]{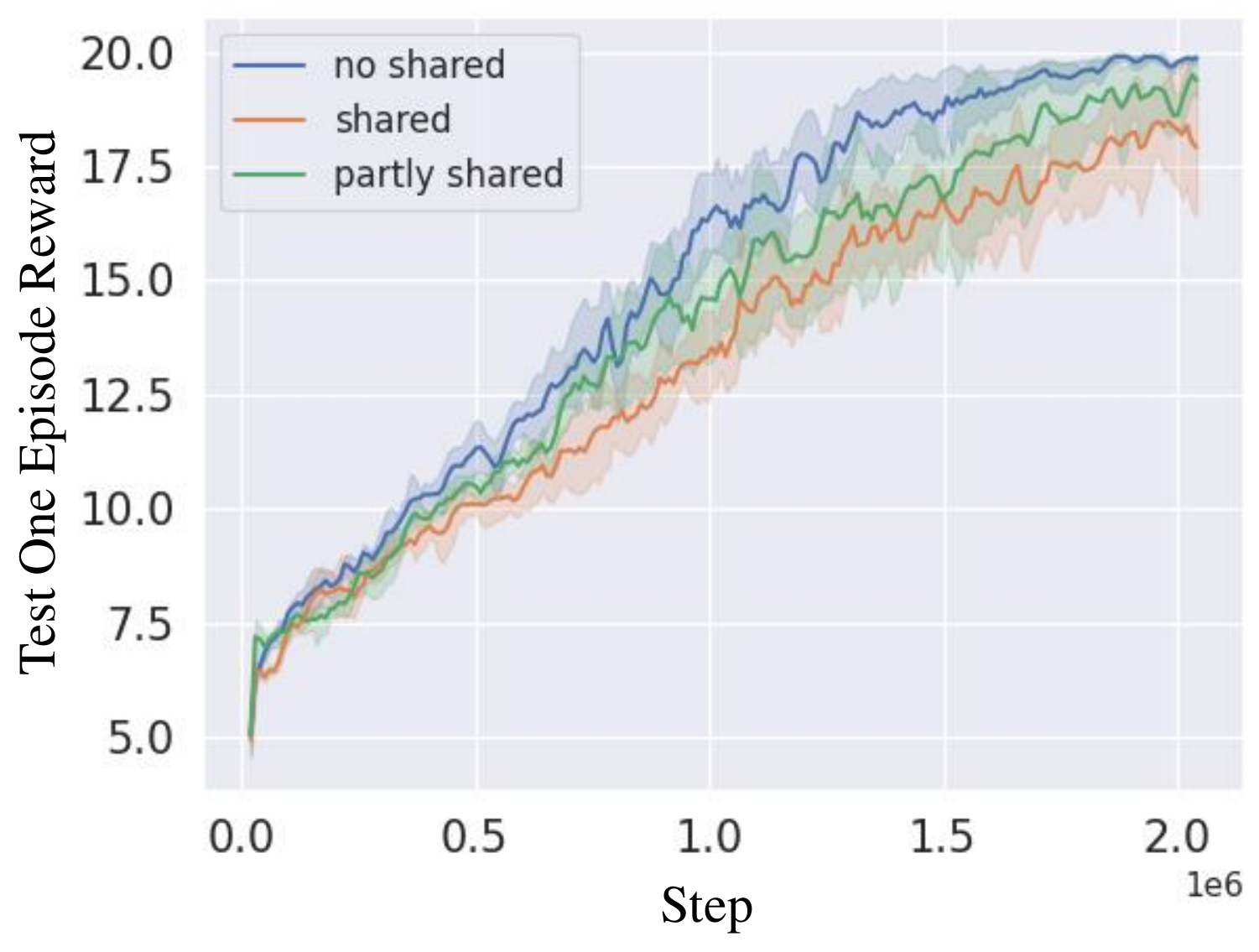}
  \caption{1s1m1h1M\_vs\_4z}
  \label{chart1:1s1m1h1M_vs_4z}
\end{subfigure}\hfil 
\begin{subfigure}{0.25\textwidth}
  \includegraphics[width=\linewidth]{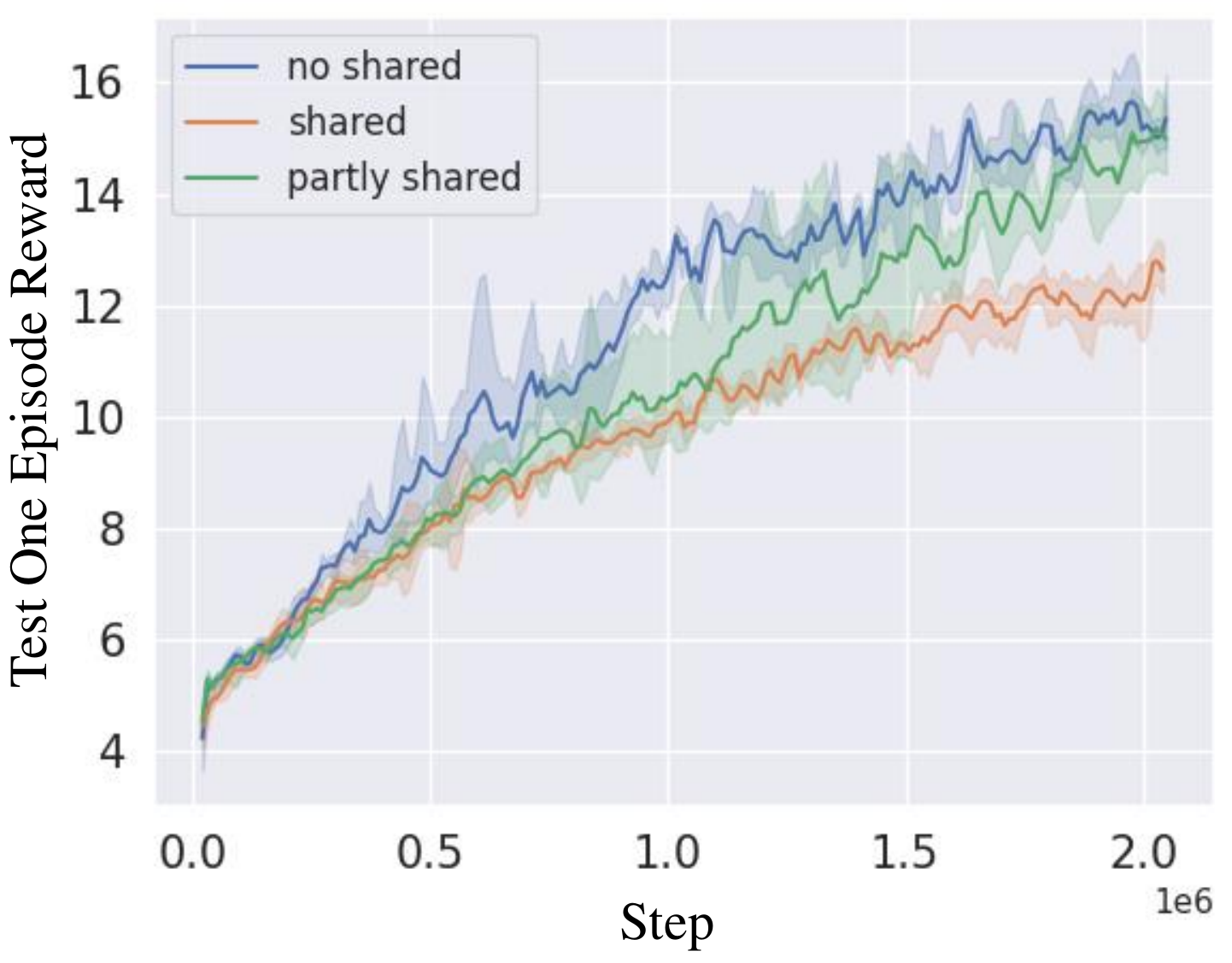}
  \caption{1s1m1h1M\_vs\_5z}
  \label{chart1:1s1m1h1M_vs_5z}
\end{subfigure}\hfil 
\caption{Policy learning curve with different parameter sharing strategies.}
\label{fig:parameter_sharing_curve}
\end{figure}

\begin{figure}[!h]
    \centering 
\begin{subfigure}{0.25\textwidth}
  \includegraphics[width=\linewidth]{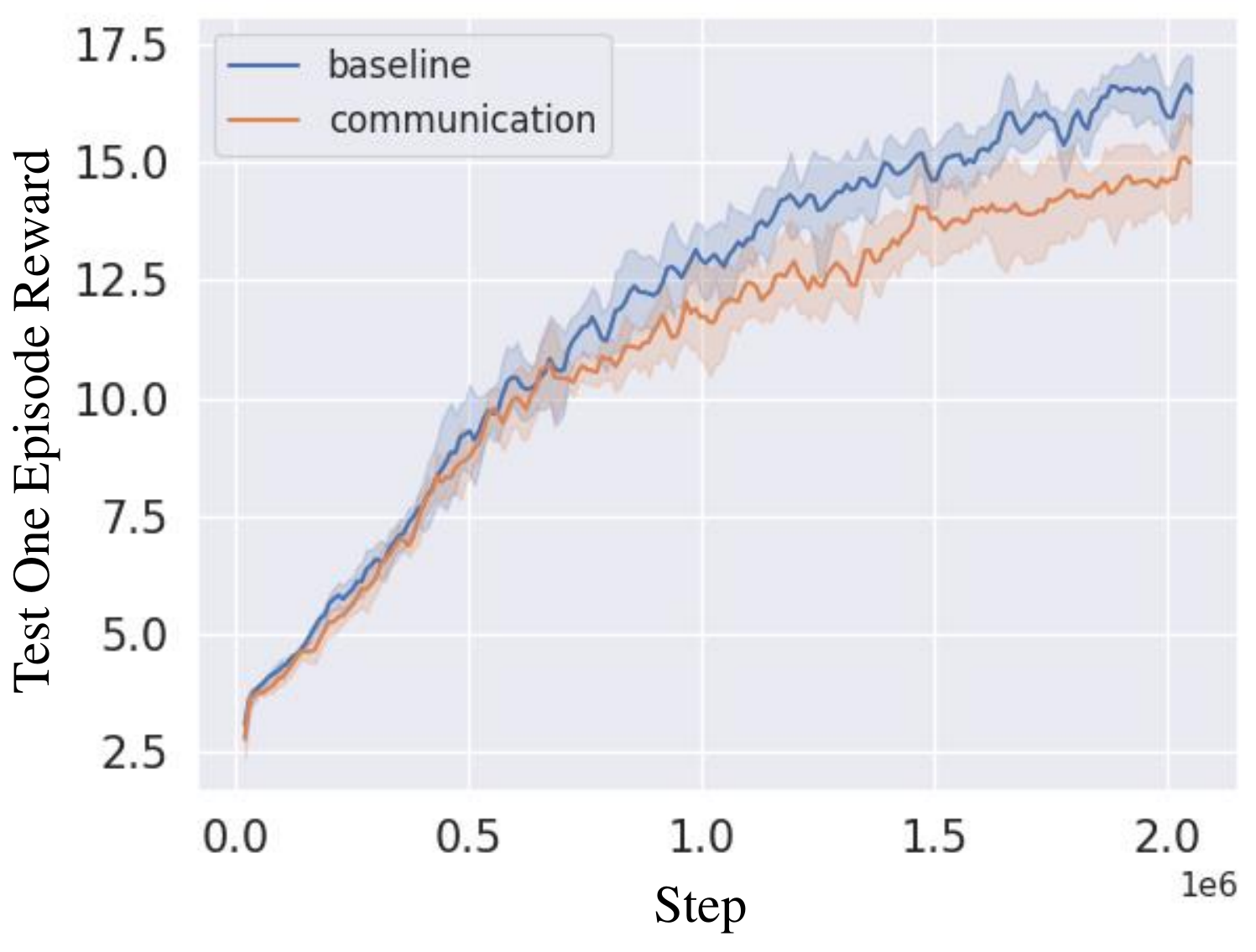}
  \caption{3s\_vs\_5z}
  \label{chart3:3s_vs_5z}
\end{subfigure}\hfil 
\begin{subfigure}{0.25\textwidth}
  \includegraphics[width=\linewidth]{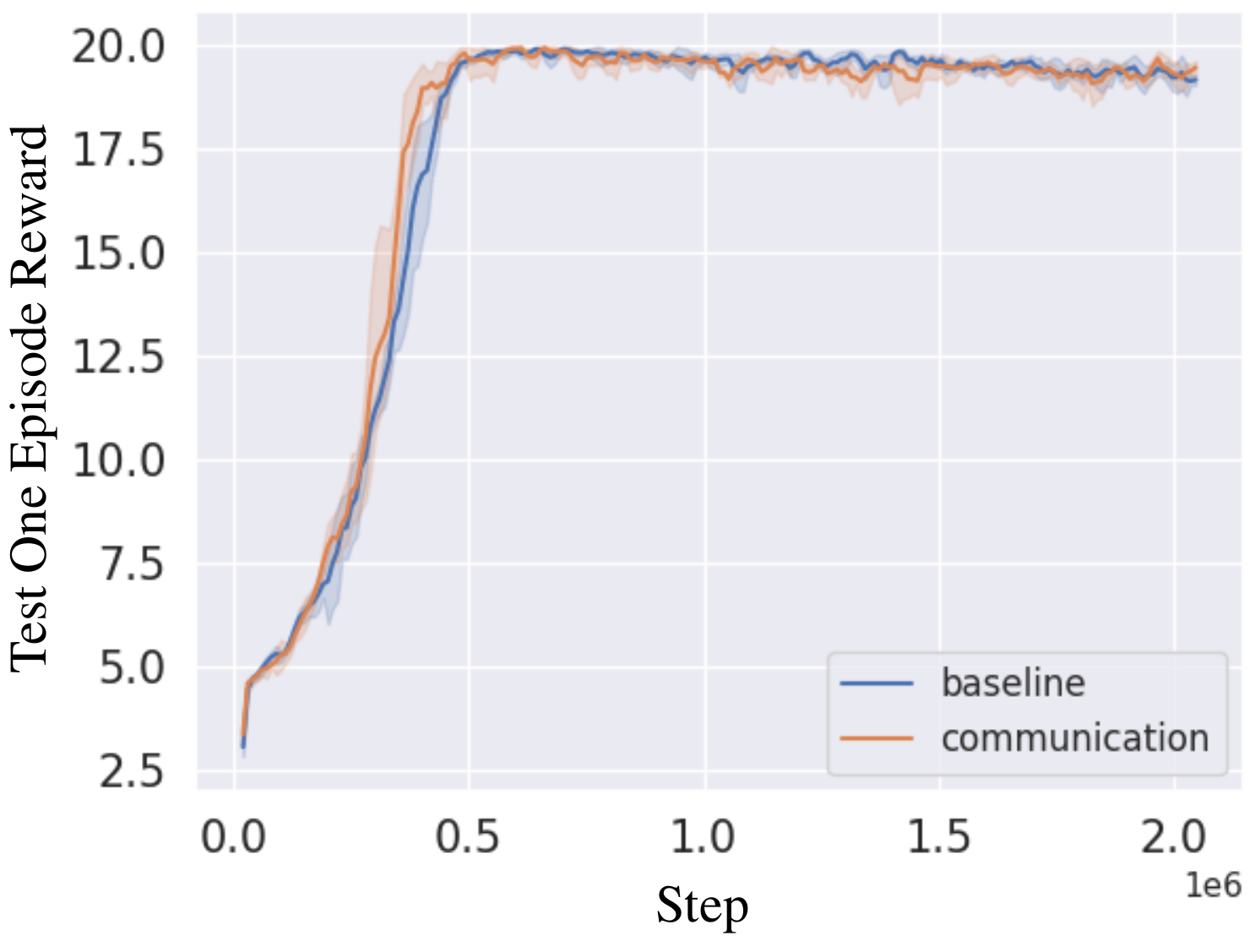}
  \caption{4m\_vs\_3z}
  \label{chart3:4m_vs_3z}
\end{subfigure}\hfil 
\begin{subfigure}{0.25\textwidth}
  \includegraphics[width=\linewidth]{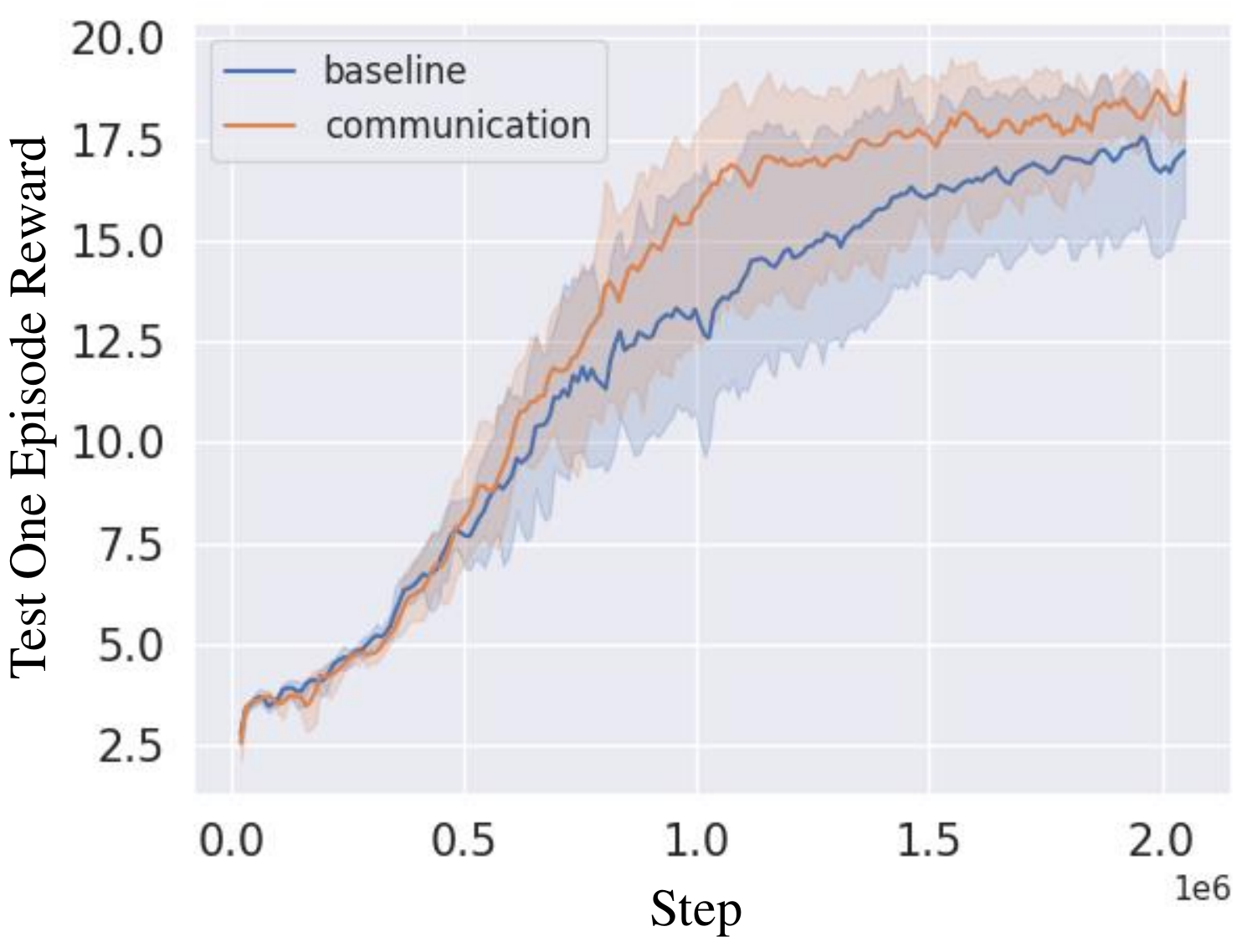}
  \caption{4m\_vs\_4z}
  \label{chart3:4m_vs_4z}
\end{subfigure}\hfil 
\begin{subfigure}{0.25\textwidth}
  \includegraphics[width=\linewidth]{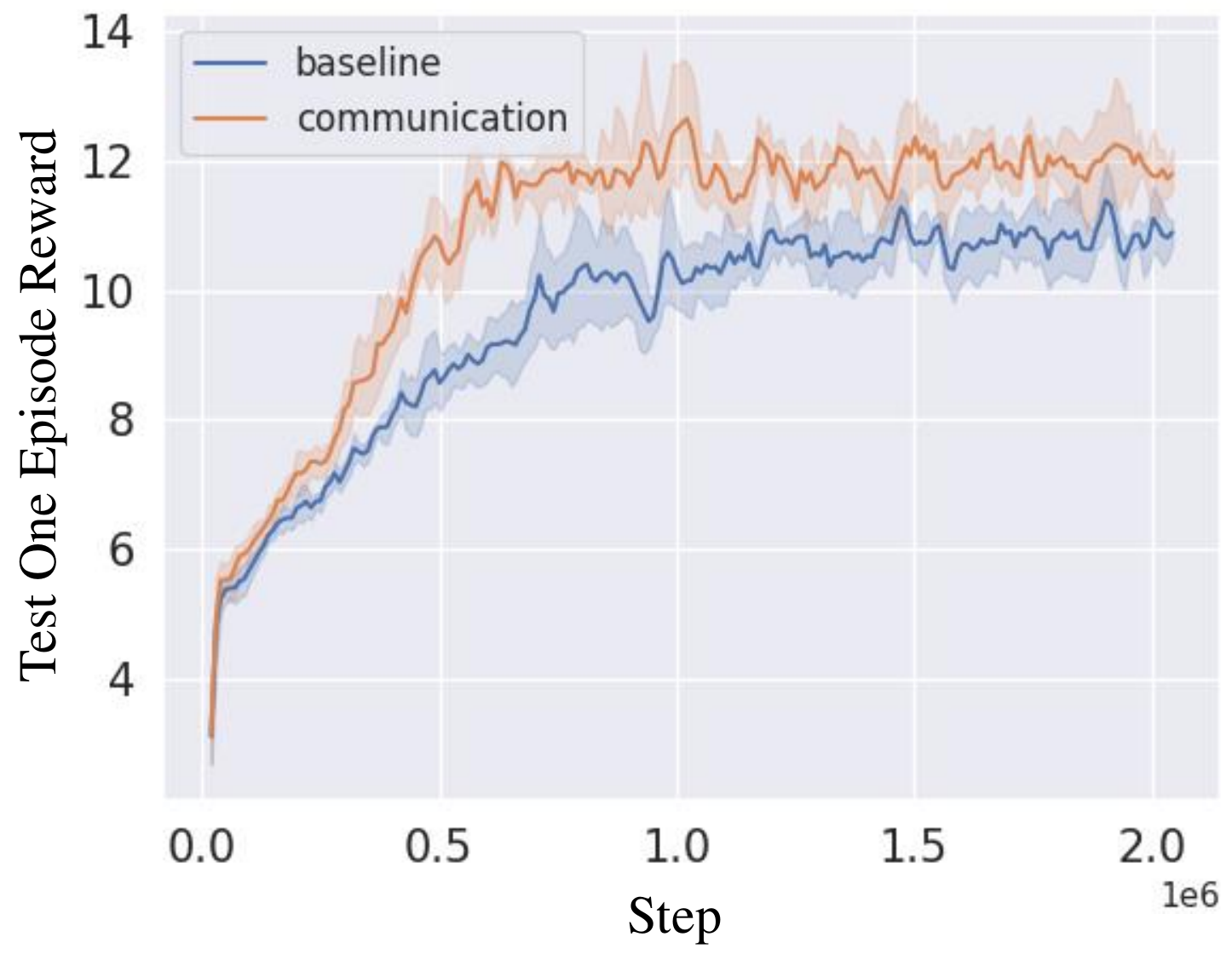}
  \caption{4m\_vs\_5m}
  \label{chart3:4m_vs_5m}
\end{subfigure}\hfil 
\medskip
\begin{subfigure}{0.25\textwidth}
  \includegraphics[width=\linewidth]{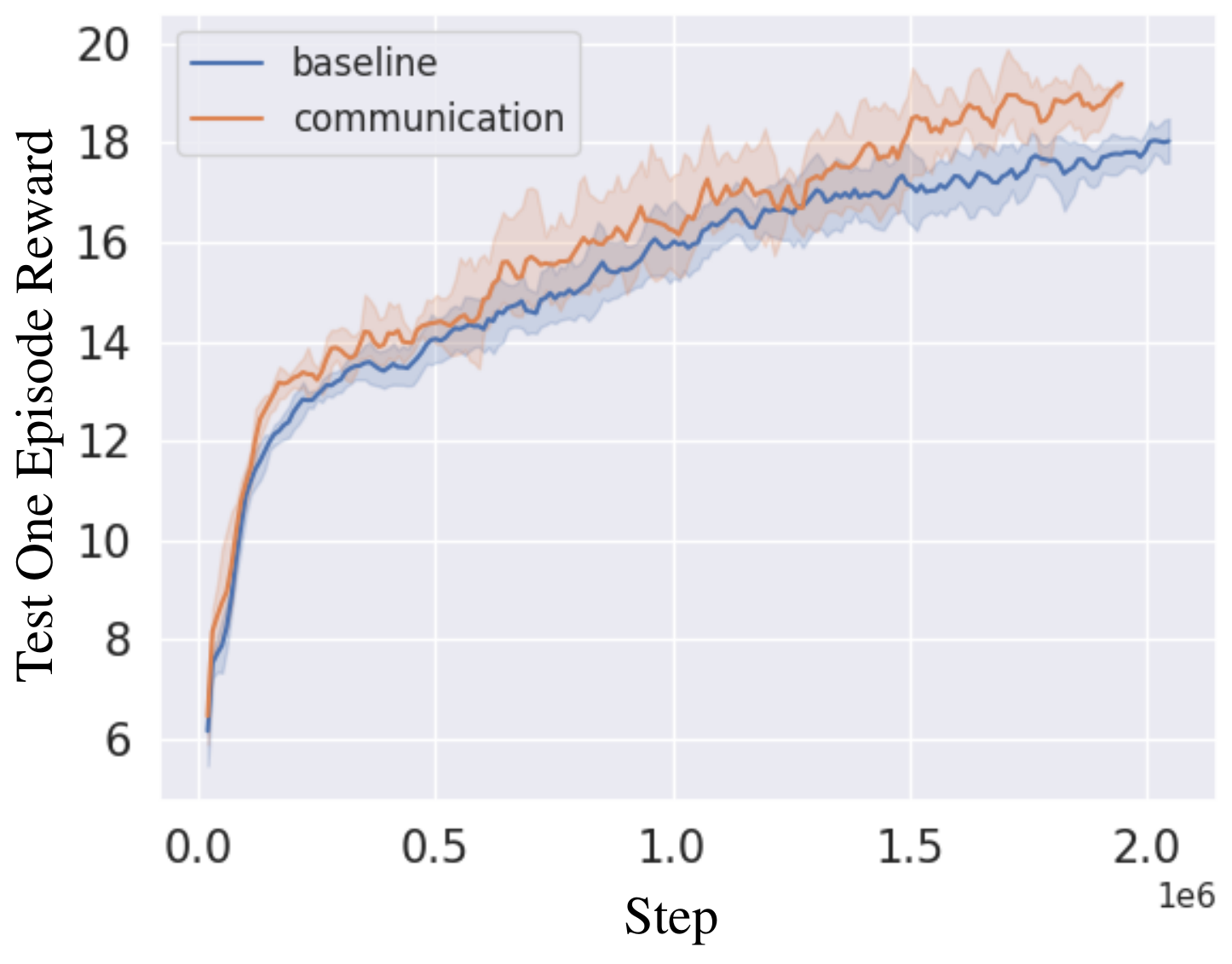}
  \caption{1c1s1z\_vs\_1c1s3z}
  \label{chart3:1c1s1z_vs_1c1s3z}
\end{subfigure}\hfil 
\begin{subfigure}{0.25\textwidth}
  \includegraphics[width=\linewidth]{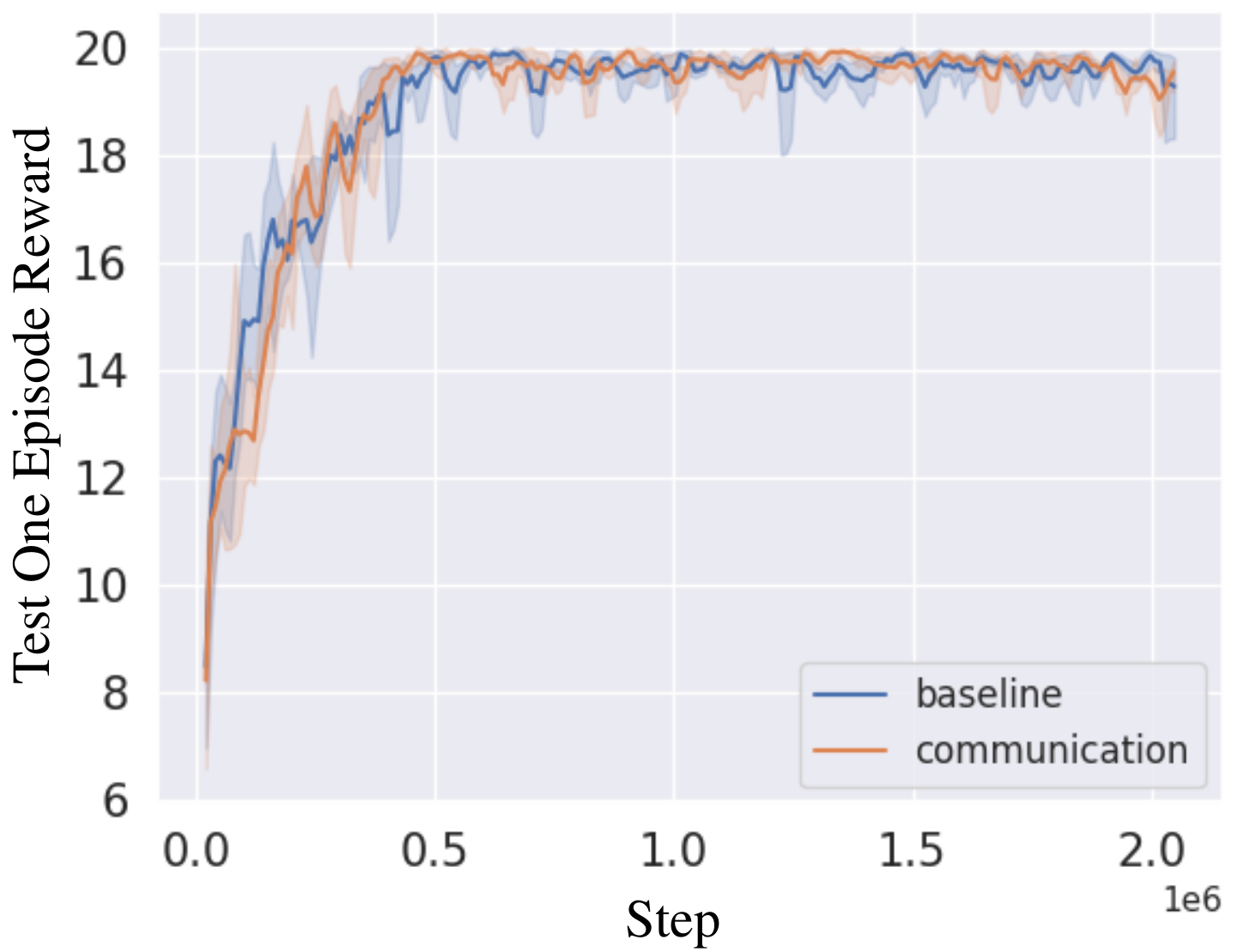}
  \caption{1s1m1h1M\_vs\_3z}
  \label{chart3:1s1m1h1M_vs_3z}
\end{subfigure}\hfil 
\begin{subfigure}{0.25\textwidth}
  \includegraphics[width=\linewidth]{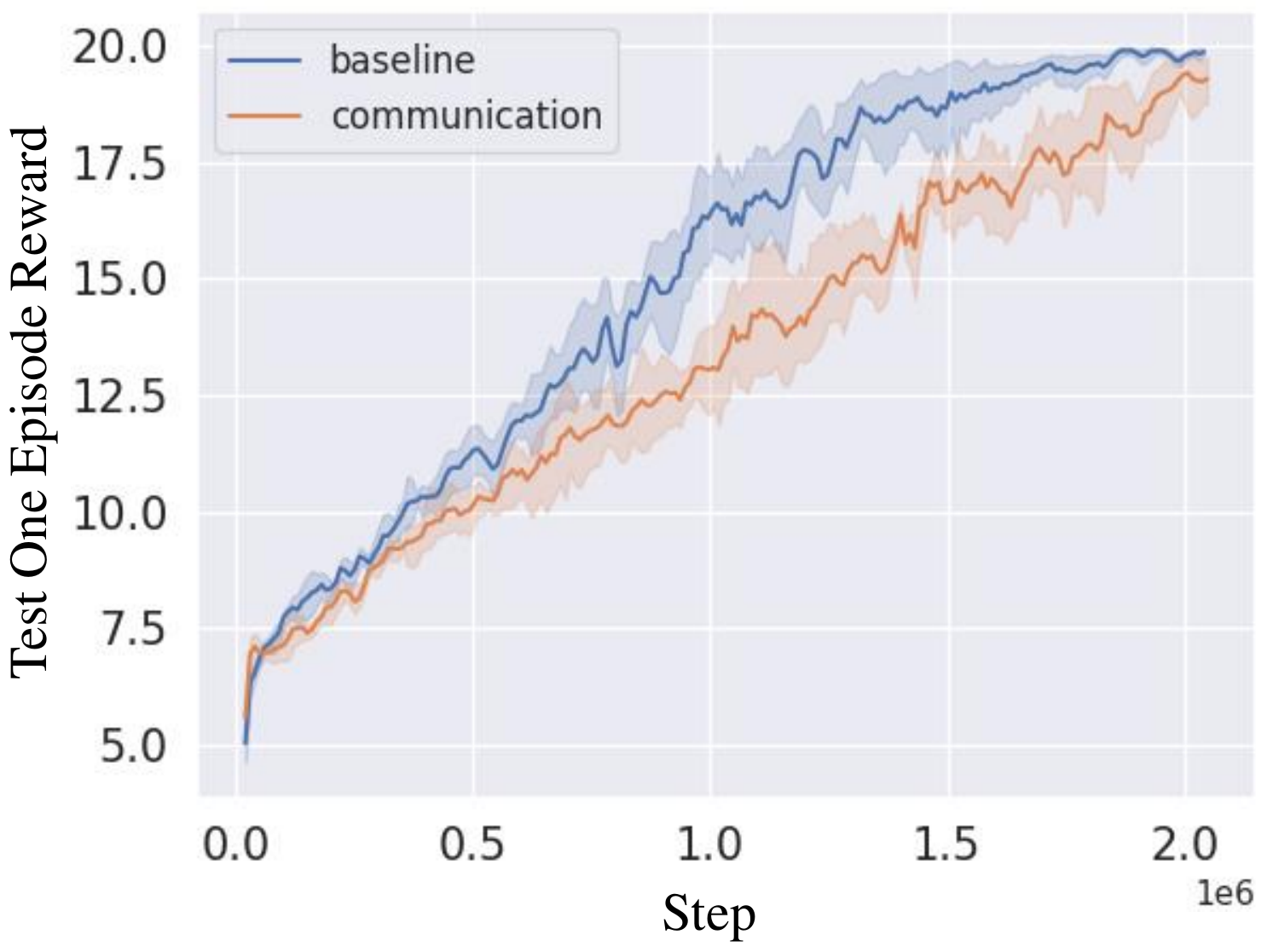}
  \caption{1s1m1h1M\_vs\_4z}
  \label{chart3:1s1m1h1M_vs_4z}
\end{subfigure}\hfil 
\begin{subfigure}{0.25\textwidth}
  \includegraphics[width=\linewidth]{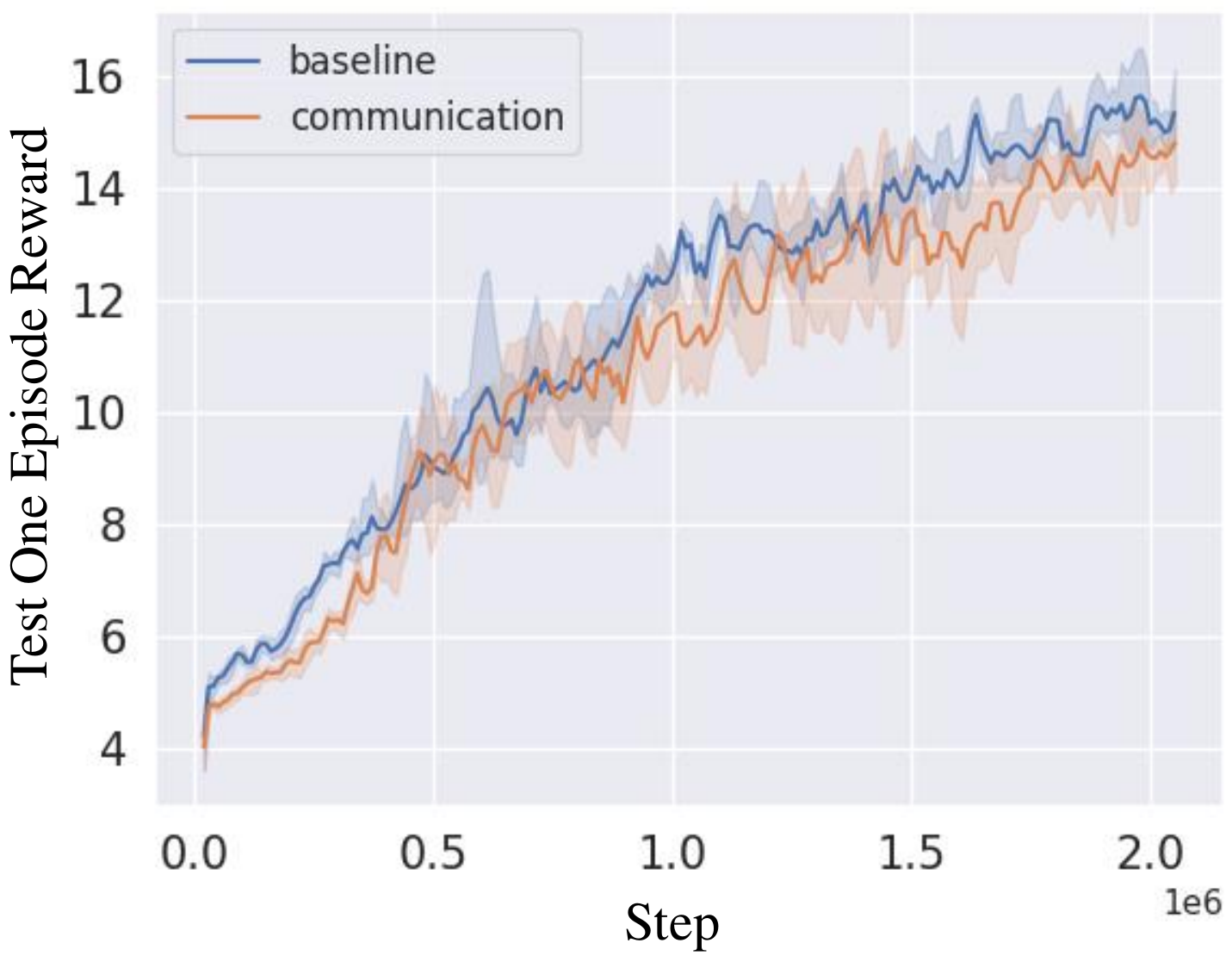}
  \caption{1s1m1h1M\_vs\_5z}
  \label{chart3:1s1m1h1M_vs_5z}
\end{subfigure}\hfil 
\caption{Policy learning curve with and without communication on different scenarios.}
  \label{fig:communication_curve}
\end{figure}

\begin{figure}[!h]
    \centering 
\begin{subfigure}{0.3\textwidth}
  \includegraphics[width=\linewidth]{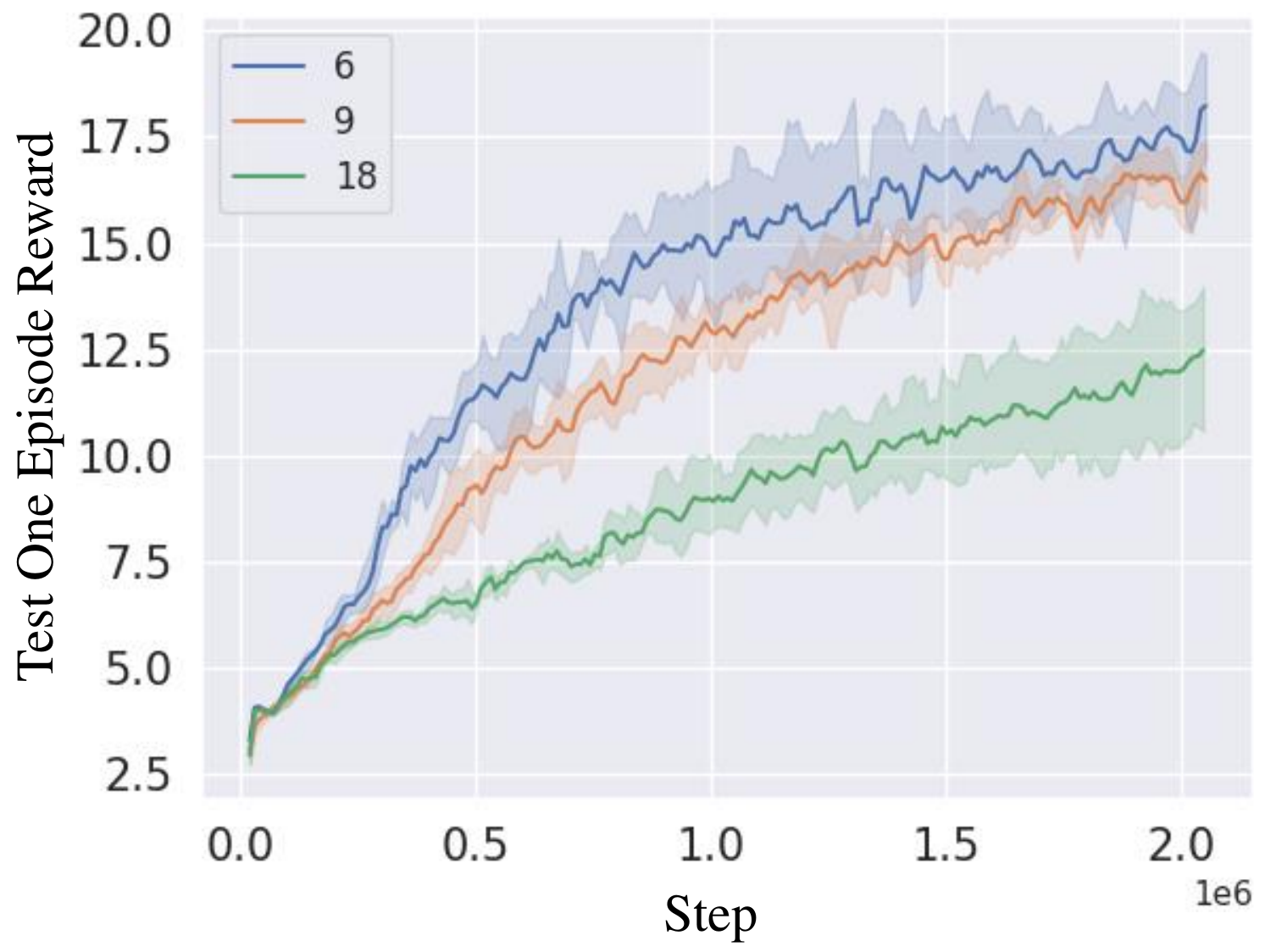}
  \caption{3s\_vs\_5z}
  \label{chart4:3s_vs_5z}
\end{subfigure}\hfil 
\begin{subfigure}{0.3\textwidth}
  \includegraphics[width=\linewidth]{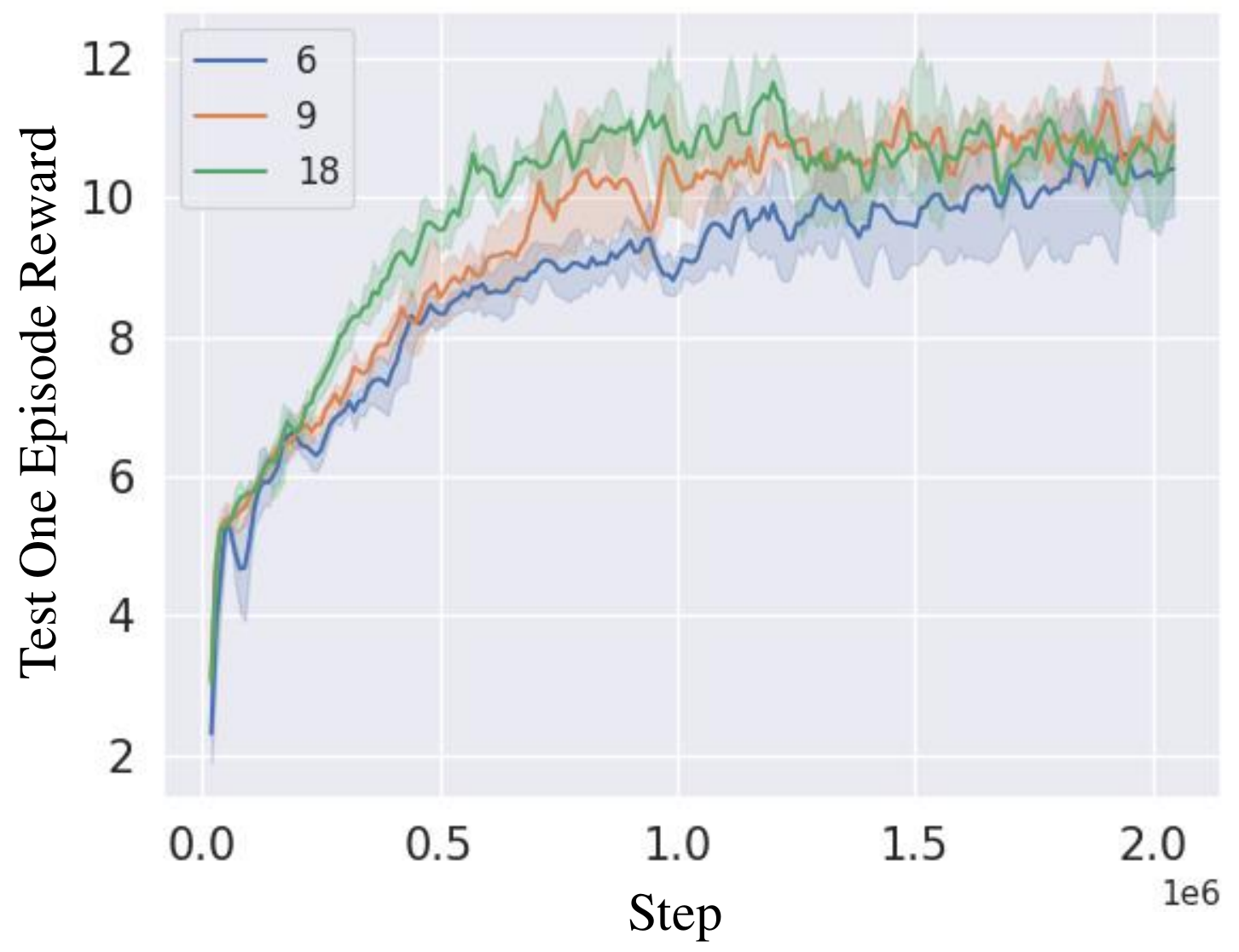}
  \caption{4m\_vs\_5m}
  \label{chart4:4m_vs_5m}
\end{subfigure}\hfil 
\begin{subfigure}{0.3\textwidth}
  \includegraphics[width=\linewidth]{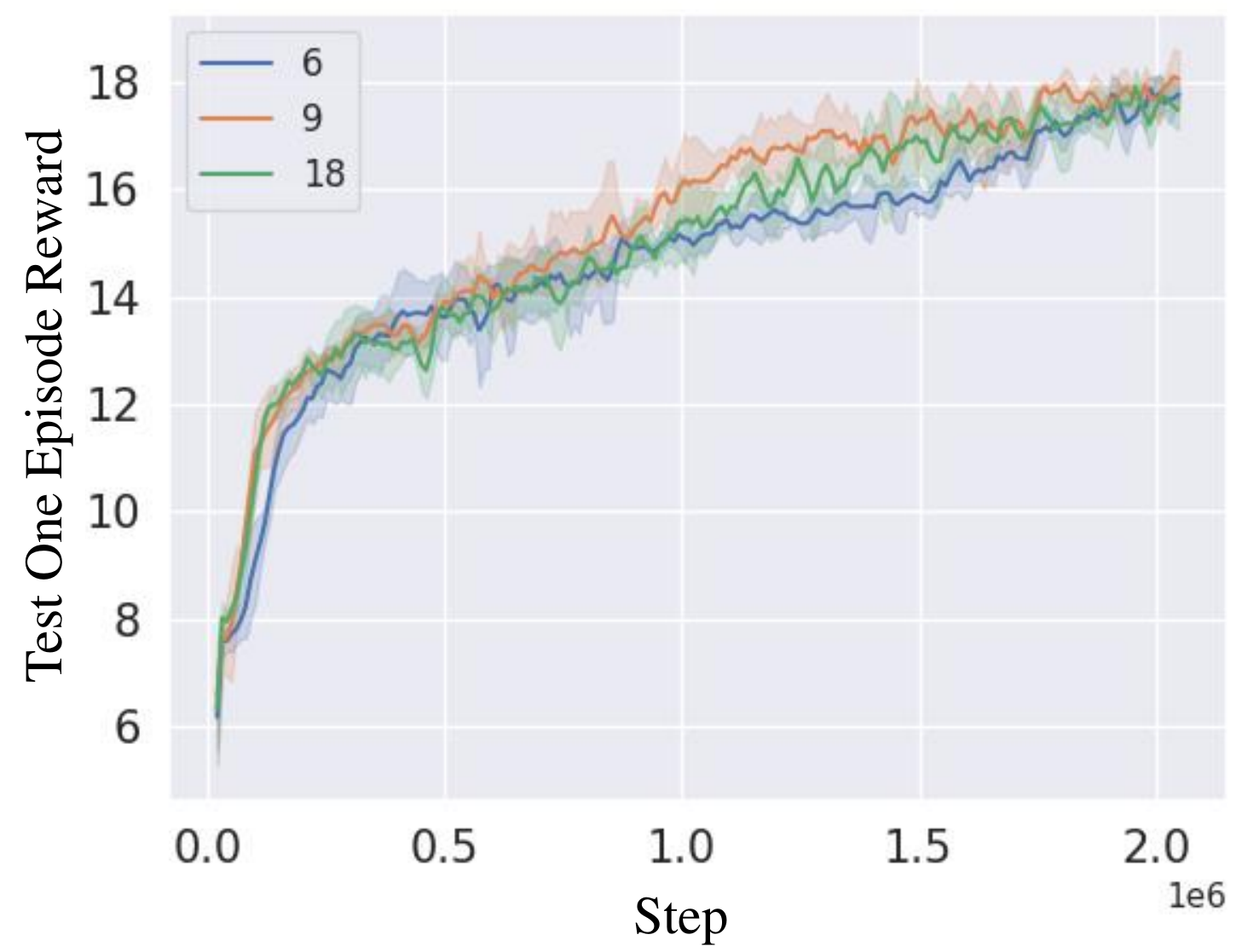}
  \caption{1c1s1z\_vs\_1c1s3z}
  \label{chart4:1c1s1z_vs_1c1s3z}
\end{subfigure}\hfil 
\medskip
\begin{subfigure}{0.3\textwidth}
  \includegraphics[width=\linewidth]{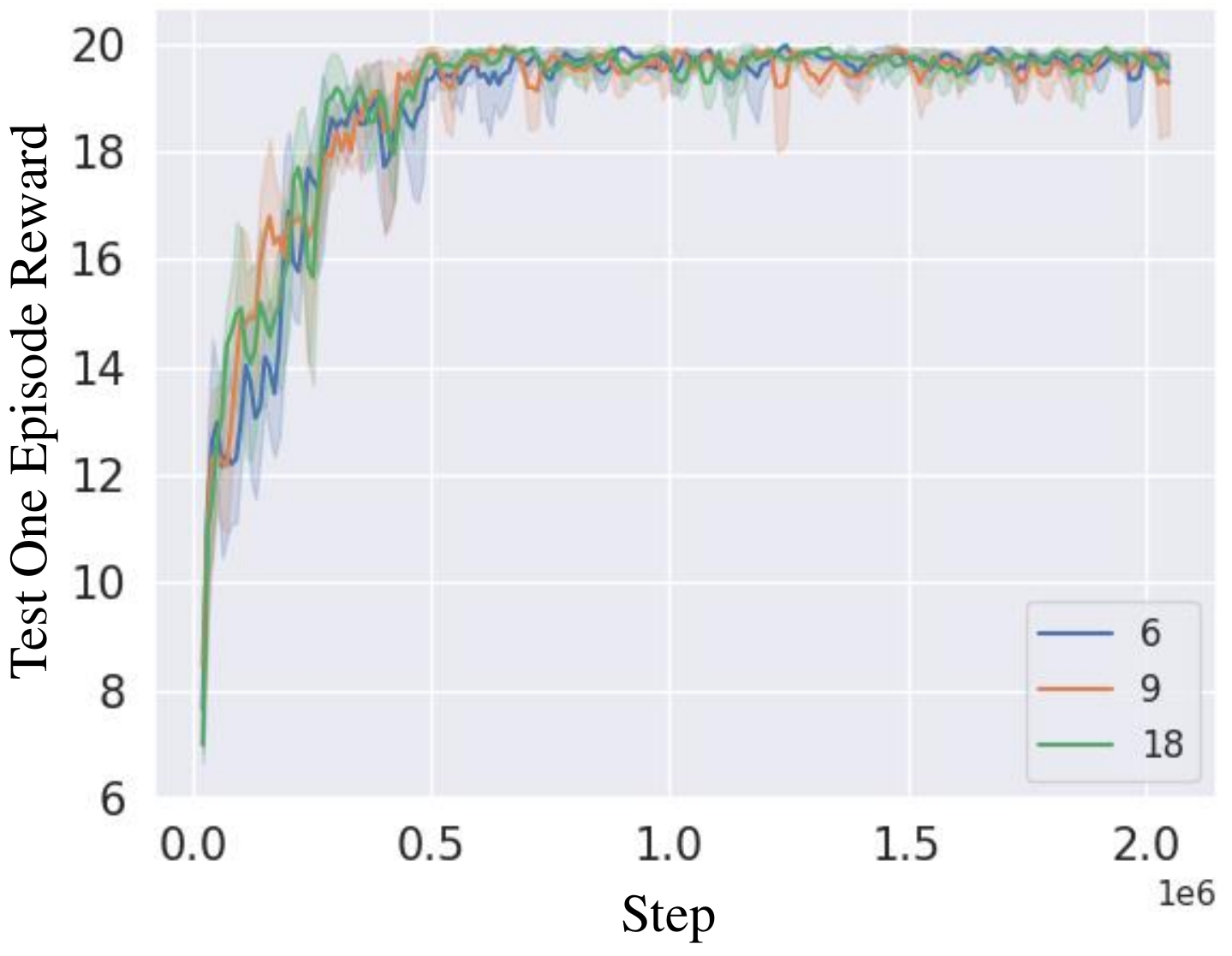}
  \caption{1s1m1h1M\_vs\_3z}
  \label{chart4:1s1m1h1M_vs_3z}
\end{subfigure}\hfil 
\begin{subfigure}{0.3\textwidth}
  \includegraphics[width=\linewidth]{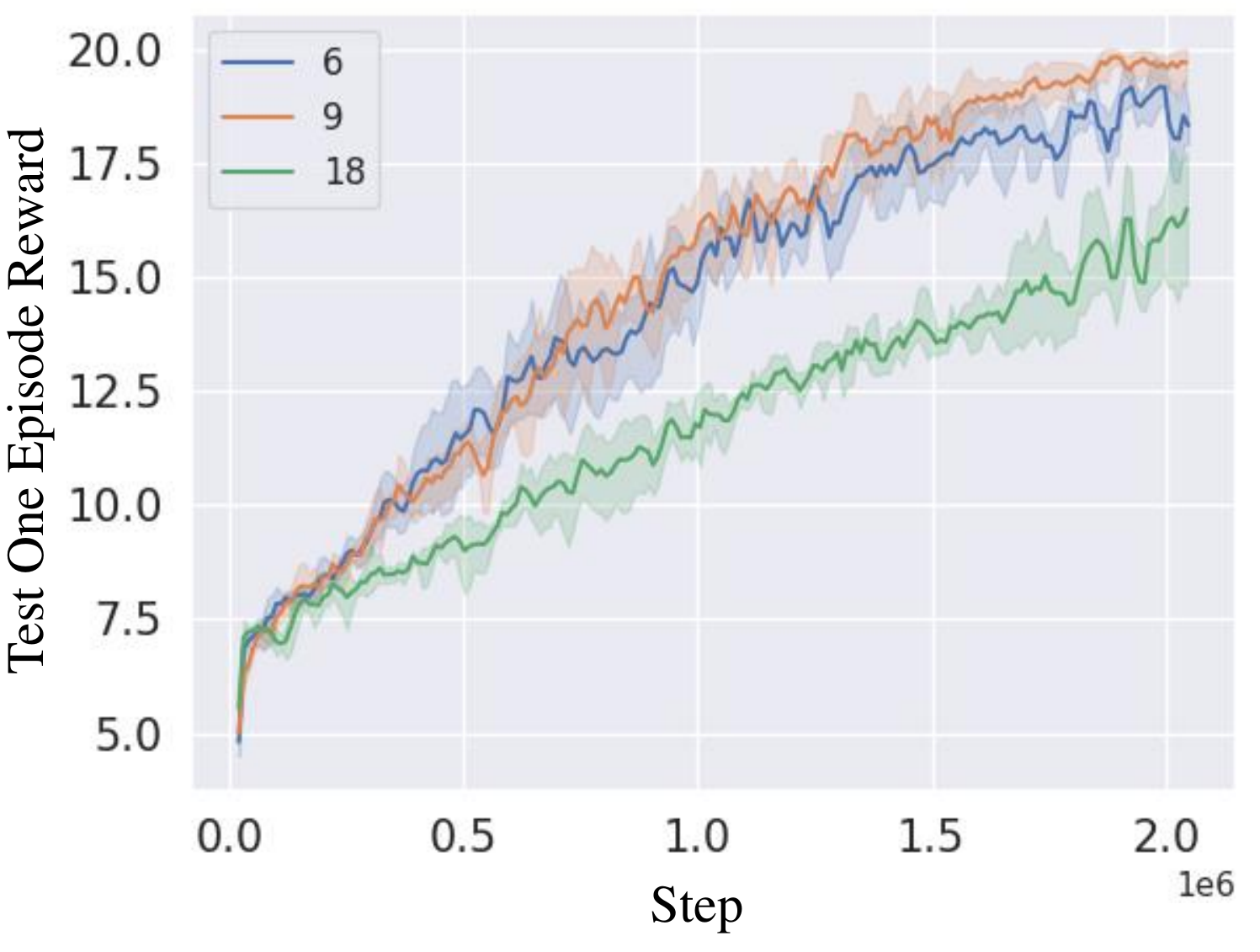}
  \caption{1s1m1h1M\_vs\_4z}
  \label{chart4:1s1m1h1M_vs_4z}
\end{subfigure}\hfil 
\begin{subfigure}{0.3\textwidth}
  \includegraphics[width=\linewidth]{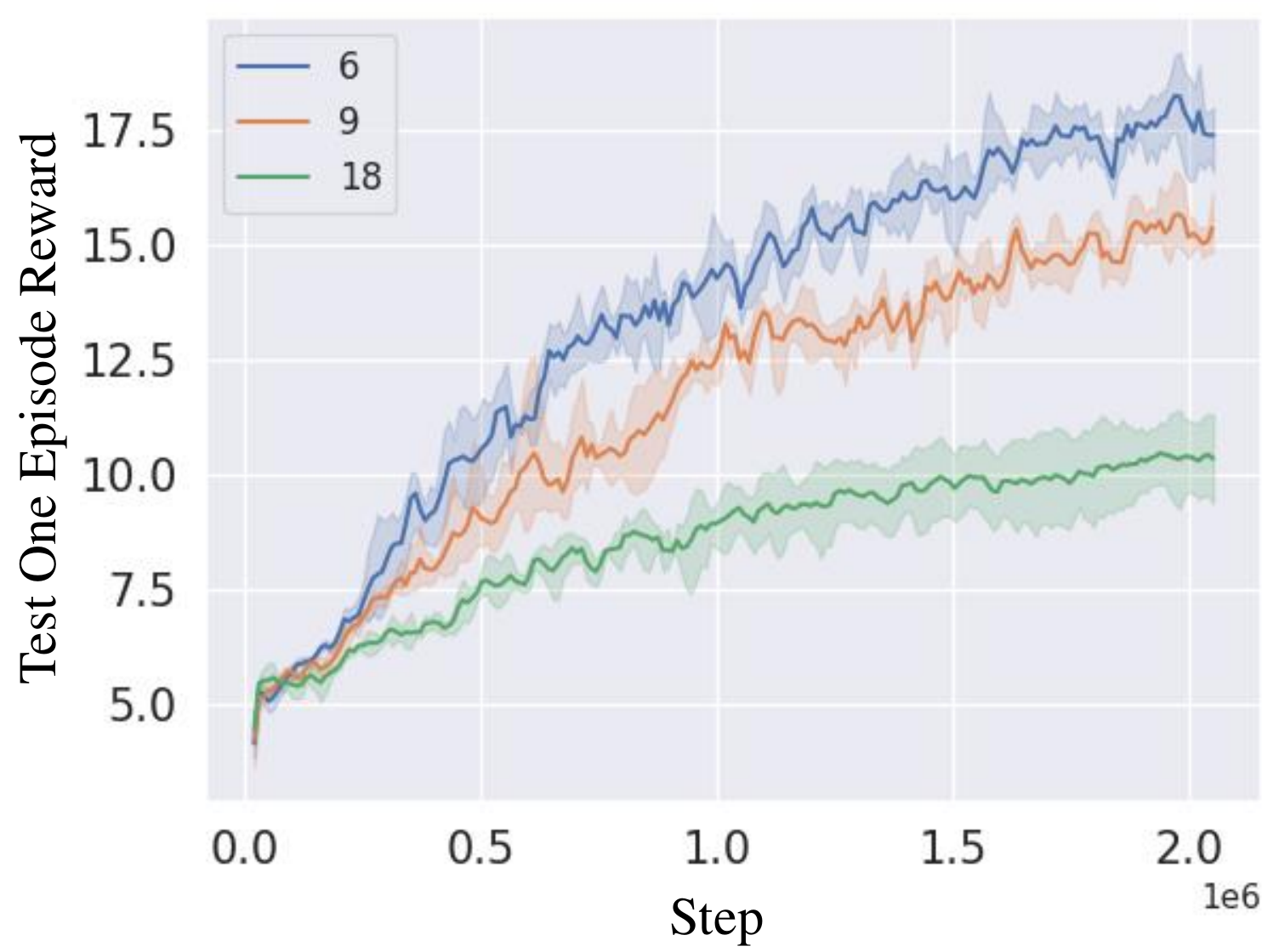}
  \caption{1s1m1h1M\_vs\_5z}
  \label{chart4:1s1m1h1M_vs_5z}
\end{subfigure}\hfil 
\caption{Policy learning curve with different vision scope (6-9-18).}
  \label{fig:vision_scope_curve}
\end{figure}

\begin{figure}[!h]
    \centering 
\begin{subfigure}{0.25\textwidth}
  \includegraphics[width=\linewidth]{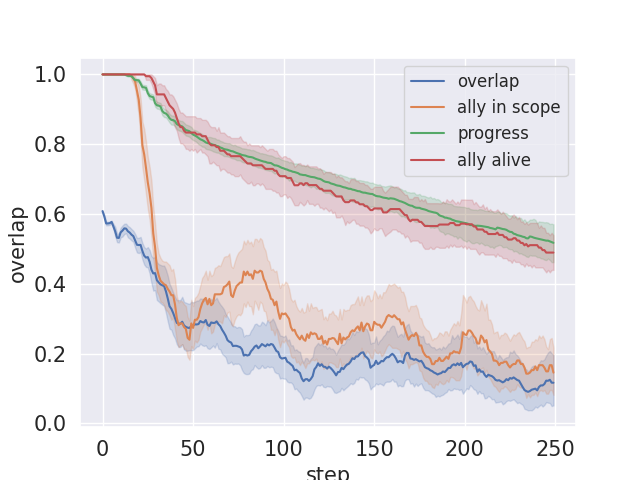}
  \caption{3s\_vs\_5z}
  \label{fig:role_diversity_curve_traj:3s_vs_5z}
\end{subfigure}\hfil 
\begin{subfigure}{0.25\textwidth}
  \includegraphics[width=\linewidth]{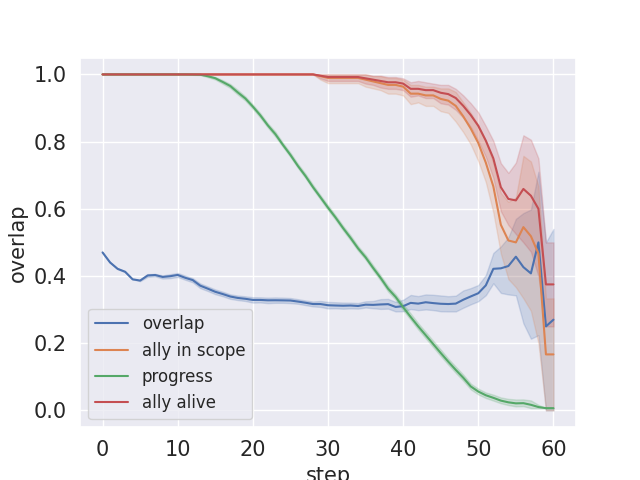}
  \caption{4m\_vs\_3z}
  \label{fig:role_diversity_curve_traj:4m_vs_3z}
\end{subfigure}\hfil 
\begin{subfigure}{0.25\textwidth}
  \includegraphics[width=\linewidth]{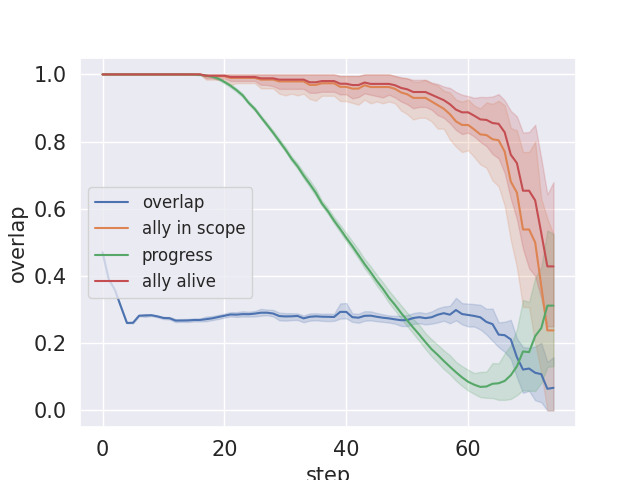}
  \caption{4m\_vs\_4z}
  \label{fig:role_diversity_curve_traj:4m_vs_4z}
\end{subfigure}\hfil 
\begin{subfigure}{0.25\textwidth}
  \includegraphics[width=\linewidth]{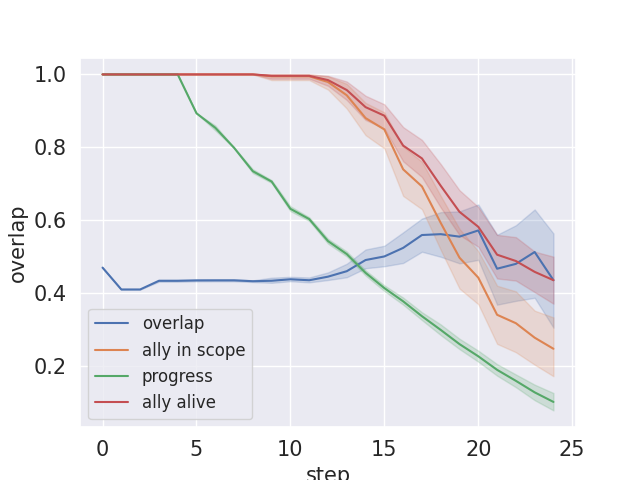}
  \caption{4m\_vs\_5m}
  \label{fig:role_diversity_curve_traj:4m_vs_5m}
\end{subfigure}\hfil 
\medskip
\begin{subfigure}{0.25\textwidth}
  \includegraphics[width=\linewidth]{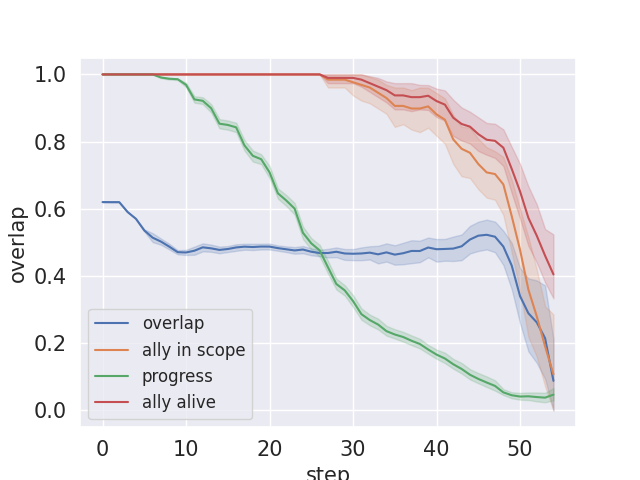}
  \caption{1c1s1z\_vs\_1c1s3z}
  \label{fig:role_diversity_curve_traj:1c1s1z_vs_1c1s3z}
\end{subfigure}\hfil 
\begin{subfigure}{0.25\textwidth}
  \includegraphics[width=\linewidth]{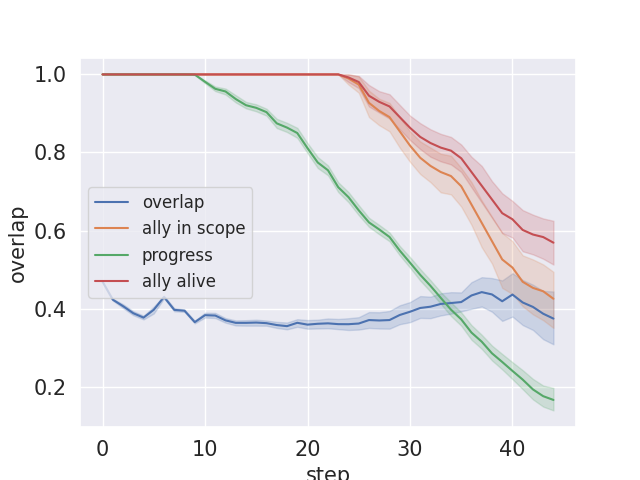}
  \caption{1s1m1h1M\_vs\_3z}
  \label{fig:role_diversity_curve_traj:1s1m1h1M_vs_3z}
\end{subfigure}\hfil 
\begin{subfigure}{0.25\textwidth}
  \includegraphics[width=\linewidth]{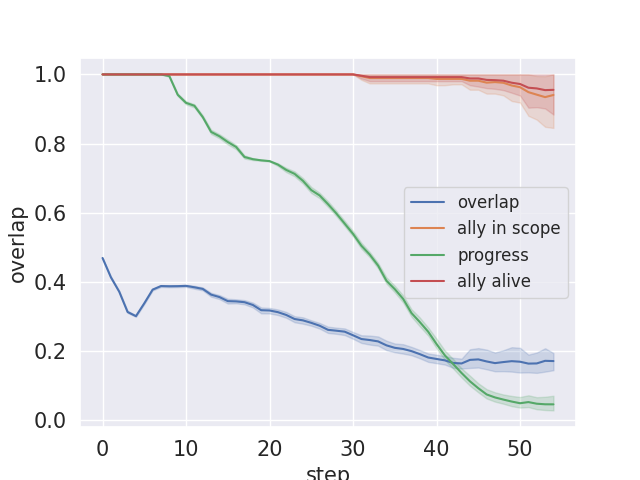}
  \caption{1s1m1h1M\_vs\_4z}
  \label{fig:role_diversity_curve_traj:1s1m1h1M_vs_4z}
\end{subfigure}\hfil 
\begin{subfigure}{0.25\textwidth}
  \includegraphics[width=\linewidth]{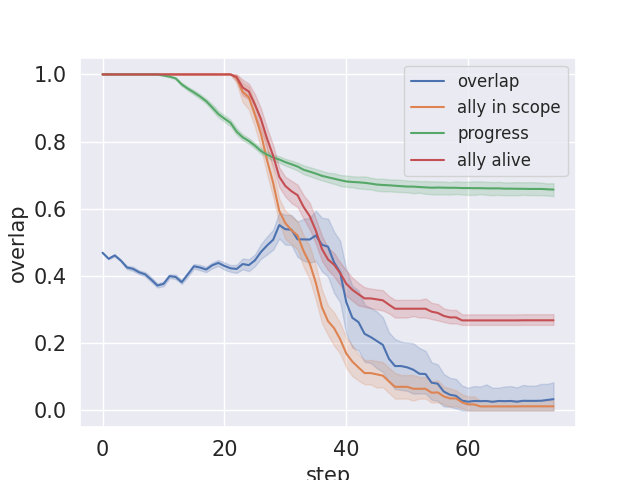}
  \caption{1s1m1h1M\_vs\_5z}
  \label{fig:role_diversity_curve_traj:1s1m1h1M_vs_5z}
\end{subfigure}\hfil 
\caption{Observation overlap curve of one episode game on different battle scenarios. The policy is trained using VDN\cite{sunehag2017value} and no parameter sharing. We also provide the curve of {\it game progress}(equals to the enemy health), {\it ally in scope} and {\it ally alive}. All values are normalized from 0 to 1.}
  \label{fig:role_diversity_curve_traj}
\end{figure}

\begin{figure}[!h]
    \centering 
\begin{subfigure}{0.25\textwidth}
  \includegraphics[width=\linewidth]{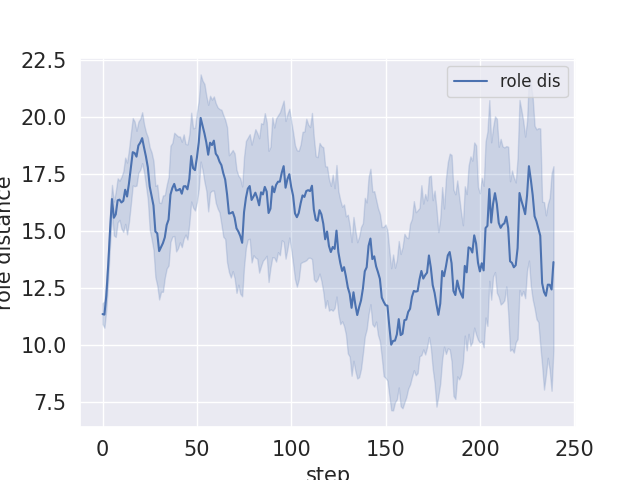}
  \caption{3s\_vs\_5z}
  \label{fig:role_diversity_curve_real_policy:3s_vs_5z}
\end{subfigure}\hfil 
\begin{subfigure}{0.25\textwidth}
  \includegraphics[width=\linewidth]{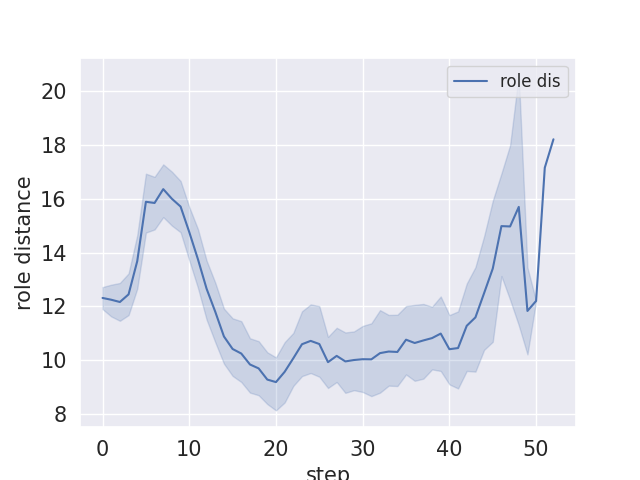}
  \caption{4m\_vs\_3z}
  \label{fig:role_diversity_curve_real_policy:4m_vs_3z}
\end{subfigure}\hfil 
\begin{subfigure}{0.25\textwidth}
  \includegraphics[width=\linewidth]{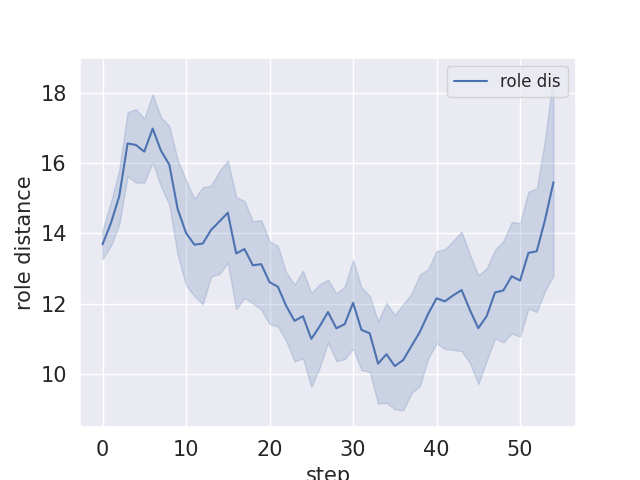}
  \caption{4m\_vs\_4z}
  \label{fig:role_diversity_curve_real_policy:4m_vs_4z}
\end{subfigure}\hfil 
\begin{subfigure}{0.25\textwidth}
  \includegraphics[width=\linewidth]{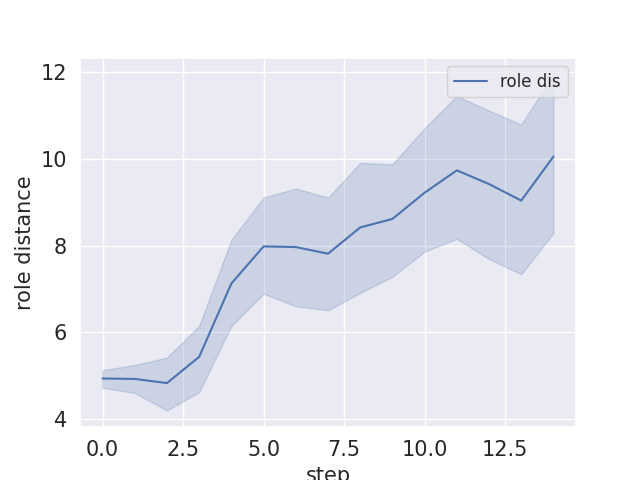}
  \caption{4m\_vs\_5m}
  \label{fig:role_diversity_curve_real_policy:4m_vs_5m}
\end{subfigure}\hfil 
\medskip
\begin{subfigure}{0.25\textwidth}
  \includegraphics[width=\linewidth]{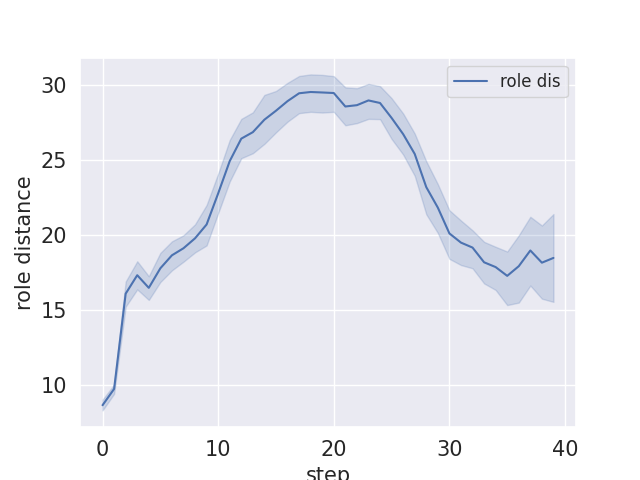}
  \caption{1c1s1z\_vs\_1c1s3z}
  \label{fig:role_diversity_curve_real_policy:1c1s1z_vs_1c1s3z}
\end{subfigure}\hfil 
\begin{subfigure}{0.25\textwidth}
  \includegraphics[width=\linewidth]{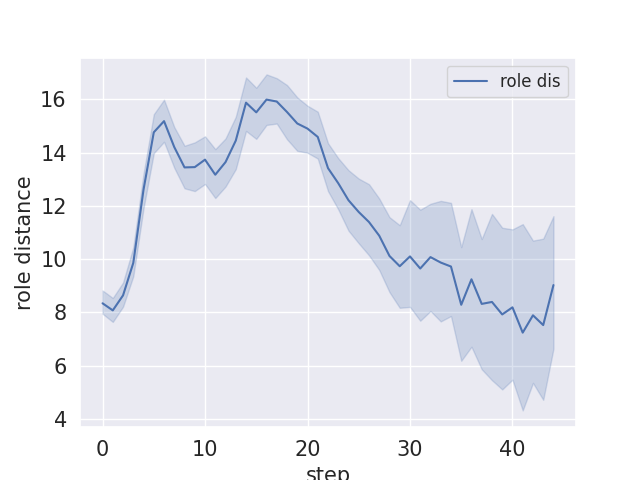}
  \caption{1s1m1h1M\_vs\_3z}
  \label{fig:role_diversity_curve_real_policy:1s1m1h1M_vs_3z}
\end{subfigure}\hfil 
\begin{subfigure}{0.25\textwidth}
  \includegraphics[width=\linewidth]{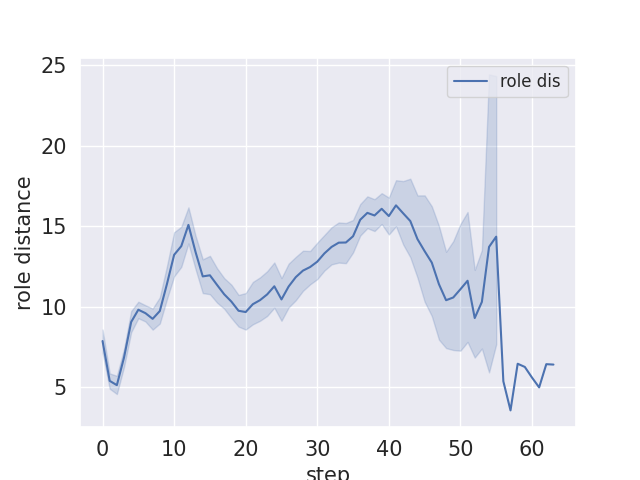}
  \caption{1s1m1h1M\_vs\_4z}
  \label{fig:role_diversity_curve_real_policy:1s1m1h1M_vs_4z}
\end{subfigure}\hfil 
\begin{subfigure}{0.25\textwidth}
  \includegraphics[width=\linewidth]{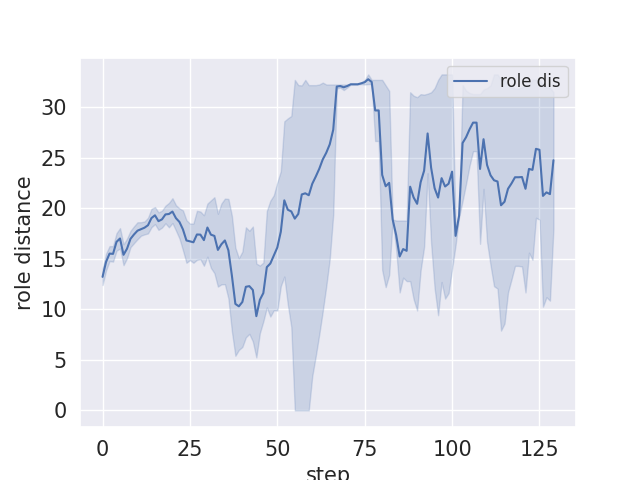}
  \caption{1s1m1h1M\_vs\_5z}
  \label{fig:role_diversity_curve_real_policy:1s1m1h1M_vs_5z}
\end{subfigure}\hfil 
\caption{Policy based role diversity(real) in one episode.}
  \label{fig:role_diversity_curve_real_policy}
\end{figure}

\begin{figure}[!h]
    \centering 
\begin{subfigure}{0.25\textwidth}
  \includegraphics[width=\linewidth]{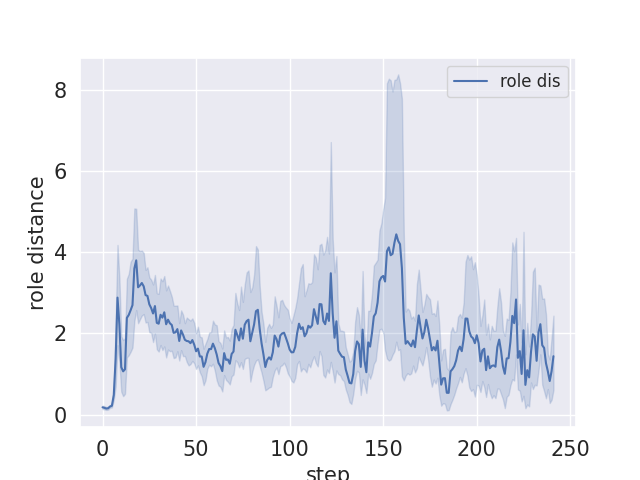}
  \caption{3s\_vs\_5z}
  \label{fig:role_diversity_curve_policy:3s_vs_5z}
\end{subfigure}\hfil 
\begin{subfigure}{0.25\textwidth}
  \includegraphics[width=\linewidth]{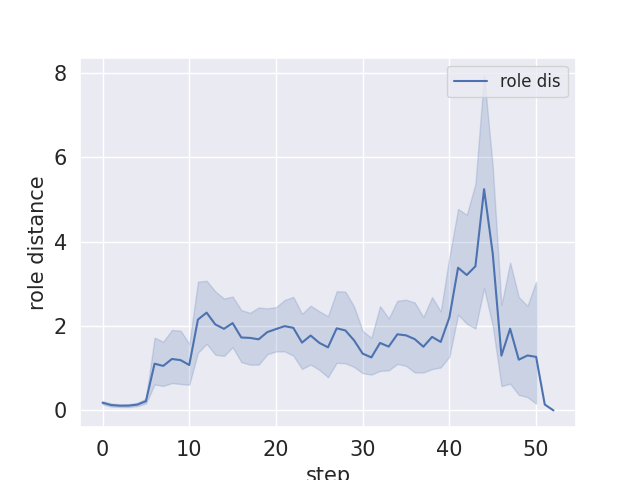}
  \caption{4m\_vs\_3z}
  \label{fig:role_diversity_curve_policy:4m_vs_3z}
\end{subfigure}\hfil 
\begin{subfigure}{0.25\textwidth}
  \includegraphics[width=\linewidth]{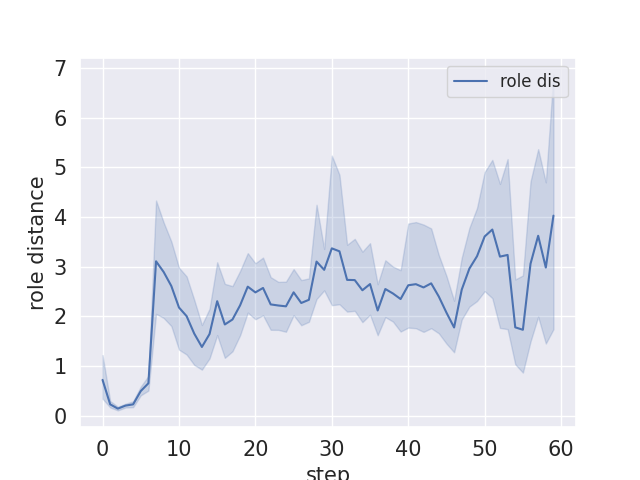}
  \caption{4m\_vs\_4z}
  \label{fig:role_diversity_curve_policy:4m_vs_4z}
\end{subfigure}\hfil 
\begin{subfigure}{0.25\textwidth}
  \includegraphics[width=\linewidth]{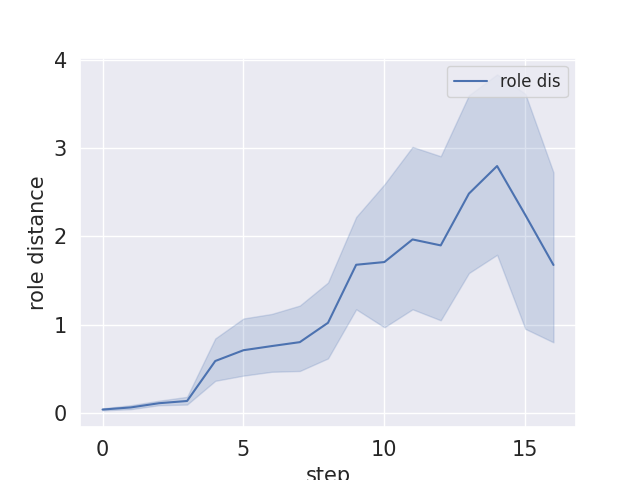}
  \caption{4m\_vs\_5m}
  \label{fig:role_diversity_curve_policy:4m_vs_5m}
\end{subfigure}\hfil 
\medskip
\begin{subfigure}{0.25\textwidth}
  \includegraphics[width=\linewidth]{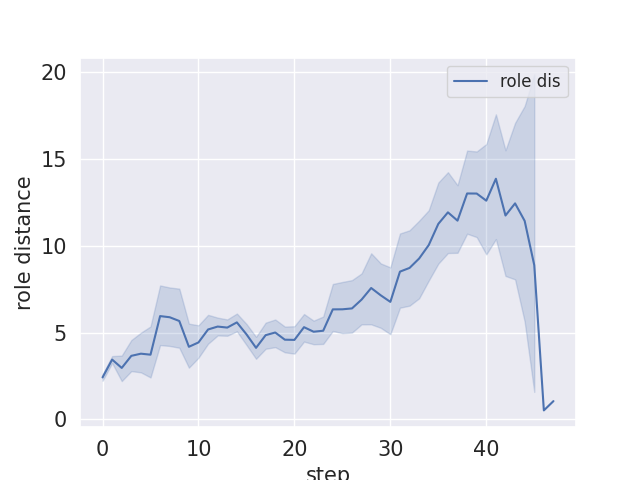}
  \caption{1c1s1z\_vs\_1c1s3z}
  \label{fig:role_diversity_curve_policy:1c1s1z_vs_1c1s3z}
\end{subfigure}\hfil 
\begin{subfigure}{0.25\textwidth}
  \includegraphics[width=\linewidth]{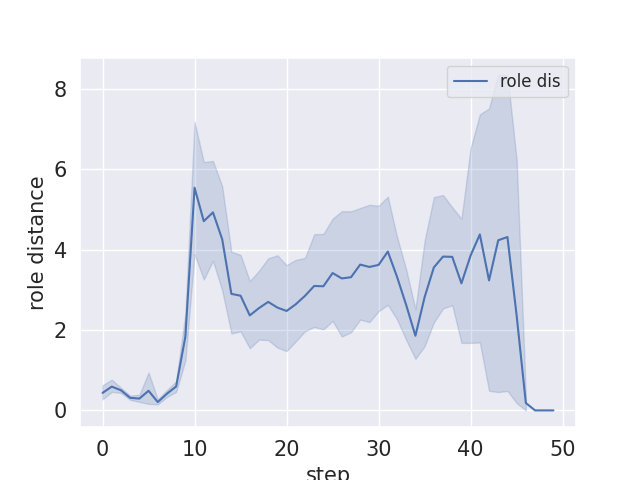}
  \caption{1s1m1h1M\_vs\_3z}
  \label{fig:role_diversity_curve_policy:1s1m1h1M_vs_3z}
\end{subfigure}\hfil 
\begin{subfigure}{0.25\textwidth}
  \includegraphics[width=\linewidth]{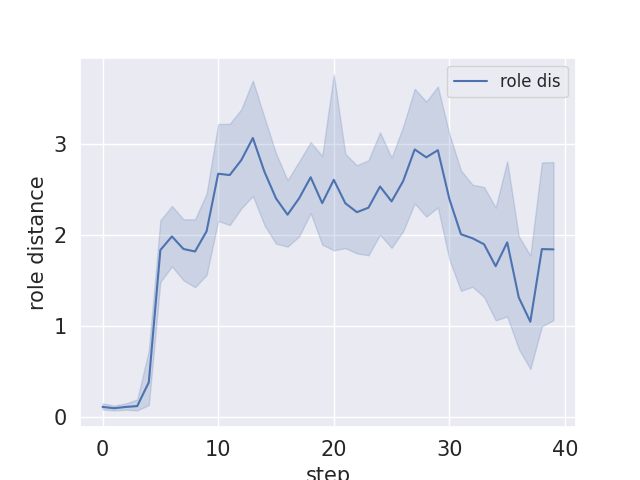}
  \caption{1s1m1h1M\_vs\_4z}
  \label{fig:role_diversity_curve_policy:1s1m1h1M_vs_4z}
\end{subfigure}\hfil 
\begin{subfigure}{0.25\textwidth}
  \includegraphics[width=\linewidth]{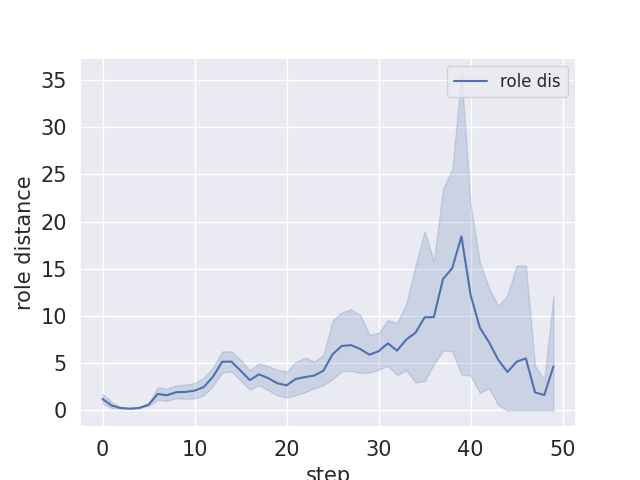}
  \caption{1s1m1h1M\_vs\_5z}
  \label{fig:role_diversity_curve_policy:1s1m1h1M_vs_5z}
\end{subfigure}\hfil 
\caption{Policy based role diversity(semantic) in one episode.}
  \label{fig:role_diversity_curve_policy}
\end{figure}

\begin{figure}[!h]
    \centering 
\begin{subfigure}{0.25\textwidth}
  \includegraphics[width=\linewidth]{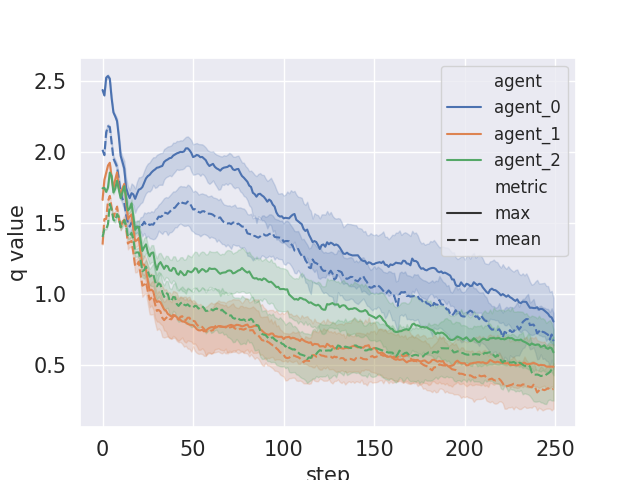}
  \caption{3s\_vs\_5z}
  \label{fig:q_diversity_curve_policy:3s_vs_5z}
\end{subfigure}\hfil 
\begin{subfigure}{0.25\textwidth}
  \includegraphics[width=\linewidth]{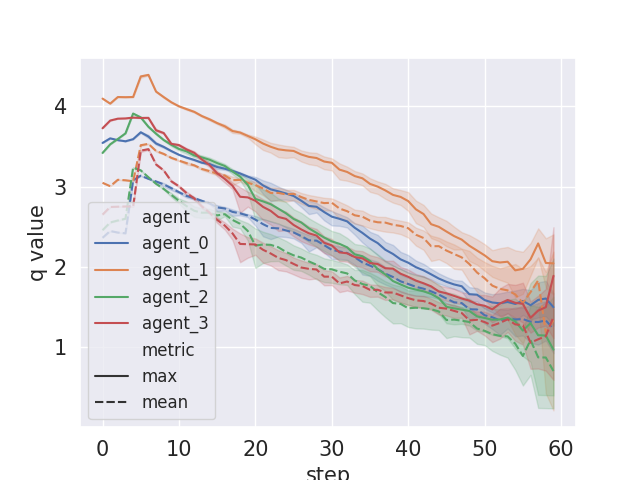}
  \caption{4m\_vs\_3z}
  \label{fig:q_diversity_curve_policy:4m_vs_3z}
\end{subfigure}\hfil 
\begin{subfigure}{0.25\textwidth}
  \includegraphics[width=\linewidth]{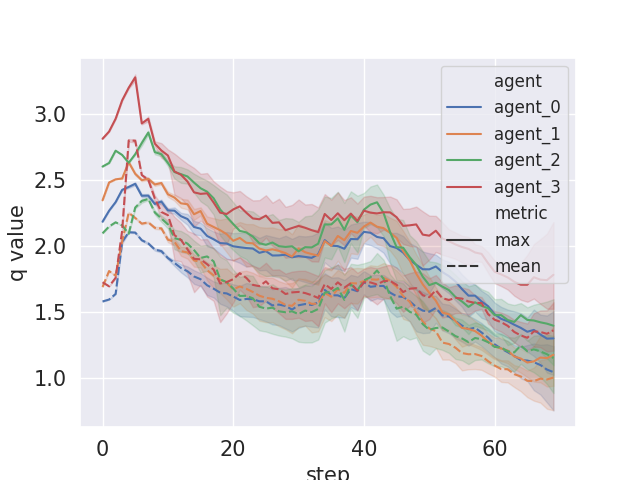}
  \caption{4m\_vs\_4z}
  \label{fig:q_diversity_curve_policy:4m_vs_4z}
\end{subfigure}\hfil 
\begin{subfigure}{0.25\textwidth}
  \includegraphics[width=\linewidth]{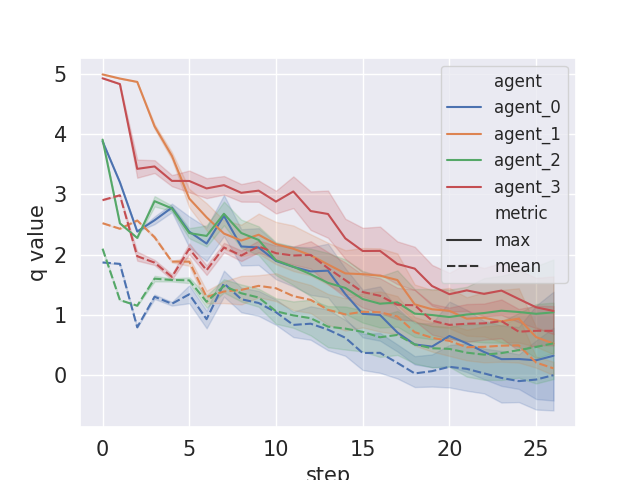}
  \caption{4m\_vs\_5m}
  \label{fig:q_diversity_curve_policy:4m_vs_5m}
\end{subfigure}\hfil 
\medskip
\begin{subfigure}{0.25\textwidth}
  \includegraphics[width=\linewidth]{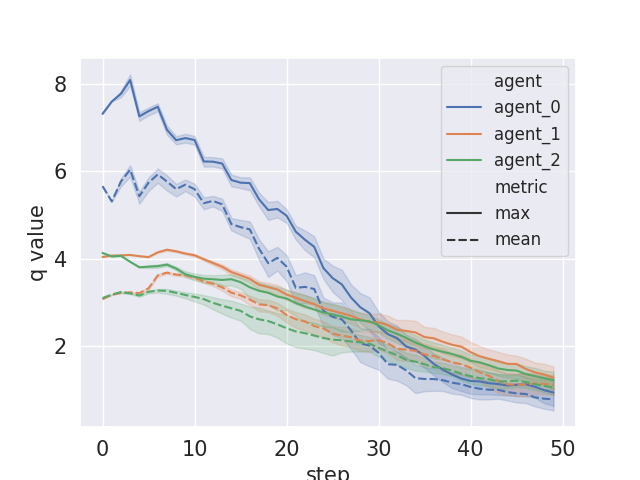}
  \caption{1c1s1z\_vs\_1c1s3z}
  \label{fig:q_diversity_curve_policy:1c1s1z_vs_1c1s3z}
\end{subfigure}\hfil 
\begin{subfigure}{0.25\textwidth}
  \includegraphics[width=\linewidth]{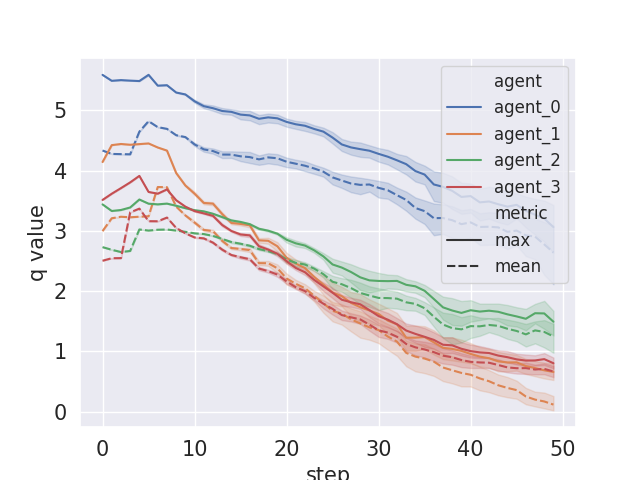}
  \caption{1s1m1h1M\_vs\_3z}
  \label{fig:q_diversity_curve_policy:1s1m1h1M_vs_3z}
\end{subfigure}\hfil 
\begin{subfigure}{0.25\textwidth}
  \includegraphics[width=\linewidth]{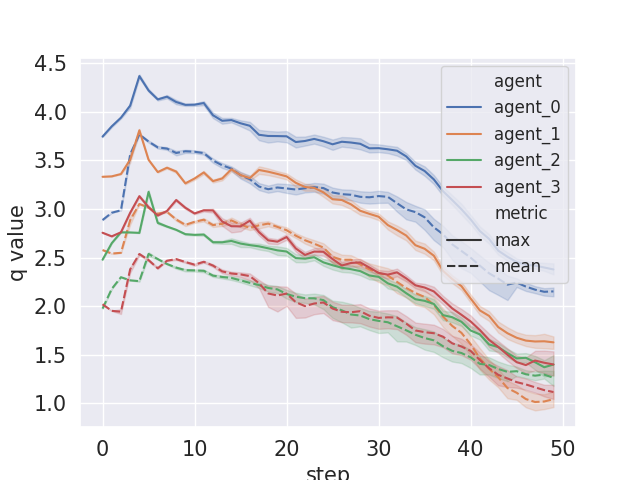}
  \caption{1s1m1h1M\_vs\_4z}
  \label{fig:q_diversity_curve_policy:1s1m1h1M_vs_4z}
\end{subfigure}\hfil 
\begin{subfigure}{0.25\textwidth}
  \includegraphics[width=\linewidth]{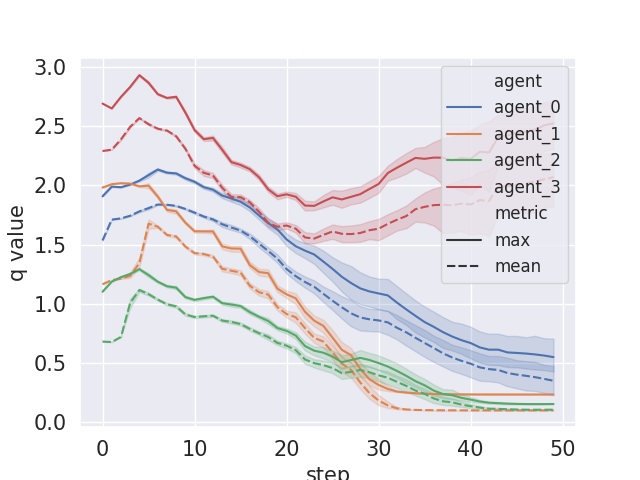}
  \caption{1s1m1h1M\_vs\_5z}
  \label{fig:q_diversity_curve_policy:1s1m1h1M_vs_5z}
\end{subfigure}\hfil 
\caption{$Q$ value curve in one episode on different scenarios.}
  \label{fig:q_diversity_curve_policy}
\end{figure}

\end{document}